\documentclass{JFM-FLM_Au}
\usepackage[utf8]{inputenc}
\usepackage{gensymb}
\usepackage{graphicx}
\usepackage{amsmath}
\usepackage[version=4]{mhchem}
\usepackage{siunitx}
\usepackage{longtable,tabularx}
\usepackage{float}
\usepackage{booktabs}
\usepackage{bm}
\usepackage{caption}
\usepackage{comment}
\usepackage{subcaption}
\usepackage{tikz}
\usetikzlibrary{shapes.geometric, arrows.meta, positioning, shadows.blur, decorations.pathmorphing}

\lefttitle{Bingfei Yan, Jennifer Franck}
\righttitle{Journal of Fluid Mechanics}

\title{Vortex gust interactions with a freely-flying rigid airfoil}

\author{Bingfei Yan\aff{1}, Jennifer A. Franck\aff{1}}

\affiliation{\aff{1}Dept. Mechanical Engineering, University of Wisconsin-Madison, Madison, WI, USA}

\corresau{Bingfei Yan, \email{byan44@wisc.edu}}

\begin{document}

\maketitle

\begin{abstract}
This study numerically investigates the interaction between an isolated vortex gust and a freely-flying airfoil, introducing a theoretical framework for interpreting the coupled lift and heave response. This complex and coupled dynamics is important for modern light-weight aircraft where gusts may easily perturb the wing, generating transient changes in trajectory and attitude.  Here, the freely-flying airfoil is modeled with a single degree-of-freedom in heave, and is impacted by an isolated vortex gust generated upstream. Computational results demonstrate that the freely-flying airfoil reaches a maximum heave displacement after vortex impingement and subsequently rebounds with a comparable magnitude. The lift coefficient is then modeled by augmenting the lift from a corresponding stationary airfoil interaction with motion induced contributions associated with the induced angle of attack and added-mass. A comparison of the modeled lift with the simulation data confirms that the dynamics of the airfoil before impingement is dominated by these two terms, however the rebound after impingement is only partially explained by the model since it is also influenced by the gust-induced vortex shedding. Comparisons across various parameters show that the pre-impingement motion depends primarily on vortex rotation direction, whereas the post-impingement and induced shedding patterns vary with respect to angle of attack and vortex transverse position. With the lift coefficient of the corresponding stationary airfoil interaction as an input, the model can successfully predict the heave trajectory, thus providing a mechanism to assess the dynamic motion of an airfoil from experimental/computational data of gusts interacting with fixed airfoils. 
\end{abstract}

\section{Introduction} \label{intro}

This research investigates the interaction between a freely-flying airfoil and an impinging vortex gust. Small and lightweight aircraft operating in unsteady environments often encounter gusts that can produce sharp aerodynamic transients. Canonical gust models, including transverse, streamwise, and vortex gusts, have been introduced to isolate the fundamental mechanisms governing unsteady airfoil force response \citep{jones2020gust}. Among these, vortex gusts are particularly challenging because they contain compact regions of concentrated vorticity and produce localized, transient, and direction-dependent disturbances to the surrounding flow. Understanding how the trajectory, strength, and orientation of such vortical disturbances influence airfoil loading is essential for predicting aerodynamic response in highly unsteady environments. This study is focused on the heave characteristics of an airfoil in response to a vortex gust, how this motion can be modeled, and identifying the primary mechanisms of the heave motion.
\par
Previous research has long sought to model the unsteady lift generated by canonical transverse and streamwise gust encounters. For periodic gusts, Greenberg-type formulations \citep{greenberg1947airfoil} and related potential-flow theories provide analytical predictions for the lift response of an airfoil subjected to periodic variations in the incoming velocity. \citet{goldstein1976complete} developed a second-order theory for the unsteady incompressible flow around an airfoil encountering a convected sinusoidal gust, explicitly accounting for the distortion of the gust by the steady potential flow around the airfoil. Experimental measurements for oscillating airfoils in sinusoidal streamwise gusts indicate that the resulting lift trends are broadly consistent with Greenberg-type predictions \citep{ma2021unsteady}. More recently, \citet{wang2024airfoil} examined the airfoil response to both periodic transverse and streamwise gusts, comparing force and flow measurements with the classical theories of Atassi and Greenberg. Low-order models based on incompressible potential flow theory have also been used to design pitch motions for mitigating transverse gust loads~\citep{andreu2023controlling}. These studies demonstrate the value of analytical and reduced-order models for connecting gust kinematics to aerodynamic forces. 

Compact vortex gusts, however, introduce additional complications beyond those present in idealized transverse or streamwise gusts. The disturbance is spatially localized, contains concentrated vorticity, and can directly interact with the airfoil boundary layer. As a result, the transient lift may depend not only on the imposed gust-induced velocity, but also on vortex splitting, vortex--boundary-layer interaction, and induced vortex shedding. Numerical studies have strongly contributed to the understanding of vortex gusts interacting with airfoils, particularly by enabling precise control of vortex strength, core size, orientation, and position. For instance, \citet{barnes2018clockwise,barnes2018counterclockwise} applied large-eddy simulations (LES) to examine the effects of Taylor vortices at a chord-based Reynolds number of \(Re_c=200{,}000\), showing that both the rotational direction and vertical position of the incoming vortex significantly influence boundary-layer development and instantaneous force. Subsequent studies on counter-clockwise vortical gust interactions showed that the angle of attack and gust core size can substantially modify boundary-layer disruption and leading-edge vortex formation \citep{barnes2020angle}. \citet{martinez2020analysis} investigated similar interactions at low Reynolds numbers using direct numerical simulation (DNS), identifying scaling relationships between vortex properties and key features of the lift response, including peak force and response timing. More recently, extreme vortex-gust encounters have been considered using LES, revealing that sufficiently strong gusts can trigger three-dimensional flow instabilities and that gust size, location, and rotational direction strongly affect the lift \citep{fukami2025extreme}. 

Experimental studies have also provided critical insight into the physical mechanisms of vortex--airfoil and blade--vortex interactions. \citet{peng2015vortex,peng2017asymmetric} implemented PIV and unsteady pressure measurements to study parallel blade--vortex interactions, showing that the aerodynamic response depends strongly on separation distance, vortex sense, and the side of the airfoil on which the vortex passes. \citet{hufstedler2019vortical} tested the generation of isolated vortical gusts using pitching- and heaving-based mechanisms, and documented the associated lift and post-gust recovery of a downstream airfoil. Comparisons between transverse- and vortex-gust encounters have shown that both can produce large lift transients, while predictions based on the gust-induced angle of attack agree well with experiments only when the induced angle of attack remains small \citep{biler2021experimental}. Related experimental studies have also considered finite-wing effects and lift mitigation strategies. For isolated vortical gusts interacting with unswept and swept wings, the peak lift was found to be closely related to the circulation and offset distance of the incoming vortex \citep{qian2022interaction}. 

A recurring observation in these studies is that vortex-gust encounters produce a highly transient lift, characterized by a rapid force peak followed by an overshoot of opposite sign as the vortex passes the airfoil and modifies the near-wall flow. For a lightweight airfoil, such rapid changes in aerodynamic loading can induce substantial heave motion, which may in turn alter the subsequent vortex interaction mechanisms. This two-way coupling motivates further investigation of vortex-gust encounters in configurations where the airfoil motion is not prescribed, but instead arises from the aerodynamic forcing. Despite extensive prior research on vortex gusts, relatively few studies have considered the passive fluid--structure interaction (FSI) associated with gust encounters, particularly the rigid-body motion and/or structural deformation induced by the gust. Among the limited studies on this topic, Chen and Jaworski \citep{chen2020aeroelastic} used potential-flow theory to investigate vortex trajectories and identified initial vortex positions that lead to direct impingement on an elastically mounted airfoil. Barnes and Visbal \citep{barnes2018gust} showed that vortex gusts can generate a sustained pitch-up moment, leading to self-excited limit-cycle oscillations of an elastically mounted airfoil. While these studies provide valuable insights, they examined only a subset of the parameter space and do not offer a systematic exploration of the interaction mechanisms.

This paper presents a two-dimensional computational investigation of isolated vortex-gust interaction with a freely heaving NACA 0012 airfoil at low Reynolds number. The airfoil is initially in equilibrium and is allowed to heave passively in response to the aerodynamic force, while the vortex gust is generated by a prescribed motion of an upstream airfoil. The primary objective is to develop and assess a modeling framework for interpreting and predicting the transient lift and heave of a freely heaving airfoil interacting with a vortex gust. The model baseline is an equivalent stationary airfoil impacted by the same vortex gust, from which the heave-induced contributions are superimposed. The mechanisms of the heave-induced terms are analyzed, while the discrepancies between the model and simulation are interpreted using the instantaneous flow fields. Finally, the framework is used to predict the gust-induced heave trajectory, and the results are compared with the fully coupled simulations to assess the usefulness and limitations of the approach.

\section{Methodology} \label{methodology}

\subsection{Configuration of freely-flying airfoil and gust generation}

\begin{figure}
\centering
\newcommand{\NACAfifteen}[5]{%
  \begin{scope}[shift={(#1,#2)}, rotate=#3, scale=#4]
    \draw[#5] plot[smooth, tension=0.6] coordinates {
      (-0.5000, 0.0000) (-0.4984, 0.0088) (-0.4935, 0.0173)
      (-0.4855, 0.0254) (-0.4743, 0.0331) (-0.4600, 0.0404)
      (-0.4427, 0.0471) (-0.4226, 0.0532) (-0.3997, 0.0586)
      (-0.3743, 0.0633) (-0.3464, 0.0673) (-0.3162, 0.0704)
      (-0.2840, 0.0728) (-0.2500, 0.0743) (-0.2143, 0.0750)
      (-0.1773, 0.0749) (-0.1391, 0.0741) (-0.1000, 0.0725)
      (-0.0603, 0.0704) (-0.0201, 0.0677) ( 0.0201, 0.0645)
      ( 0.0603, 0.0610) ( 0.1000, 0.0570) ( 0.1391, 0.0529)
      ( 0.1773, 0.0485) ( 0.2143, 0.0440) ( 0.2500, 0.0395)
      ( 0.2840, 0.0350) ( 0.3162, 0.0305) ( 0.3464, 0.0262)
      ( 0.3743, 0.0220) ( 0.3997, 0.0181) ( 0.4226, 0.0145)
      ( 0.4427, 0.0113) ( 0.4600, 0.0084) ( 0.4743, 0.0060)
      ( 0.4855, 0.0041) ( 0.4935, 0.0027) ( 0.4984, 0.0019)
      ( 0.5000, 0.0000)
      ( 0.4984,-0.0019) ( 0.4935,-0.0027) ( 0.4855,-0.0041)
      ( 0.4743,-0.0060) ( 0.4600,-0.0084) ( 0.4427,-0.0113)
      ( 0.4226,-0.0145) ( 0.3997,-0.0181) ( 0.3743,-0.0220)
      ( 0.3464,-0.0262) ( 0.3162,-0.0305) ( 0.2840,-0.0350)
      ( 0.2500,-0.0395) ( 0.2143,-0.0440) ( 0.1773,-0.0485)
      ( 0.1391,-0.0529) ( 0.1000,-0.0570) ( 0.0603,-0.0610)
      ( 0.0201,-0.0645) (-0.0201,-0.0677) (-0.0603,-0.0704)
      (-0.1000,-0.0725) (-0.1391,-0.0741) (-0.1773,-0.0749)
      (-0.2143,-0.0750) (-0.2500,-0.0743) (-0.2840,-0.0728)
      (-0.3162,-0.0704) (-0.3464,-0.0673) (-0.3743,-0.0633)
      (-0.3997,-0.0586) (-0.4226,-0.0532) (-0.4427,-0.0471)
      (-0.4600,-0.0404) (-0.4743,-0.0331) (-0.4855,-0.0254)
      (-0.4935,-0.0173) (-0.4984,-0.0088) (-0.5000, 0.0000)
    };
  \end{scope}
}

\newcommand{\NACAtwelve}[5]{%
  \begin{scope}[shift={(#1,#2)}, rotate=#3, scale=#4]
    \draw[#5] plot[smooth, tension=0.6] coordinates {
      (-0.5000, 0.0000) (-0.4984, 0.0070) (-0.4935, 0.0138)
      (-0.4855, 0.0203) (-0.4743, 0.0265) (-0.4600, 0.0323)
      (-0.4427, 0.0376) (-0.4226, 0.0425) (-0.3997, 0.0469)
      (-0.3743, 0.0507) (-0.3464, 0.0538) (-0.3162, 0.0563)
      (-0.2840, 0.0582) (-0.2500, 0.0594) (-0.2143, 0.0600)
      (-0.1773, 0.0599) (-0.1391, 0.0592) (-0.1000, 0.0580)
      (-0.0603, 0.0563) (-0.0201, 0.0542) ( 0.0201, 0.0516)
      ( 0.0603, 0.0488) ( 0.1000, 0.0456) ( 0.1391, 0.0423)
      ( 0.1773, 0.0388) ( 0.2143, 0.0352) ( 0.2500, 0.0316)
      ( 0.2840, 0.0280) ( 0.3162, 0.0244) ( 0.3464, 0.0210)
      ( 0.3743, 0.0176) ( 0.3997, 0.0145) ( 0.4226, 0.0116)
      ( 0.4427, 0.0090) ( 0.4600, 0.0067) ( 0.4743, 0.0048)
      ( 0.4855, 0.0033) ( 0.4935, 0.0022) ( 0.4984, 0.0015)
      ( 0.5000, 0.0013)
      ( 0.4984,-0.0015) ( 0.4935,-0.0022) ( 0.4855,-0.0033)
      ( 0.4743,-0.0048) ( 0.4600,-0.0067) ( 0.4427,-0.0090)
      ( 0.4226,-0.0116) ( 0.3997,-0.0145) ( 0.3743,-0.0176)
      ( 0.3464,-0.0210) ( 0.3162,-0.0244) ( 0.2840,-0.0280)
      ( 0.2500,-0.0316) ( 0.2143,-0.0352) ( 0.1773,-0.0388)
      ( 0.1391,-0.0423) ( 0.1000,-0.0456) ( 0.0603,-0.0488)
      ( 0.0201,-0.0516) (-0.0201,-0.0542) (-0.0603,-0.0563)
      (-0.1000,-0.0580) (-0.1391,-0.0592) (-0.1773,-0.0599)
      (-0.2143,-0.0600) (-0.2500,-0.0594) (-0.2840,-0.0582)
      (-0.3162,-0.0563) (-0.3464,-0.0538) (-0.3743,-0.0507)
      (-0.3997,-0.0469) (-0.4226,-0.0425) (-0.4427,-0.0376)
      (-0.4600,-0.0323) (-0.4743,-0.0265) (-0.4855,-0.0203)
      (-0.4935,-0.0138) (-0.4984,-0.0070) (-0.5000, 0.0000)
    };
  \end{scope}
}

\begin{tikzpicture}[>=Stealth, thick, font=\normalsize]

\draw[->, gray!60, dashed, line width=1.2pt] (5, 0) -- (7, 0) node[right] {$x$};
\draw[->, gray!60, dashed, line width=1.2pt]  (5, 0) -- (5, 2) node[above] {$y$};

\NACAfifteen{-0.5}{-2.5}{0}{1}{fill=violet!70!black, draw=violet!70!black, thick}

\draw[<->, gray] (-1,1.15) -- (5.0,1.15) node[midway, below] {$5c$};

\foreach \ypos/\ang/\cx in {
    -2.30 / -22.0 / -0.5,
    -2 /-45.0 / -0.5,
    -1.5 / -60.0 / -0.5,
     -0.75 /-60.0 / -0.5,
     0. / -45.0 / -0.5,
     0.75 / -45 / -0.5,
     1.5 /-22.0 / -0.5,
     2.2 / -10 / -0.5,
     2.5 / 0 / -0.5
}{
  \NACAfifteen{\cx}{\ypos}{\ang}{1}{violet!70!black, thin}
  \NACAfifteen{\cx}{\ypos}{\ang}{1}{fill=violet!70!black, draw=violet!70!black, opacity=0.25}
}

\NACAtwelve{5.5}{0}{0}{1}{fill=blue, draw=blue!60!black, thick}

\draw[<->, gray] (5.0, 0.33) -- (6.0, 0.33) node[midway, above] {$c$};

\node[right] at (4, -0.75) {Freely-flying airfoil with vortex gust};

\draw[<->, gray] (-1.0,-2.75) -- (0.0,-2.75) node[midway, below] {$c$};

\node[right] at (0.1,-2.4) {Vortex-generating airfoil};

\draw[<->, gray] (-1.2, 0) -- (-1.2,-2.5) node[midway, left] {$2.5c$};

\draw[->, gray] (-3.5,0.4) -- (-2.5,0.4) node[midway, below] {};
\draw[->, gray] (-3.5,0) -- (-2.5,0) node[midway, below] {};
\draw[->, gray] (-3.5,-0.4) -- (-2.5,-0.4) node[midway, below] {$U_{\infty}$};

\shade[inner color=red!90!black, outer color=red!5, opacity=0.95]
    (0.05, -0.10) circle (0.25);

\shade[inner color=red!90!black, outer color=red!5, opacity=0.95]
    (4.7, -0.06) circle (0.25);

\draw[->, red, dotted, opacity=0.5] (0.20,-0.1) -- (4.5,-0.06);

\end{tikzpicture}
\caption{Schematic of the configuration. An upstream airfoil generates a vortex through a prescribed heaving and pitching motion, which subsequently interacts with a downstream airfoil free to heave in the transverse flow direction.}
\label{fig:geo}
\end{figure}
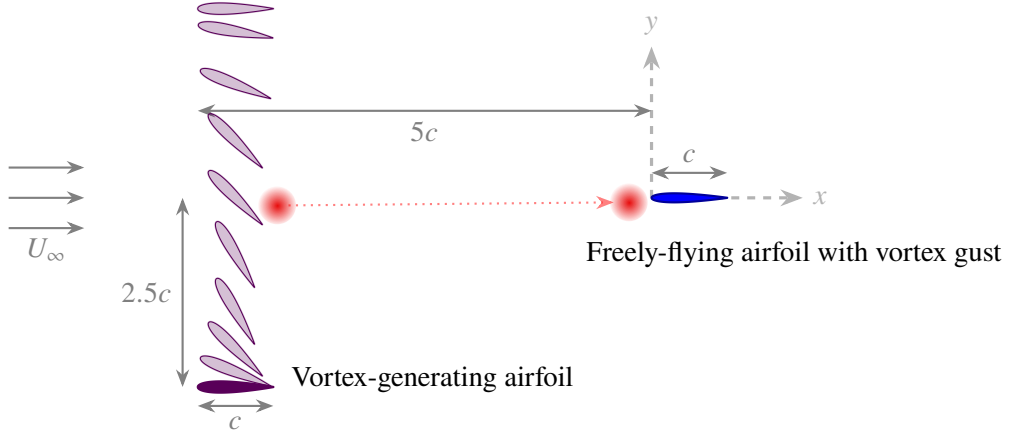

The computational domain consists of two airfoils, as illustrated in Figure \ref{fig:geo}, with a constant freestream flow aligned with the \(x\)-direction. The investigation is primarily focused on the instantaneous heave and lift of the downstream NACA 0012 airfoil, with chord length \(c\), initially positioned at \(y/c=0\). This airfoil is considered "freely-flying" since it is unconstrained in the transverse, \(y\), direction, and will thus have a dynamic heave in response to gust disturbances. For configurations where a positive angle of attack is prescribed, the airfoil is rotated about its leading edge so that the leading edge starts at \(y/c=0\).
\par
The freely-flying airfoil is modeled as a single-degree-of-freedom rigid body. The nondimensional heave displacement, velocity, acceleration, time, and transverse force per unit span are defined as

\begin{equation}
h=\frac{h^*}{c}, \qquad
\dot{h}=\frac{\dot{h^*}}{U_\infty}, \qquad
\ddot{h}=\frac{\ddot{h^*}c}{U_\infty^2}, \qquad
t=\frac{t^* U_\infty}{c},\qquad
F_y=\frac{F_y^*}{\rho_f U_\infty^2 c }, 
\end{equation}
where the asterisk subscript indicates a dimensional quantity, $h^*$ is the heave amplitude, $F_y^*$ is the vertical force, and $\rho_f$ is the density of the fluid. The heave motion is then governed by the following equation of motion

\begin{equation}
\label{eqn:eom}
m \ddot{h}(t) + c_h \dot{h}(t) + k_h h(t) = F_y(t),
\end{equation}
where

\begin{equation}
m
=\frac{A}{c^2}\frac{\rho_s}{\rho_f},
\end{equation}
with \(A\) denoting the airfoil cross-sectional area (or volume per unit span), and \(\rho_s\) the airfoil density. For the NACA0012 airfoil \(A/c^2 \approx 0.0822\). The density ratio is \(\rho_s/\rho_f=10\) unless stated otherwise. 
The coefficients \(c_h\) and \(k_h\) are the nondimensional damping and stiffness in the heave direction, however both are set to zero in the present study, corresponding to a freely-flying airfoil. The force \(F_y\) includes both the fluid force (pressure and shear) and a constant force introduced to maintain equilibrium (steady flight) before the gust encounter, $F_0$.

Vortex gusts are generated by an upstream NACA 0015 airfoil of chord length \(c\), positioned \(5c\) upstream of the freely-flying airfoil. To produce an isolated vortex while minimizing the influence of the continuous wake, the gust-generating airfoil undergoes a prescribed unidirectional heaving and pitching motion as it traverses the domain, following the method introduced in \cite{yan2026generation}. This motion produces a compact vortex that convects downstream and interacts with the freely-flying airfoil. Because the prescribed heave motion rapidly moves the generating airfoil away from the vortex after its formation, the continuous airfoil wake remains spatially separated from the primary vortex and has limited influence on the vortex-airfoil interaction. In contrast to superimposing a vortex profile on the freestream, this approach ensures that the vortex is fully consistent with the background flow and can be implemented directly in experiments. By adjusting the kinematic parameters, vortices of various direction and strength can be generated at different vertical positions. The detailed kinematic equations and parameters used to generate the vortices used in this study are provided in Appendix~\ref{app:vortex_generation}.

For the results reported in this study, \(t=0\) denotes the onset of the gust-generating maneuver of the upstream airfoil. This time origin is chosen so that the vortex impingement time is consistent among simulations with the same vortex orientation, enabling meaningful quantitative comparisons of the transient lift and heave trajectories.
\subsection{Numerical Solver}

The simulations solve the incompressible Navier--Stokes equations at a chord-based Reynolds number of $Re=1000$ with the OpenFOAM open-source libraries. Spatial discretization is performed using a second-order finite-volume method. Time advancement is achieved using a second-order backward method for all cases except the lowest density-ratio, for which a first-order Euler method is used to enhance numerical stability, since the added-mass is increased relative to the structural inertia. Pressure--velocity coupling is handled using the pressure-implicit split operator (PISO) algorithm.

Both of the airfoils in Fig.~\ref{fig:geo} are in motion relative to one another, and relative to the fixed coordinate system. To account for the two-body motion a mesh motion solver is implemented that is governed by the following equation for cell displacement field \(\bm{d}\),

\begin{equation}
\nabla \cdot \left( 2 D_f \nabla \bm{d} \right)
+ \nabla \cdot \left( 
    D_f \left[ 
        \bm{n} \cdot \left( (\nabla \bm{d})^T - \nabla \bm{d} \right) 
        - \bm{n} \, \mathrm{tr}(\nabla \bm{d}) 
    \right]
\right) = 0,
\end{equation}
where the diffusivity constant \(D_f\) controls the rigidity of the mesh deformation. The diffusivity \(D_f\) is scaled inversely with the distance to the airfoil surfaces to preserve mesh structure and quality near the bodies. This mesh morphing technique has been applied and validated in numerous previous studies \citep{ribeiro2021wake}. The motion of both airfoils is imposed through boundary conditions on the mesh displacement field. The vortex-generating airfoil has prescribed motion whereas the freely-flying airfoil heave motion is passively determined by its two-way interaction with the fluid forces at its boundary. 

\begin{figure}[!htbp]
\centering
\resizebox{0.65\textwidth}{!}{%
\tikzset{
  every node/.style={font=\sffamily}
}
\tikzset{
  startstop/.style={
    rectangle,
    rounded corners=10pt,
    draw=orange!70!black,
    very thick,
    fill=orange!25,
    minimum width=4.8cm,
    minimum height=1.2cm,
    align=center,
    blur shadow
  },
  process/.style={
    rectangle,
    rounded corners=8pt,
    draw=blue!70!black,
    very thick,
    fill=cyan!20,
    text width=4.8cm,
    minimum height=1.25cm,
    align=center,
    blur shadow
  },
  decision/.style={
    diamond,
    draw=olive!60!black,
    very thick,
    fill=yellow!25,
    aspect=2.2,
    text width=3.2cm,
    align=center,
    inner sep=1pt,
    blur shadow
  },
  arrow/.style={
    very thick,
    -{Latex[length=4mm,width=2.5mm]},
    draw=black!75
  }
}
\begin{tikzpicture}[node distance=1.2cm and 1.6cm]
\node (structure) [process]
{Solve for heave $h$\\ of the freely-flying airfoil\\ Eq.~(2)};
\node (bcmesh) [process, below=of structure]
{Update foil boundary conditions for mesh motion $\bm{d}^k_{b}$ and flow solver ${\bm{u}^k_b}$};
\node (mesh) [process, below=of bcmesh]
{Solve mesh motion\\ equation, Eq.~(4), for\\ cell displacement $\bm{d}^k$};
\node (flow) [process, right=of mesh]
{Solve Navier-Stokes for the pressure $p^k$ and\\ velocity $\bm{u}^k$};
\node (force) [process, above=of flow]
{Compute the total\\ heave force $F^k_y$ ex-\\ erted on the airfoil};
\node (check) [decision, right=1.8cm of structure, xshift=-0.35cm]
{Convergence: $|F_y^k-F_y^{k-1}|<$ tol?};
\node (start) [startstop] at ([yshift=1.5cm]current bounding box.north)
{Advance timestep};
% Arrows
\draw [arrow] (start.west) -- node[midway, above, xshift=-5pt] {$F_y^{t-1}$} (structure.north);
\draw [arrow] (structure) -- (bcmesh);
\draw [arrow] (bcmesh) -- (mesh);
\draw [arrow] (mesh) -- (flow);
\draw [arrow] (flow) -- (force);
\draw [arrow] (force.north) -- (check.south);
\draw [arrow] (check.west) -- node[midway, above] {\normalsize No} node[midway, below, xshift=4pt] {advance $k$} (structure.east);
\draw [arrow] (check.north) -- node[midway, above, xshift=10pt] {\normalsize Yes} (start.east);
\end{tikzpicture}
}
\caption{Flowchart of the tightly coupled fluid–structure interaction (FSI) algorithm. At each timestep, the fluid and structural solvers iterate until the lift force converges before advancing the simulation.}
\label{fig:coupling}
\end{figure}
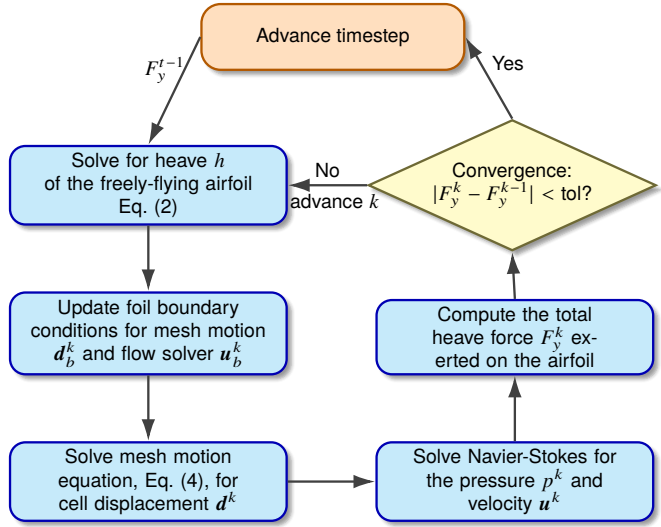

Fluid--structure coupling is achieved through a partitioned, tightly coupled approach. Within each time step, the rigid-body equation of motion is solved with the Newmark-Beta method for displacement, which is imposed as a boundary conditions on the mesh displacement field to update the mesh. The heave motion is also applied to the fluid through the no-slip condition at the airfoil surface. The fluid solver computes the net vertical force on the body \(F_y\), which is fed back to the rigid-body solver. The fluid and rigid-body solvers are iterated until the change in \(F_y\) between successive coupling iterations is less than \(10^{-6}\). The overall coupling procedure is illustrated schematically in Fig.~\ref{fig:coupling}. The numerical implementation has been validated against benchmark cases of vortex-induced vibrations of a circular cylinder, with details provided in Appendix~\ref{app:validation}.

\subsection{Computational domain and mesh independence}
\begin{figure}
     \centering
     \begin{subfigure}[b]{0.26\textwidth}
         \centering
         \includegraphics[width=\textwidth]{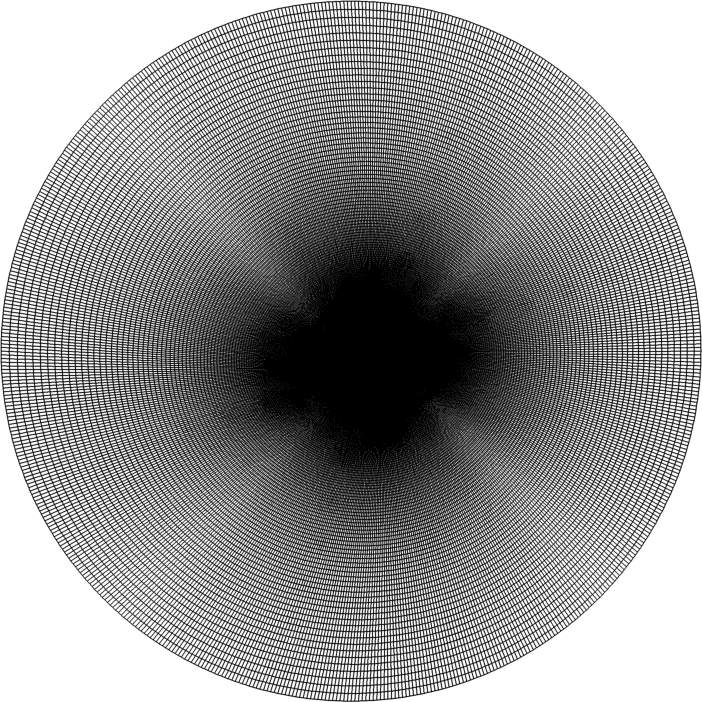}
         \caption{Overall domain}
     \end{subfigure}
     \hfill
     \begin{subfigure}[b]{0.34\textwidth}
         \centering
         \includegraphics[width=\textwidth]{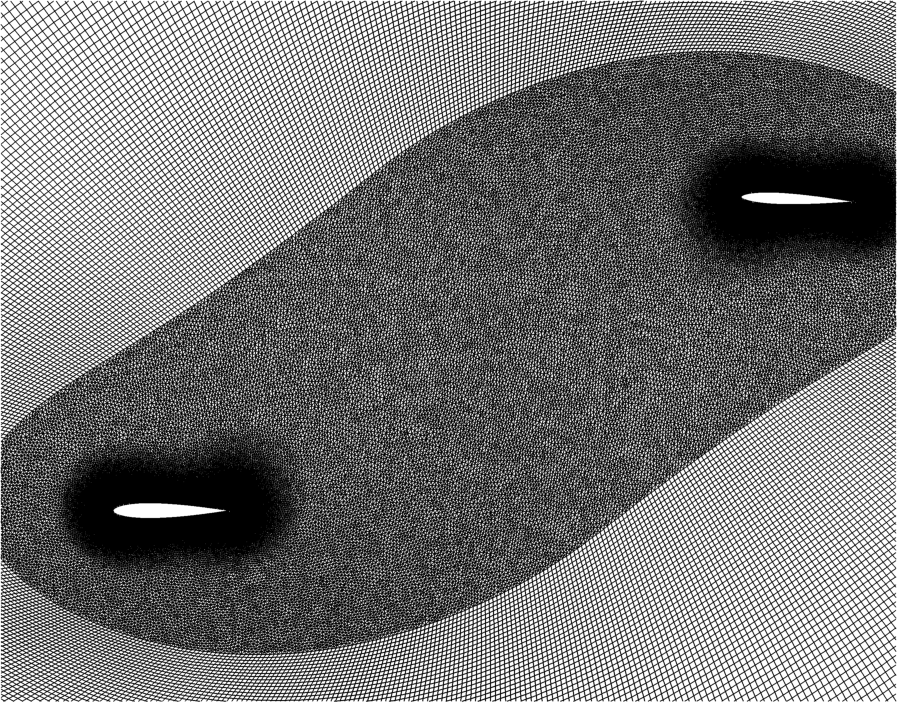}
         \caption{Intermediate mesh region}
     \end{subfigure}
     \hfill
     \begin{subfigure}[b]{0.34\textwidth}
         \centering
         \includegraphics[width=\textwidth]{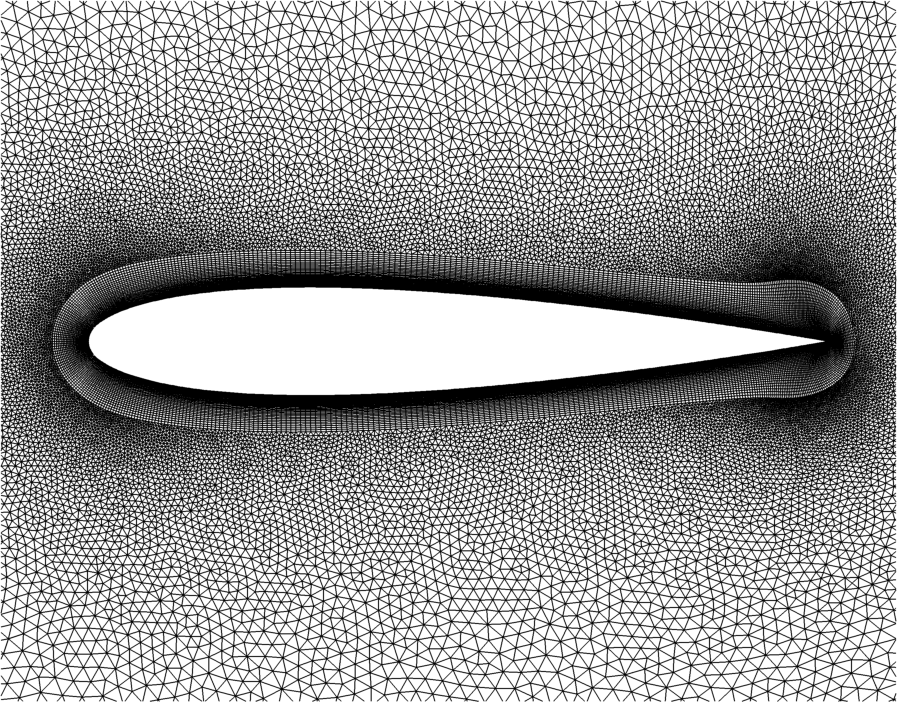}
         \caption{Near-airfoil mesh}
     \end{subfigure}
    \caption{Meshing strategy for the two airfoil domain includes structured body-fitted mesh layers on the airfoils and a structured far-field mesh connected with an unstructured mesh between the airfoils where maximum relative mesh motion occurs.}
    \label{fig:mesh}
\end{figure}

The computational domain contains the two airfoils from Fig.~\ref{fig:geo} in a circle with radius \(50c\) from the vortex-generating airfoil. A uniform inflow condition is prescribed on the upstream portion of the boundary, while a pressure outlet condition is applied on the downstream portion. The mesh, shown in Fig.~\ref{fig:mesh}, is carefully constructed to resolve the flow field between the airfoils and accommodate their relative motion without excessive cell distortion. A structured body-fitted grid is employed on both airfoils to accurately capture boundary-layer behavior and near-field flow features. A separate structured grid is used along the outer boundary to improve numerical stability and allow for mesh stretching in the far-field. These structured regions are connected by an unstructured mesh, which provides flexibility for handling the geometry and mesh deformation.
\par
A mesh resolution study was performed to assess the sensitivity of the solution to the choice of mesh. Five meshes of varying refinement levels were considered, as summarized in Tab.~\ref{tab:grid}. Mesh3 is taken as the reference configuration. Relative to this baseline, Mesh1 and Mesh4 modify the resolution in the unstructured region, $\Delta_{\mathrm{wake}}/c$, between the airfoils, while Mesh2 and Mesh5 vary the resolution of the body-fitted layers, $\Delta y/c$, surrounding the airfoils. All were simulated using the same vortex-generating kinematics over a duration of 20 convective time units. To compare the mesh performance, the maximum and minimum lift coefficient, as well as the maximum absolute heave displacement of the freely-flying airfoil, are recorded and reported in Tab.~\ref{tab:grid}.

\begin{table}
\centering
\setlength{\tabcolsep}{5pt}
\renewcommand{\arraystretch}{1.05}

\begin{tabular}{lccccccc}
\toprule
Mesh
& $N$
& $\Delta_{\mathrm{wake}}/c$
& $N_{\theta}$
& $\Delta y/c$
& $C_{l,\max}$
& $C_{l,\min}$
& $|h|_{\max}/c$ \\
\hline
Mesh 1 & $1.67 \times 10^{5}$ & $4.59 \times 10^{-2}$ & 250 & $8.00 \times 10^{-4}$ & 0.615 & $-0.158$ & 0.226 \\
Mesh 2 & $2.10 \times 10^{5}$ & $3.05 \times 10^{-2}$ & 150 & $1.20 \times 10^{-3}$ & 0.628 & $-0.171$ & 0.228 \\
Mesh 3 & $2.54 \times 10^{5}$ & $3.05 \times 10^{-2}$ & 250 & $8.00 \times 10^{-4}$ & 0.624 & $-0.173$ & 0.228 \\
Mesh 4 & $4.26 \times 10^{5}$ & $2.08 \times 10^{-2}$ & 250 & $8.00 \times 10^{-4}$ & 0.630 & $-0.181$ & 0.228 \\
Mesh 5 & $3.03 \times 10^{5}$ & $3.05 \times 10^{-2}$ & 350 & $6.00 \times 10^{-4}$ & 0.624 & $-0.172$ & 0.227 \\
\bottomrule
\end{tabular}
\caption{Mesh parameters and aerodynamic quantities from the grid-independence study. Here, $N$ is the total number of cells, $\Delta_{\mathrm{wake}}$ is the characteristic cell size between the airfoils, $N_{\theta}$ is the number of grid points along the upper airfoil surfaces, and $\Delta y$ is the first-layer mesh height.}
\label{tab:grid}
\end{table}
\par
The results show that variations in the maximum and minimum lift coefficients across Mesh2–Mesh5 remain within $0.01$, with similarly small differences observed in the maximum heave amplitude. This indicates that the freely-flying airfoil response is effectively grid-independent beyond the resolution of Mesh3. Accordingly, Mesh3 is adopted for all simulations presented in this study as it balances between accuracy and efficiency.

\subsection{Simulation parameters}

The input parameters to the simulations that are varied include the rotation direction and vertical position of the  vortex disturbance, and the angle of attack of the freely-flying foil, and are summarized in Tab.~\ref{tab:config}. The angle of attack of the freely-flying airfoil is varied between \(0^\circ\) and \(5^\circ\). The vortex orientation is prescribed to be either clockwise (CW) or counterclockwise (CCW). The vortex position is classified as 'below', 'direct' or 'above', indicating where the core passes relative to the airfoil. In the 'above' and 'below' configurations the vortex core is approximately \(+0.2c\) and \(-0.2c\), respectively, relative to the leading edge. As a baseline, a corresponding set of simulations is performed with the downstream airfoil held stationary (not freely-flying). 

\begin{table}
\centering

\setlength{\tabcolsep}{8pt}
\renewcommand{\arraystretch}{1.05}

\begin{tabular}{@{}lccc@{}}
\toprule
Label & AoA ($^\circ$) & Vortex orientation & Vortex position \\
\hline 
\multicolumn{4}{@{}l}{{\bf Freely-flying airfoil}}: \vspace{0.05in}
\\
CW0      & 0.0 & CW  & Direct impingement \\
CCW0     & 0.0 & CCW & Direct impingement \\
CW2.5    & 2.5 & CW  & Direct impingement \\
CW2.5a   & 2.5 & CW  & Above foil \\
CW2.5b   & 2.5 & CW  & Below foil \\
CCW2.5   & 2.5 & CCW & Direct impingement \\
CCW2.5a  & 2.5 & CCW & Above foil \\
CCW2.5b  & 2.5 & CCW & Below foil \\
CW5      & 5.0 & CW  & Direct impingement \\
CCW5     & 5.0 & CCW & Direct impingement \\
\hline
\multicolumn{4}{@{}l}{{\bf Stationary  airfoil:}} \vspace{0.05in} \\
CW0-S      & 0.0 & CW  & Direct impingement \\
CCW0-S     & 0.0 & CCW & Direct impingement \\
CW2.5-S    & 2.5 & CW  & Direct impingement \\
CCW2.5-S   & 2.5 & CCW & Direct impingement \\
CW5-S      & 5.0 & CW  & Direct impingement \\
CCW5-S     & 5.0 & CCW & Direct impingement \\
\bottomrule
\end{tabular}
\caption{Summary of vortex--airfoil configurations.}
\label{tab:config}
\end{table}

\begin{figure}
\centering
	\begin{subfigure}[t]{0.48\textwidth}
    \centering
        \includegraphics[width=0.9\textwidth]{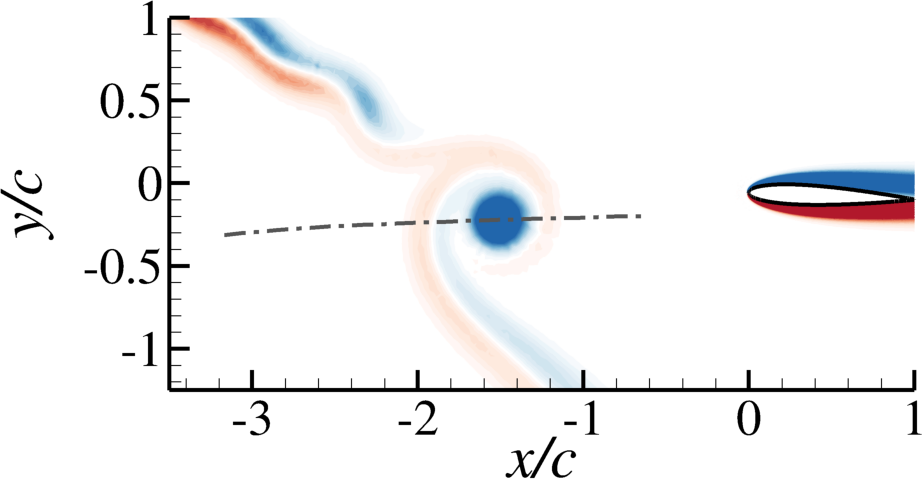}
        \caption{CW2.5: clockwise vortex.}
    \end{subfigure}
    \hspace{0.02\textwidth}
    \begin{subfigure}[t]{0.48\textwidth}
	\centering
        \includegraphics[width=0.9\textwidth]{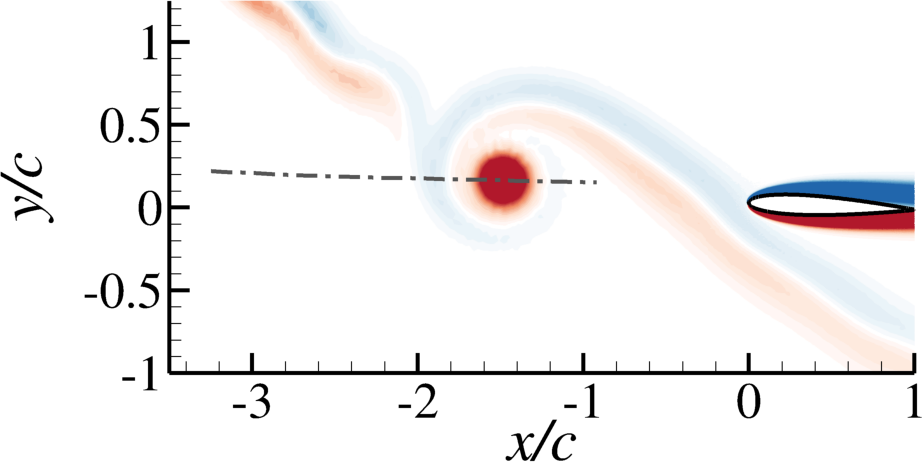}
        \caption{CCW2.5: counter-clockwise vortex.}
	\end{subfigure}
    \\[0.2cm]   % space between rows
    \begin{subfigure}[b]{0.75\textwidth}
	\centering
        \includegraphics[width=1\textwidth]{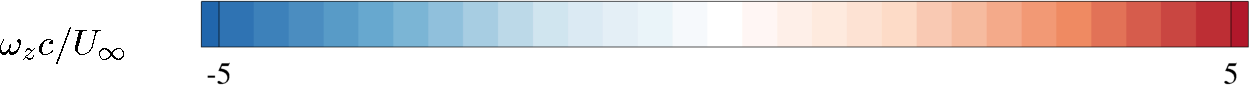}
	\end{subfigure}
\caption{Spanwise vorticity contours showing the vortex gust and freely-flying airfoil prior to impingement. The dash-dot lines correspond to vortex trajectories.}
\label{f:vap}	
\end{figure}

Fig.~\ref{f:vap} shows representative clockwise and counter-clockwise vortex gusts using CW2.5 and CCW2.5 as examples. Each gust consists of a coherent primary vortex core accompanied by a weaker oblique wake structure. The gust ratio is defined as

\begin{equation}
    \mathrm{GR}=\frac{U_v}{U_\infty},
\end{equation}
where \(U_v\) is the maximum velocity induced by the vortex. The generated vortices maintain similar structure, size, and strength, with \(\mathrm{GR} = 0.5-0.6\), and the nondimensional circulation of the vortex core is approximately \(0.60\pm0.02\), and the core radius is approximately \(0.18c\pm0.01c\). Complete details on the vortex characterization are provided in \cite{yan2026generation}.

\par

\section{Analysis of the freely-flying lift and heave from stationary foil data}
\label{sec:theory}

This section develops a framework for analyzing and predicting the lift and heave of a freely-flying airfoil during direct vortex impingement. The starting point is the time-dependent lift of the corresponding stationary airfoil, which represents the aerodynamic response to an equivalent incoming vortex in the absence of airfoil motion.

The framework is used in two ways. First, the lift is modeled using the heave velocity and acceleration obtained from the freely-flying simulation. This diagnostic reconstruction is used to determine how much of the simulated freely-flying lift can be explained by the stationary airfoil response superimposed with motion induced effects. Secondly, the force model is incorporated into the heave equation of motion to predict the airfoil dynamics using only the stationary lift as an input.  

\subsection{Modelling the lift coefficient}

The induced angle of attack due to the heaving motion is defined as

\begin{equation}
    \alpha_{\mathrm{i}}(t) = -\arctan\left(\dot{h}(t)\right)
    \;\approx\;-\dot{h}(t),
\end{equation}
where the approximation holds for small $\dot{h}$. The corresponding lift contribution is estimated as

\begin{equation}
\label{eqn:cli}
    C_{L,i}(t) \approx a\,\alpha_{\mathrm{i}}(t) \approx -a\dot{h}(t),
\end{equation}
which represents the change in lift associated with the instantaneous shift in effective angle of attack. The lift slope, $a$, is lower than $2\pi$ due to viscous and low Reynolds number effects. It is computed to be $a=2.47$ for the NACA 0012 at $Re=1000$.

In addition, the airfoil acceleration gives rise to an added-mass force. This is modeled by

\begin{equation}
\label{eqn:cla}
    C_{L,a}(t) = -2\,m_{\mathrm{a}} \ddot{h}(t),
\end{equation}
where $m_{\mathrm{a}}$ is a nondimensional added-mass coefficient. Impulsive heave simulations are performed to reveal $m_{\mathrm{a}}\approx\pi/4$, which is consistent with Theodorsen's value for thin airfoils \citep{theodorsen1949general}. 

Using the above motion induced lift contributions, a diagnostic reconstruction of the freely-flying lift coefficient is given by 

\begin{equation}
\label{eqn:liftrecon}
    \widetilde{C}_L(t)
    =
    C_L^{\mathrm{stat}}(t) - C_{L,0}
    + C_{L,i}(t)
    + C_{L,a}(t),
\end{equation}
where $C_{L,0}$ is the steady-state lift coefficient, \(C_L^{\mathrm{stat}}\) is the transient lift coefficient of the stationary airfoil impacted by the same gust, and $\widetilde{C}_L$ denotes the modeled freely-flying lift coefficient (distinct from $C_L$, the lift coefficient obtained directly from the freely-flying simulation). 

To complete this reconstruction, the heave velocity and acceleration are extracted from the simulation data. The modeled lift therefore does not represent an independent prediction; instead, it is used to assess how the motion influences the lift force. Agreement between $\widetilde{C}_L$ and $C_L$  indicates that the freely-flying lift response can be largely explained by the stationary vortex--airfoil interaction plus the instantaneous effects of heave velocity and acceleration. Discrepancies indicate additional contributions not captured by the reconstruction, such as coupling between the airfoil motion and the vortex--airfoil interaction, changes in the vortex--boundary-layer interaction, and interaction-induced vortex shedding.

\subsection{Prediction of the heave trajectory}
To obtain a prediction of the heave trajectory the motion induced force model is incorporated into the equation of motion. In this formulation, the only aerodynamic input is the instantaneous stationary airfoil lift \(C_L^{\mathrm{stat}}\). Starting with the equation of motion for the freely-flying airfoil, Eqn.~\ref{eqn:eom}, and substituting Eqn.~\ref{eqn:liftrecon} as an approximation for the right-hand-side forcing term yields,

\begin{equation}
    m \ddot{h}(t) = \frac{1}{2} \left(
    C_L^{\mathrm{stat}}(t) - C_{L,0}
    + C_{L,i}(t)
    + C_{L,a}(t)\right).
    \label{eqn:eom1}
\end{equation}
Here, $C_{L,i}$ and $C_{L,a}$ are functions of the predicted heave motion rather than being prescribed from the freely-flying simulation. Substituting the expressions for the motion induced terms into Eqn.~\ref{eqn:eom1} gives

\begin{equation}
\label{eqn:eom2}
    m\ddot{h}(t)
    = \frac{1}{2}
    \left(C_L^{\mathrm{stat}}(t) - C_{L,0}
    -a\dot{h}(t)
    -2m_a\ddot{h}(t)\right).
\end{equation}
Rearranging and defining an effective mass as $M=m+m_a$, the final equation of motion is
\begin{equation}
\label{eq:freely_flying_model}
    M\ddot{h}(t)
    +\frac{a}{2}\dot{h}(t)
    = \frac{1}{2}
    \left(C_L^{\mathrm{stat}}(t) - C_{L,0}\right) .
\end{equation}
Eqn.~\ref{eq:freely_flying_model} is a linear second-order equation for \(h(t)\) forced by the stationary airfoil lift. The heave trajectory is then obtained as

\begin{equation}
    h(t) =
    h_0
    + \frac{2M\dot{h}_0}{a}
    \left(1-e^{-\frac{a}{2M}(t-t_0)}\right)
    + \frac{1}{a}\int_{t_0}^{t}
    \left(1-e^{-\frac{a}{2M}(t-\tau)}\right)
    \left[C_L^{\mathrm{stat}}(\tau) - C_{L,0}\right]\,d\tau,
\end{equation}
where \(h_0\) and \(\dot{h}_0\) are the initial displacement and velocity. For an airfoil initially in equilibrium, \(h_0=\dot{h}_0=0\), the heave trajectory is

\begin{equation}
    h(t) =
    \frac{1}{a}\int_{t_0}^{t}
    \left(1-e^{-\frac{a}{2M}(t-\tau)}\right)
    \left[C_L^{\mathrm{stat}}(\tau) - C_{L,0}\right]\,d\tau.
\end{equation}

\section{Results and discussion}

The response of the downstream airfoil is characterized using the heave displacement \(h(t)\) and the unsteady lift coefficient, and accompanied by vorticity and pressure flow fields. 
Throughout this section, \textit{airfoil} refers to the freely-flying airfoil, and \textit{vortex} refers to the primary vortex generated by the upstream airfoil, unless stated otherwise.

\subsection{Overview of airfoil response}
\begin{figure}
 \centering
     \begin{subfigure}[b]{0.495\textwidth}
	\centering
        \includegraphics[width=0.99\textwidth]{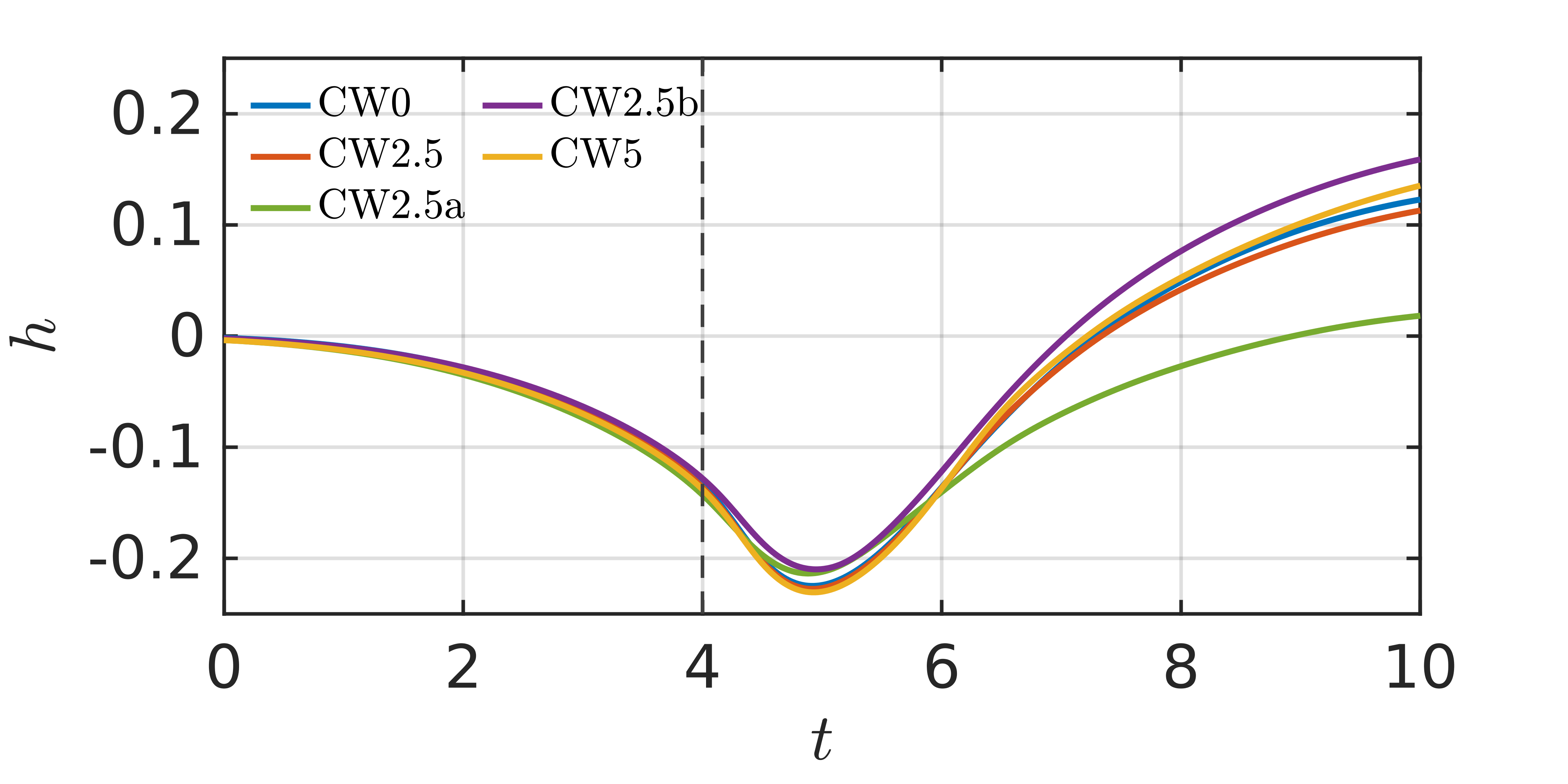}
        \caption{CW vortex interactions}
	\end{subfigure}
    \begin{subfigure}[b]{0.495\textwidth}
	\centering
        \includegraphics[width=0.99\textwidth]{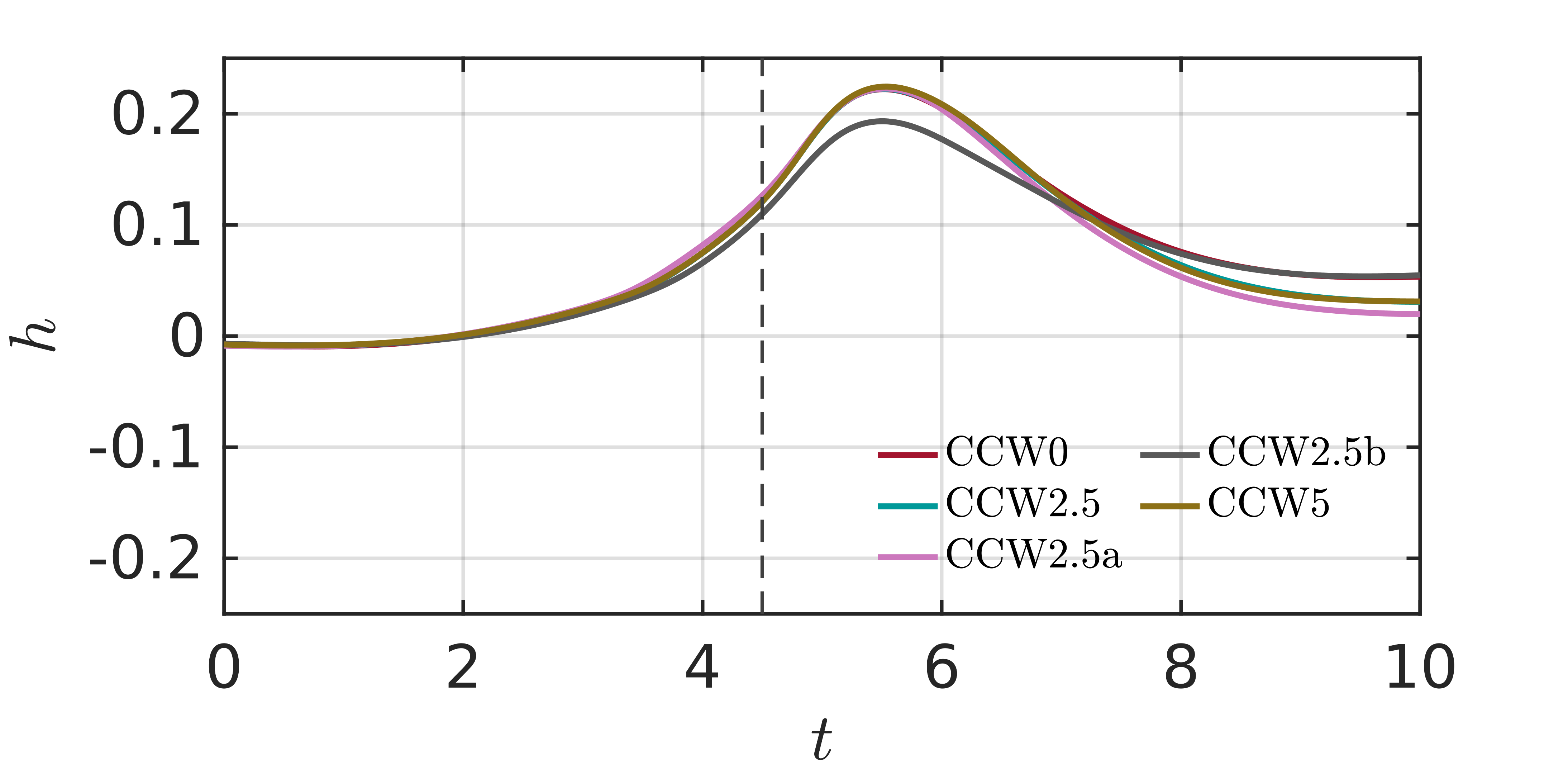}
        \caption{CCW vortex interactions}	
    \end{subfigure}
    \begin{subfigure}[b]{0.495\textwidth}
	\centering
        \includegraphics[width=0.99\textwidth]{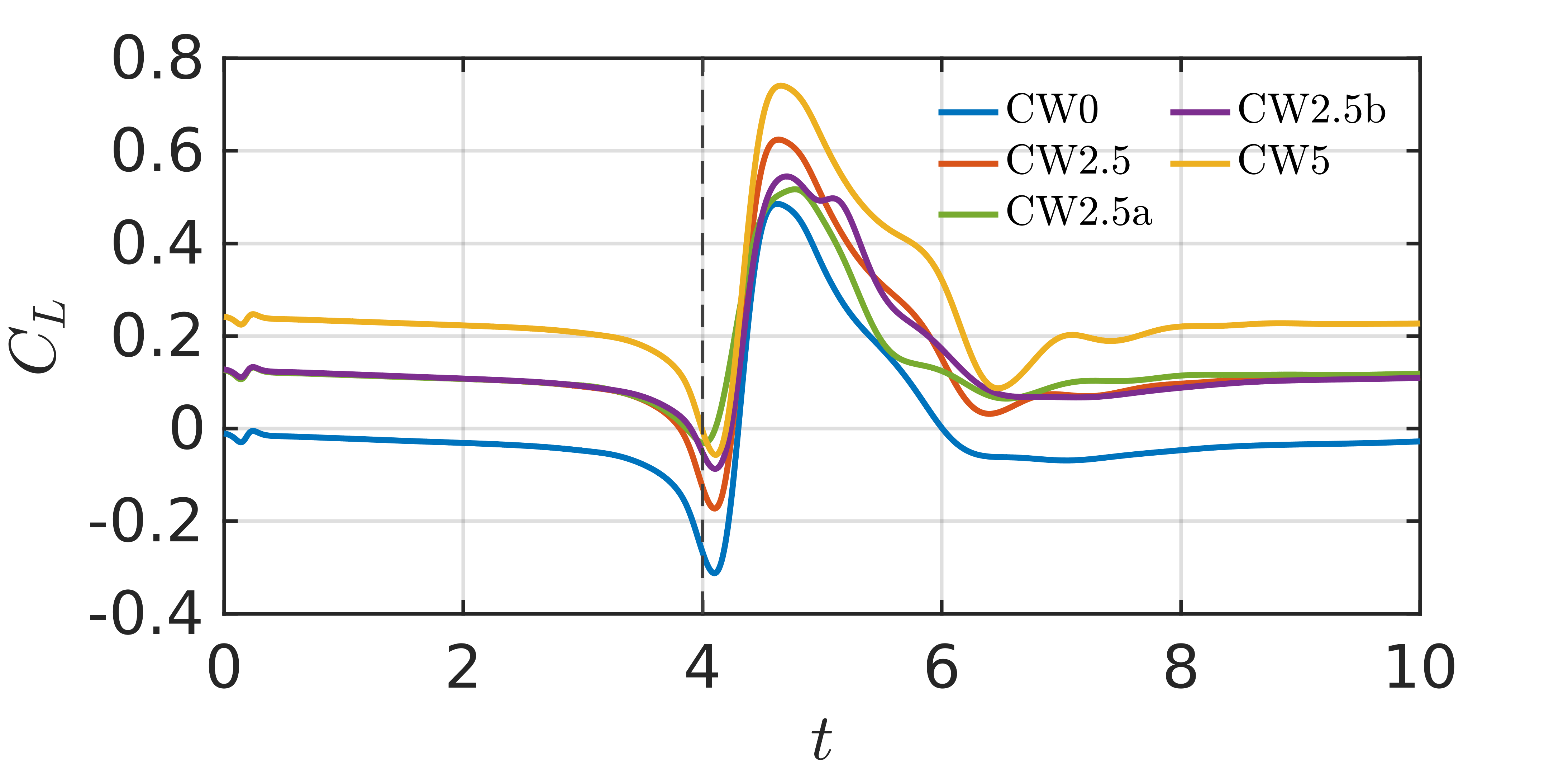}
        \caption{CW vortex interactions}
	\end{subfigure}
    \begin{subfigure}[b]{0.495\textwidth}
	\centering
        \includegraphics[width=0.99\textwidth]{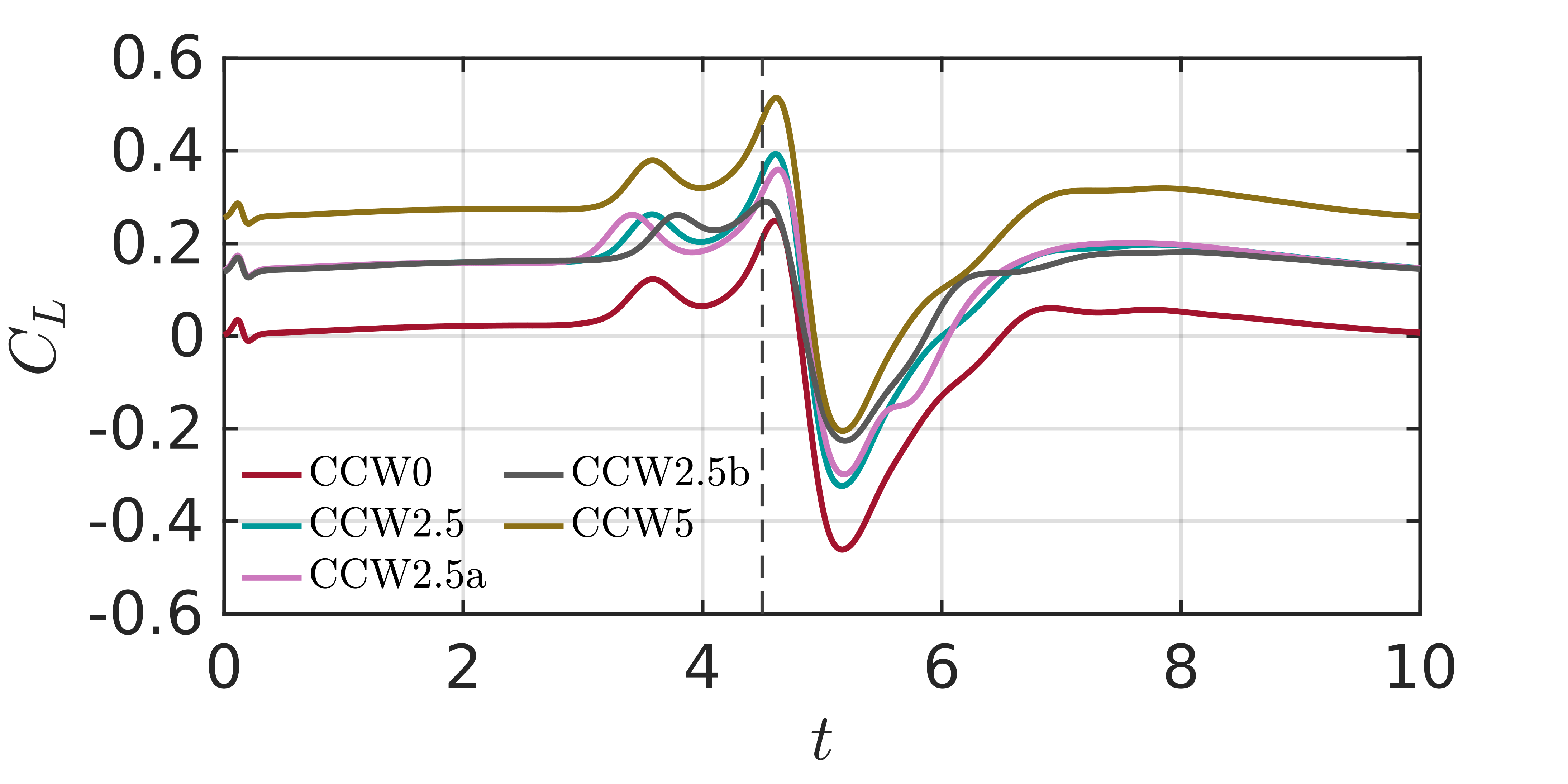}
        \caption{CCW vortex interactions}
	\end{subfigure}
\caption{Airfoil's heave and lift coefficient in response to vortex interactions. Dashed lines correspond to approximate vortex impingement time.}
 \label{f:hclt}	
 \end{figure}

The heave and lift are reported over the entire interaction process in Fig.~\ref{f:hclt}, with the approximate vortex impingement time represented by the dashed vertical line. Prior to vortex impingement, the vortex is upstream of the airfoil and weakly influencing the flow around the airfoil. Nevertheless, the airfoil undergoes a gradual upward or downward motion depending on the vortex rotation, which can be interpreted as the upwash or downwash induced by the vortex. Within each rotation group, the airfoil trajectories are all similar prior to impingement. 

\par
During vortex impingement, the aerodynamic response becomes significantly stronger as the vortex interacts directly with the airfoil boundary layer. For the CW vortices, the lift coefficient \(C_L\) first reaches a negative peak and then rapidly changes sign, rising quickly to a positive peak. A similar but opposite response is observed for the CCW vortices. This behavior is reflected in the heave response, although a phase difference exists between the lift and displacement: the airfoil continues to accelerate in its initial direction when the first lift peak occurs and only begins to decelerate as \(\Delta C_L=C_L-C_{L,0}\) changes sign.

Following the peak heave displacement, all airfoils exhibit a rebound motion. For the CW vortices, the airfoil overshoots its initial position, resulting in a pronounced rebound, with the only exception being the non-direct impingement Case CW2.5a, whereas for the CCW vortices the rebound is milder and the airfoil returns closer to its initial position. 
It is also observed that CW0 and CCW0 exhibit asymmetric heave and lift profiles, despite the apparent symmetry of the configuration. The asymmetry may be related to differences in vortex structure, such as the oblique wake necessary for CCW vortices (see Fig.~\ref{f:vap}(b)), as well as weak effects from the upstream airfoil in the computational domain.
 
\subsection{Modeling the physics of a freely-flying and stationary airfoil}
\begin{figure}
 \centering
     \begin{subfigure}[b]{0.49\textwidth}
	\centering
        \includegraphics[width=0.99\textwidth]{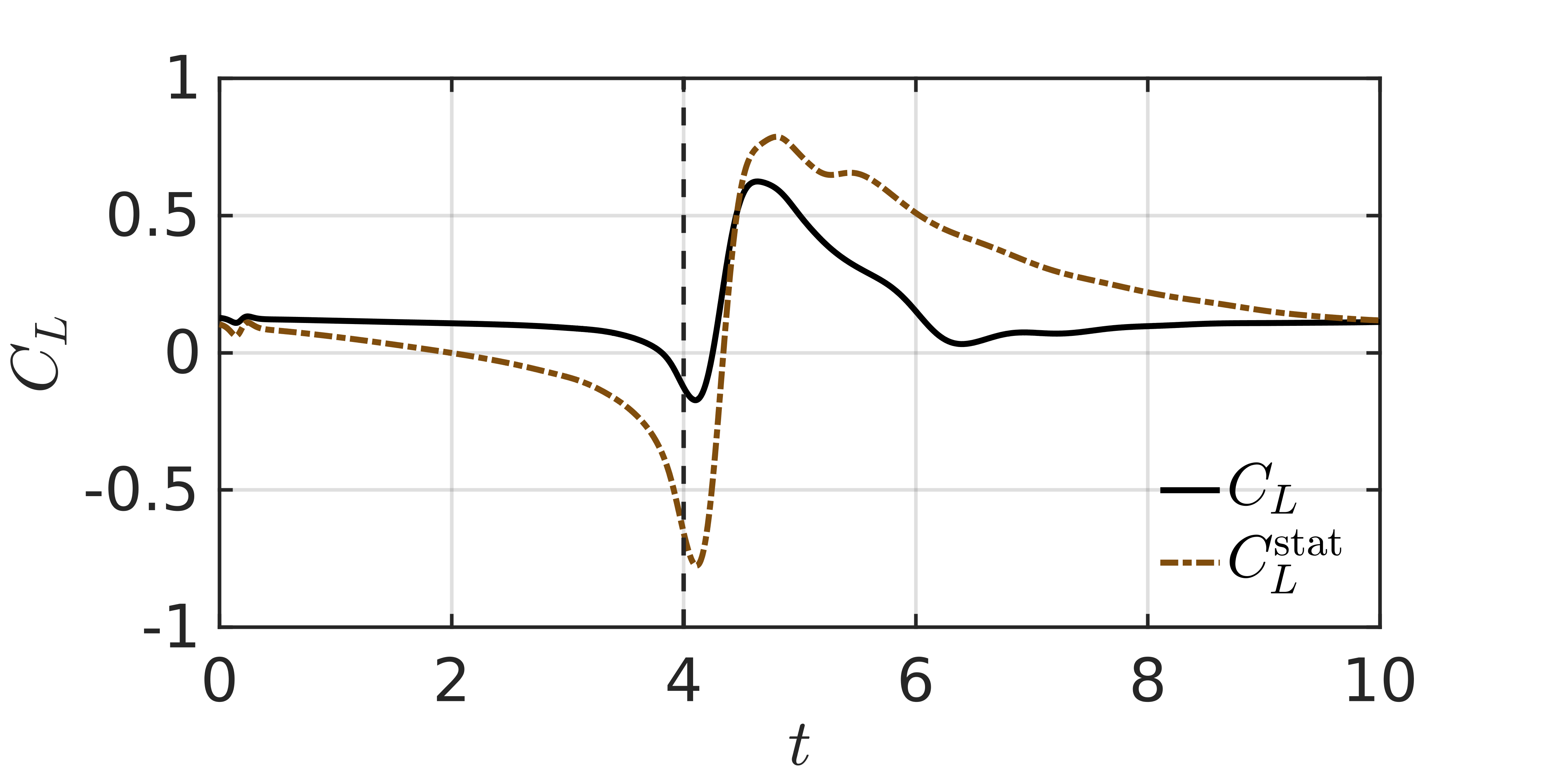}
        \caption{Freely-flying $C_L$ vs. stationary $C_L$}
	\end{subfigure}
    \begin{subfigure}[b]{0.49\textwidth}
	\centering
        \includegraphics[width=0.99\textwidth]{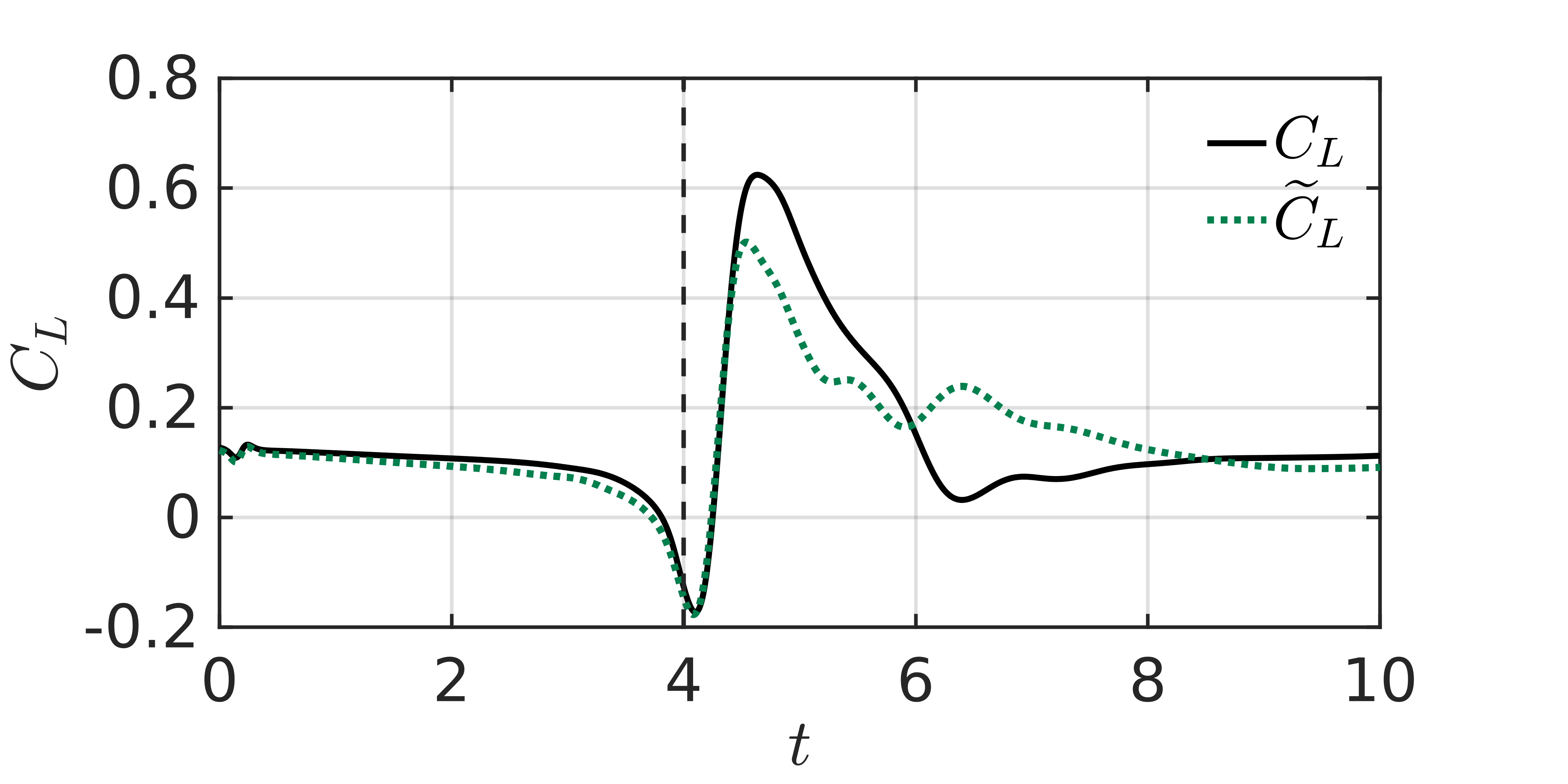}
        \caption{Freely-flying $C_L$ vs. modeled $\widetilde{C}_L$}
	\end{subfigure}
    \begin{subfigure}[b]{0.49\textwidth}
	\centering
        \includegraphics[width=0.99\textwidth]{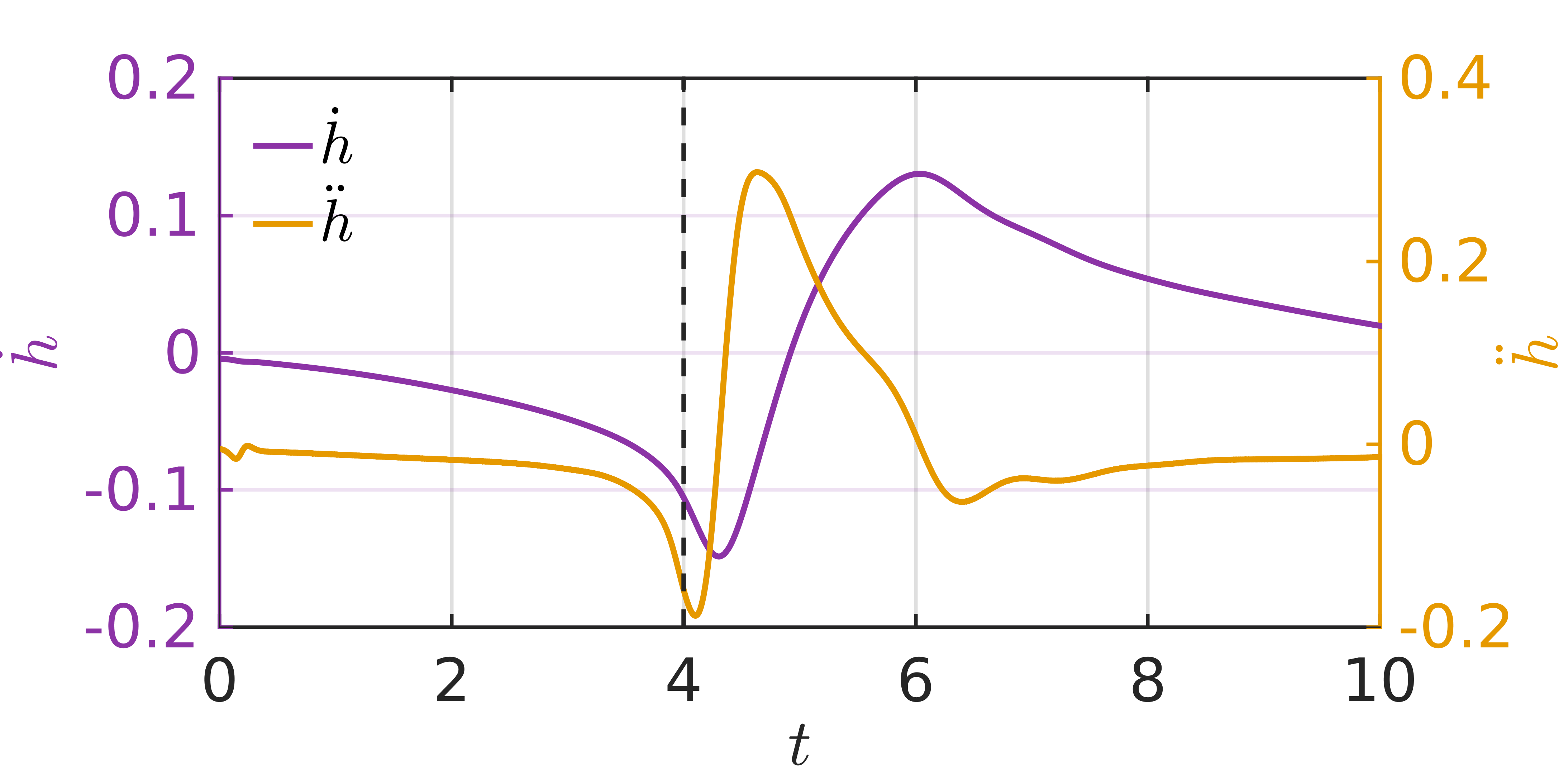}
        \caption{Heave velocity and acceleration}
	\end{subfigure}
         \begin{subfigure}[b]{0.49\textwidth}
	\centering
        \includegraphics[width=0.99\textwidth]{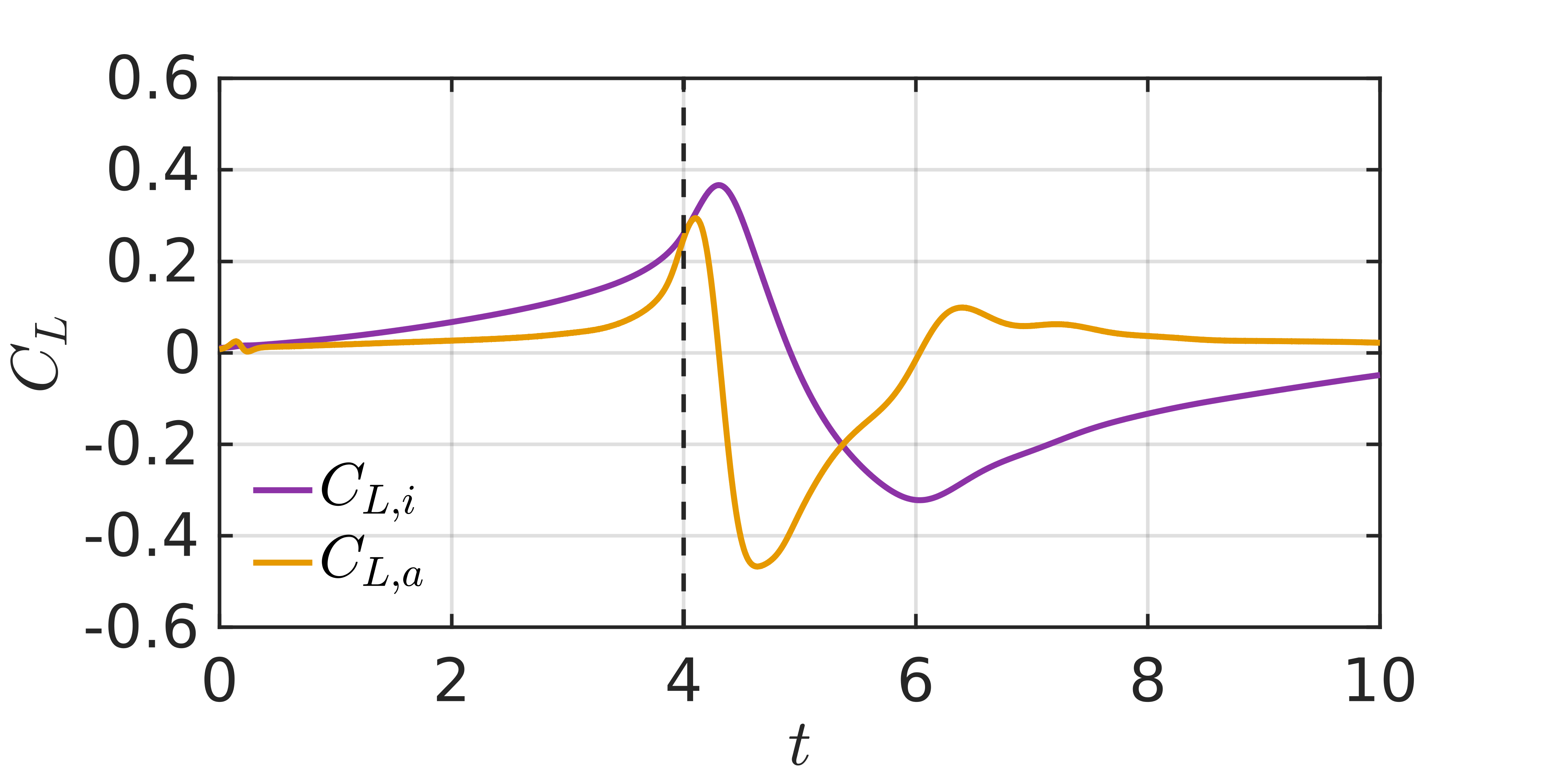}
        \caption{Motion induced contributions to $C_L$}
	\end{subfigure}
\caption{Heave motion contributions to $C_L$ for CW2.5. Dashed lines correspond to approximate impingement time.}
 \label{f:reconCW2.5}	
 \end{figure}

\begin{figure}
 \centering
     \begin{subfigure}[b]{0.49\textwidth}
	\centering
        \includegraphics[width=0.99\textwidth]{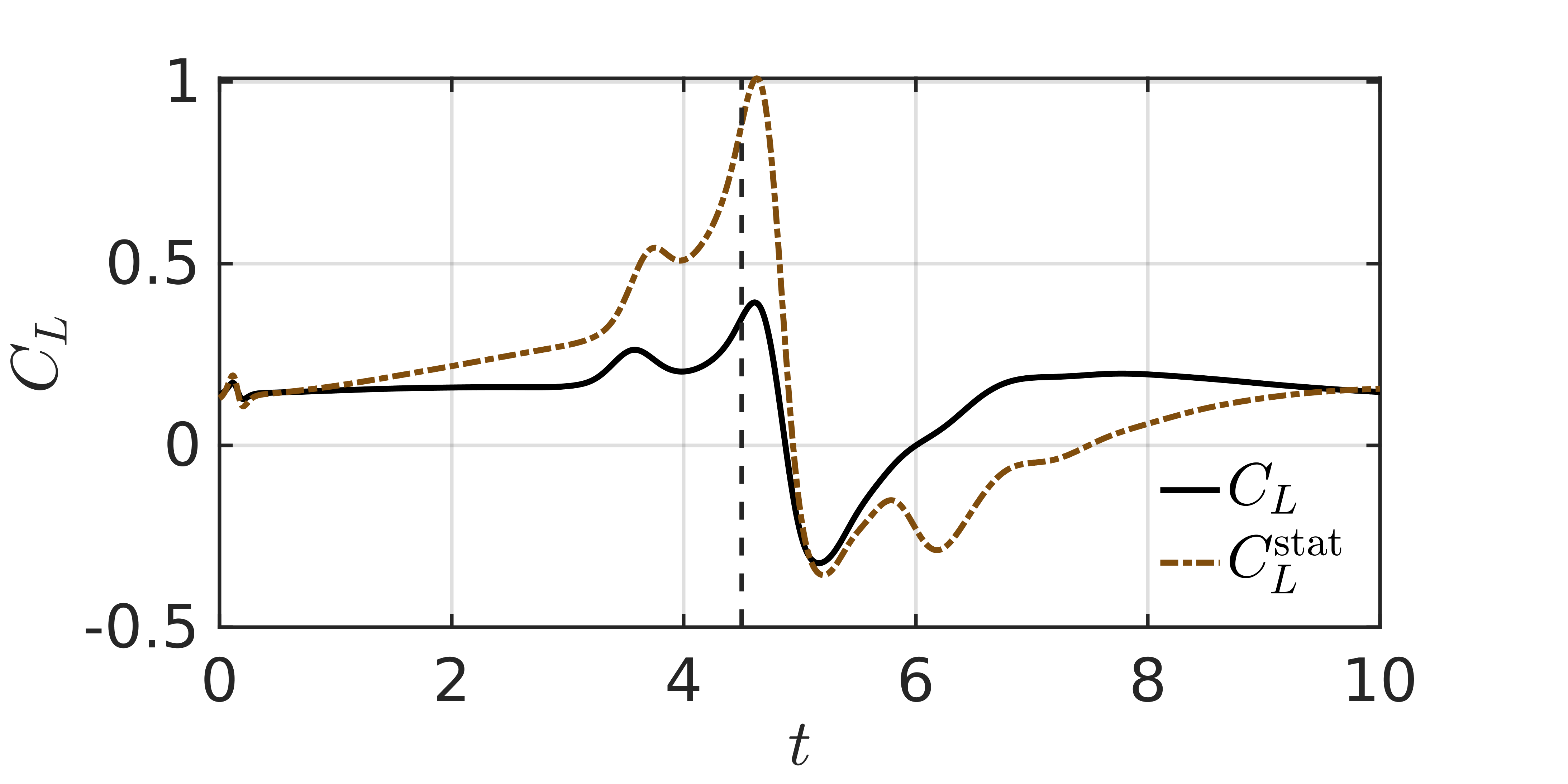}
        \caption{Freely-flying $C_L$ vs. stationary $C_L$}
	\end{subfigure}
    \begin{subfigure}[b]{0.49\textwidth}
	\centering
        \includegraphics[width=0.99\textwidth]{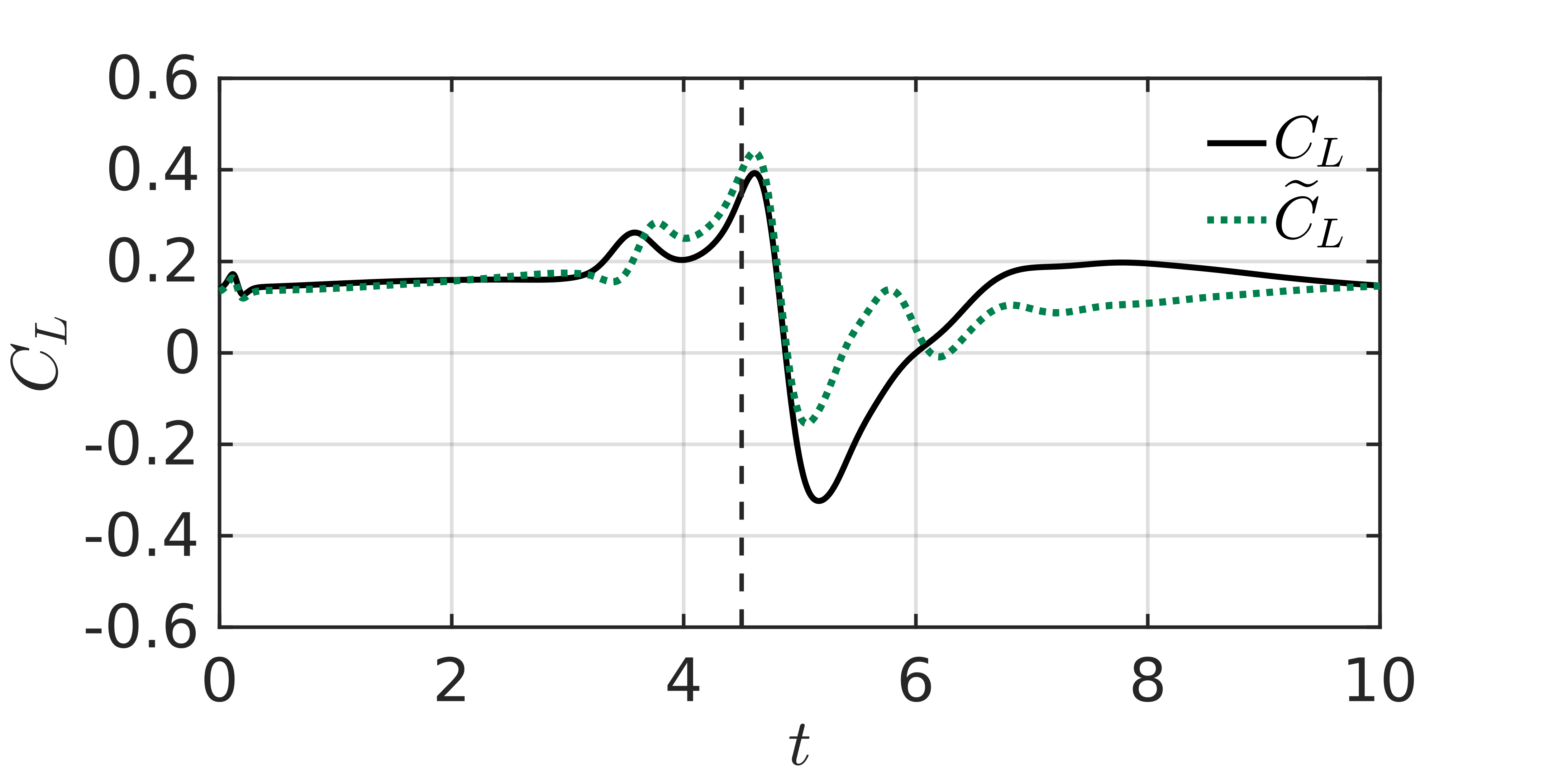}
        \caption{Freely-flying $C_L$ vs. modeled $\widetilde{C}_L$}
	\end{subfigure}
    \begin{subfigure}[b]{0.49\textwidth}
	\centering
        \includegraphics[width=0.99\textwidth]{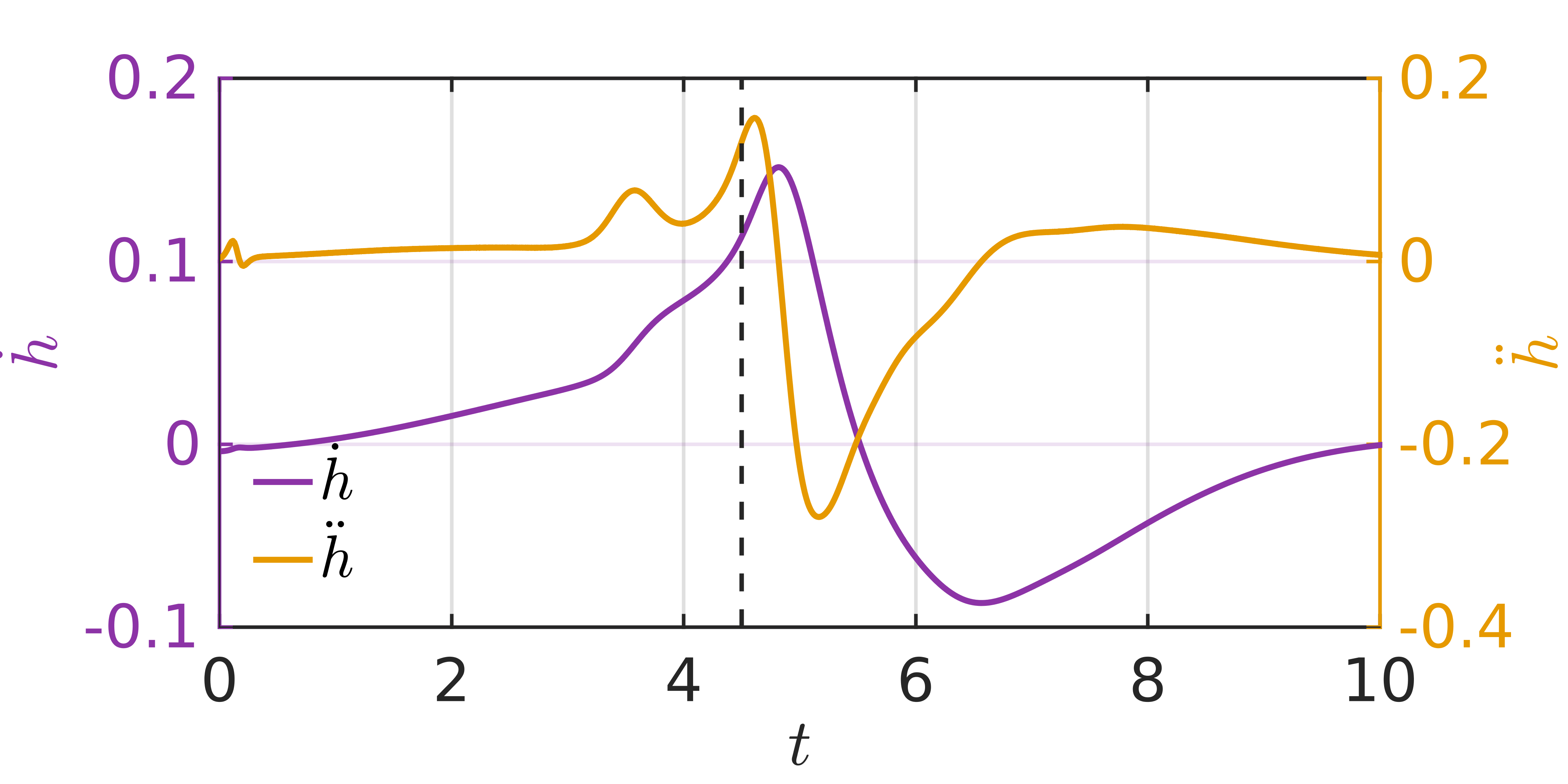}
        \caption{Heave velocity and acceleration}
	\end{subfigure}
         \begin{subfigure}[b]{0.49\textwidth}
	\centering
        \includegraphics[width=0.99\textwidth]{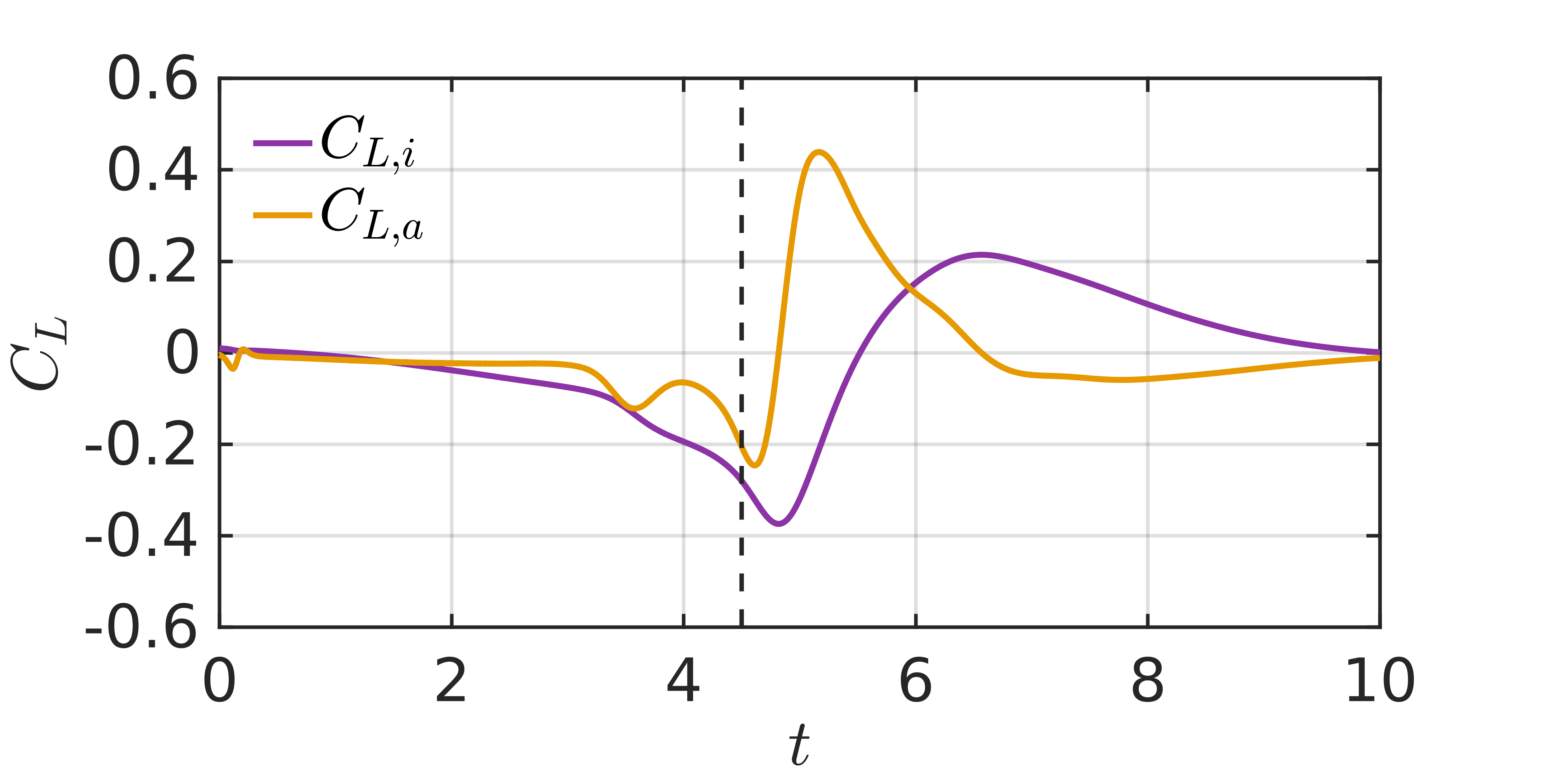}
        \caption{Motion induced contributions to $C_L$}
	\end{subfigure}
\caption{Heave motion contributions to $C_L$ for CCW2.5. Dashed lines correspond to approximate impingement time.}
 \label{f:reconCCW2.5}	
 \end{figure}

Figs.~\ref{f:reconCW2.5} and \ref{f:reconCCW2.5} compare and constrast CW2.5, CCW2.5, against their stationary counterparts. This analysis serves two purposes. First, the estimated $\widetilde{C}_L$ provides a baseline for interpreting how the motion induced contributions associated with heave velocity and acceleration shape the freely-flying response, including the rebound behavior. Second, differences between the estimated and simulated freely-flying $C_L$, together with the corresponding flow-field snapshots, are used to identify when and how coupled vortex--airfoil dynamics modify this response.
\par
Figs.~\ref{f:reconCW2.5}(a) and \ref{f:reconCCW2.5}(a) compare the freely-flying and stationary lift curves for the CW2.5 and CCW2.5 interactions, respectively. In Fig.~\ref{f:reconCW2.5}(a), the first negative peak for CW2.5 is significantly attenuated relative to CW2.5-S, whereas the subsequent positive peak is only mildly affected by the heaving motion. Fig.~\ref{f:reconCCW2.5}(a) shows the same trend with reversed sign. Thus, in both vortex orientations, the freely-flying motion primarily reduces the first impingement-induced lift peak, while the subsequent opposite-signed peak remains closer to the stationary response. 

This change in the relative strength of the two lift peaks is important for the heave dynamics. If the stationary $C_L^{\mathrm{stat}}$ is considered in isolation, the inferred heave would suggest an initial motion driven by the first lift peak, followed by a gradual recovery as the lift changes sign, with limited or even no rebound. In the freely-flying cases, however, the attenuation of the first peak limits the initial heave response, while the subsequent opposite-signed peak remains relatively strong. The resulting imbalance favors a faster reversal of the heave motion and therefore a stronger rebound.
\par
Figures~\ref{f:reconCW2.5}(b) and \ref{f:reconCCW2.5}(b) compare the modeled $\widetilde{C}_L$ with the simulated freely-flying lift response for the CW2.5 and CCW2.5 interactions, respectively. In both vortex orientations, the modeled lift captures the main trend observed in the freely-flying airfoils. The first impingement-induced peak is attenuated, while the subsequent opposite-signed peak has a higher relative amplitude. The agreement between $\widetilde{C}_L$ and $C_L$ is particularly good before and near impingement, when the vortex has not yet strongly affected the boundary layer of the airfoil. A minor exception occurs in the CCW interactions ($t\approx3.7$ in Fig.~\ref{f:reconCCW2.5}(b)), where the small peak associated with shear-layer impingement appears at a shifted time. This feature precedes the main impingement-induced lift peak and is therefore distinct from the post-impingement discrepancy discussed below.

The freely-flying lift can be interpreted through the motion induced contributions up to $0.5$ time units after initial vortex impingement. The role of these motion induced contributions can be understood from Figs.~\ref{f:reconCW2.5}(c,d) and \ref{f:reconCCW2.5}(c,d). The heave velocity and acceleration shown in panels (c) determine the induced angle of attack and added-mass contributions shown in panels (d). Around the first lift peak, these contributions act to reduce the magnitude of the vortex-induced lift, thereby limiting the initial acceleration of the airfoil. As the lift changes sign and approaches the subsequent opposite-signed peak, the heave velocity has decreased substantially and has not yet fully reversed. As a result, the induced angle of attack contribution no longer strongly attenuates the lift and may instead weakly reinforce the opposite-signed response. This asymmetry in the motion induced correction attenuates the first peak more strongly than the second, producing the correct trend toward a rebound in the freely-flying response.
\par
At later times, beginning $0.5$ time units after impingement, $\widetilde{C}_L$ deviates from $C_L$. Although $\widetilde{C}_L$ captures the peak imbalance responsible for the basic rebound tendency, it underpredicts the magnitude of the subsequent opposite-signed peak. Therefore, the modeled lift implies a slower rebound than that observed in the simulation. This difference suggests that additional coupled vortex--airfoil dynamics become important after impingement, especially through modifications of the vortex--boundary-layer interaction and the induced vortex-shedding process. These coupled-flow effects further enhance the opposite-signed lift peak, producing a more positive lift for CW interactions and a more negative lift for CCW interactions, and thereby contribute to the faster rebound observed in the freely-flying cases. The detailed flow features responsible for these differences are examined in the next subsection.

\subsection{Vortex interaction and flow physics}
\begin{figure}
\centering
	\begin{subfigure}[b]{0.24\textwidth}
	\centering
        \includegraphics[width=0.95\textwidth]{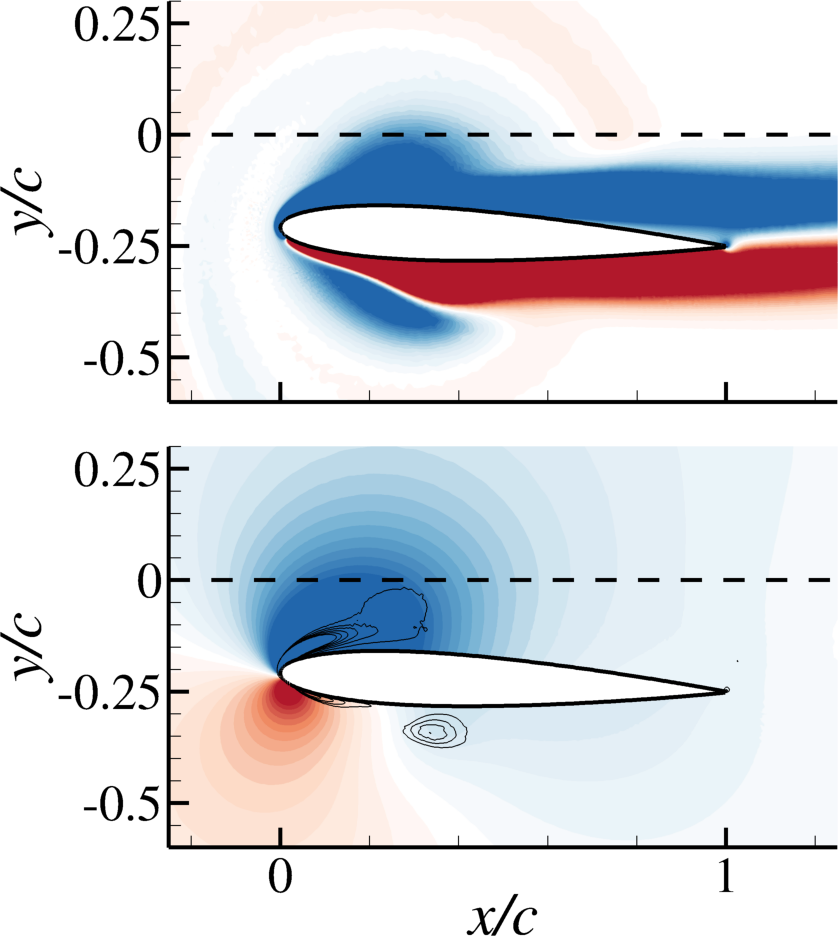}
        \caption{CW2.5,~$t=4.54$}
	\end{subfigure}
    	\begin{subfigure}[b]{0.24\textwidth}
	\centering
        \includegraphics[width=0.95\textwidth]{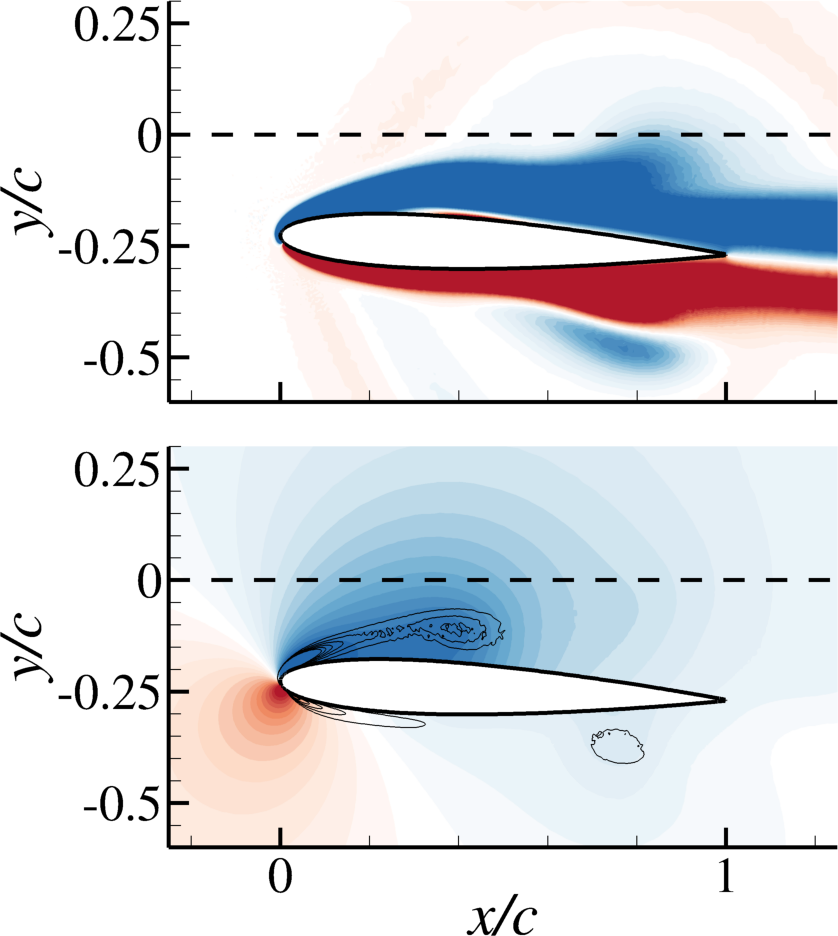}
        \caption{CW2.5,~$t=5.04$}
	\end{subfigure}
    \begin{subfigure}[b]{0.24\textwidth}
	\centering
        \includegraphics[width=0.95\textwidth]{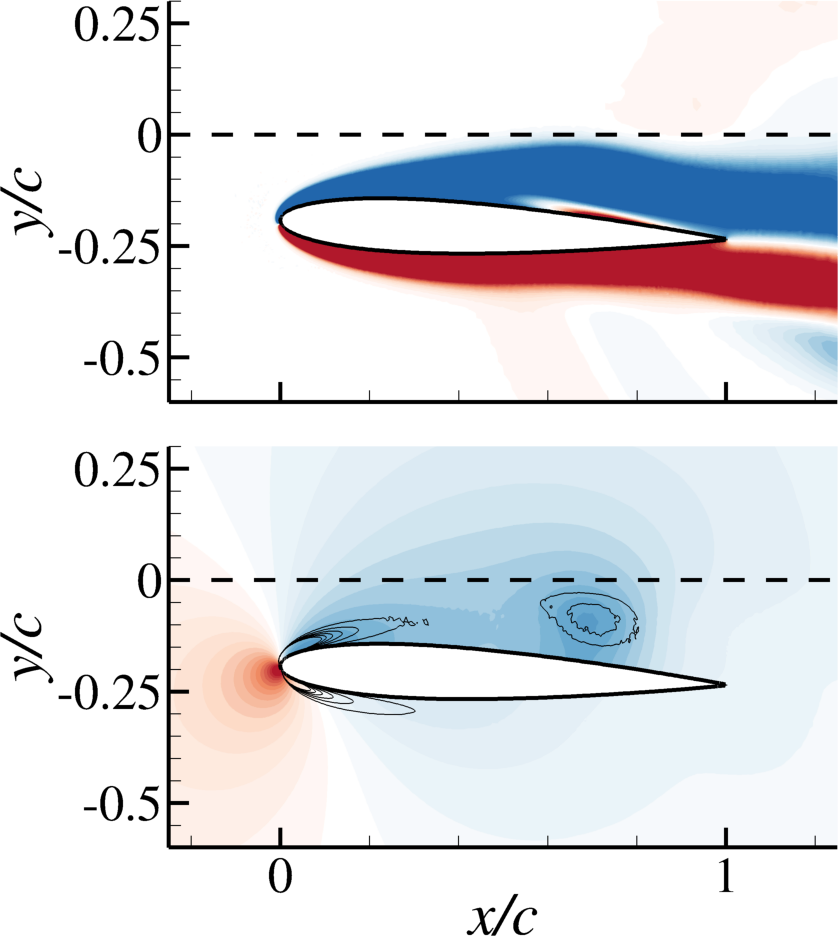}
        \caption{CW2.5,~$t=5.54$}
	\end{subfigure}
        \begin{subfigure}[b]{0.24\textwidth}
	\centering
        \includegraphics[width=0.95\textwidth]{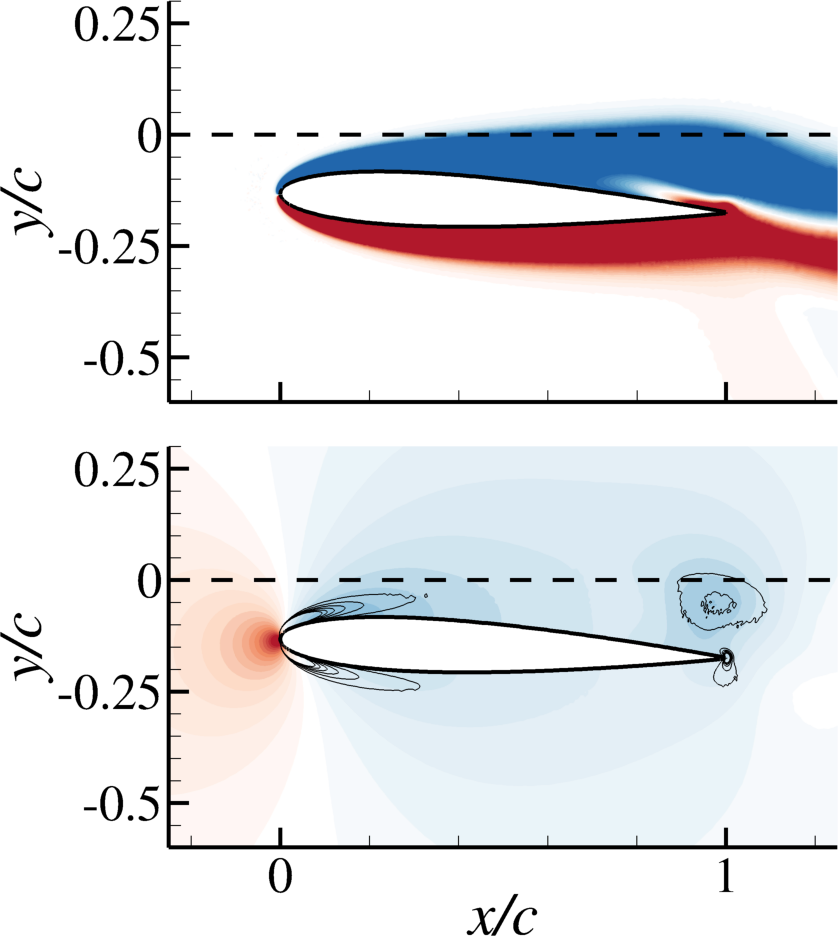}
        \caption{CW2.5,~$t=6.04$}
	\end{subfigure}
    \begin{subfigure}[b]{0.24\textwidth}
	\centering
        \includegraphics[width=0.95\textwidth]{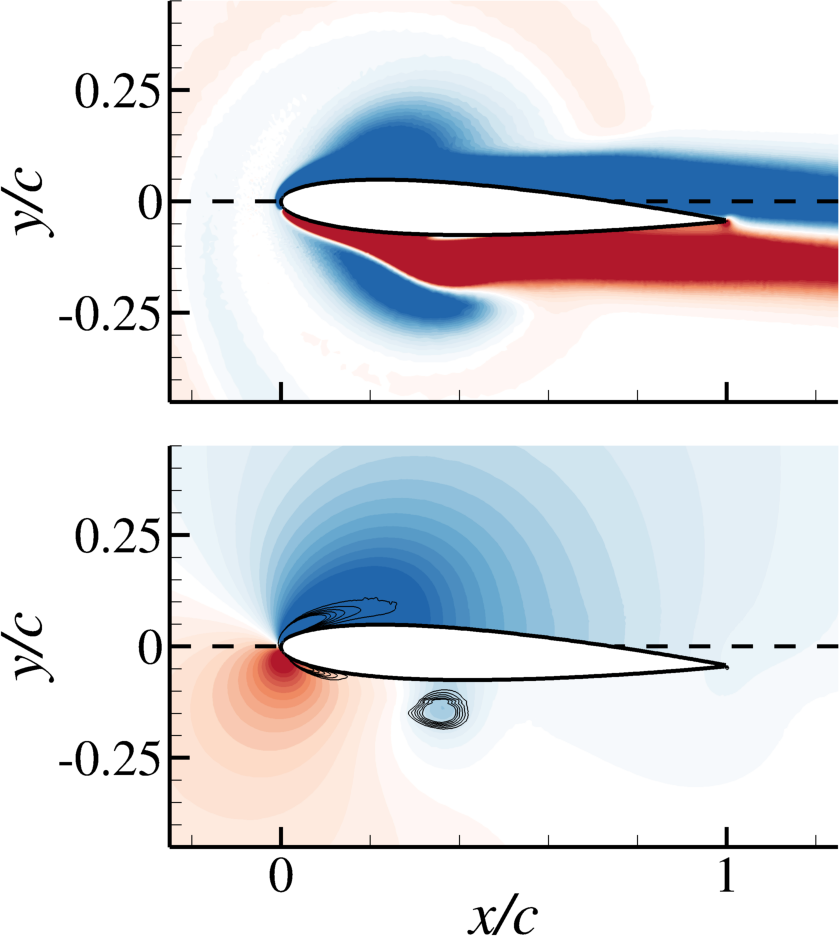}
        \caption{CW2.5-S,~$t=4.54$}
	\end{subfigure}
    	\begin{subfigure}[b]{0.24\textwidth}
	\centering
        \includegraphics[width=0.95\textwidth]{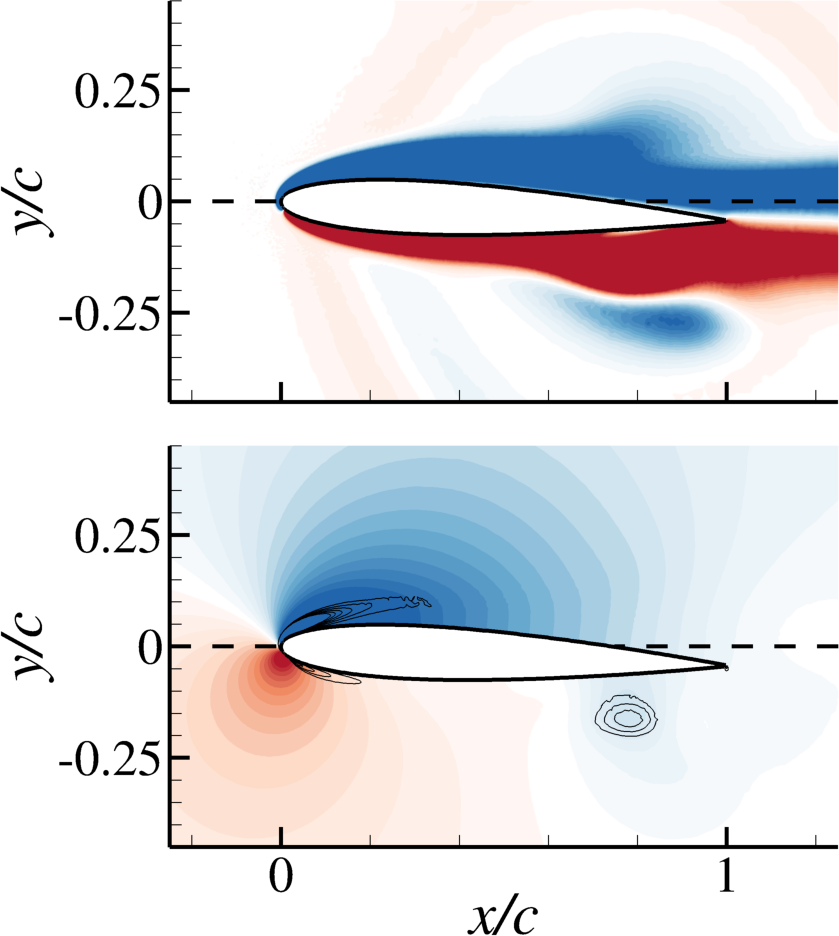}
        \caption{CW2.5-S,~$t=5.04$}
	\end{subfigure}
    	\begin{subfigure}[b]{0.24\textwidth}
	\centering
        \includegraphics[width=0.95\textwidth]{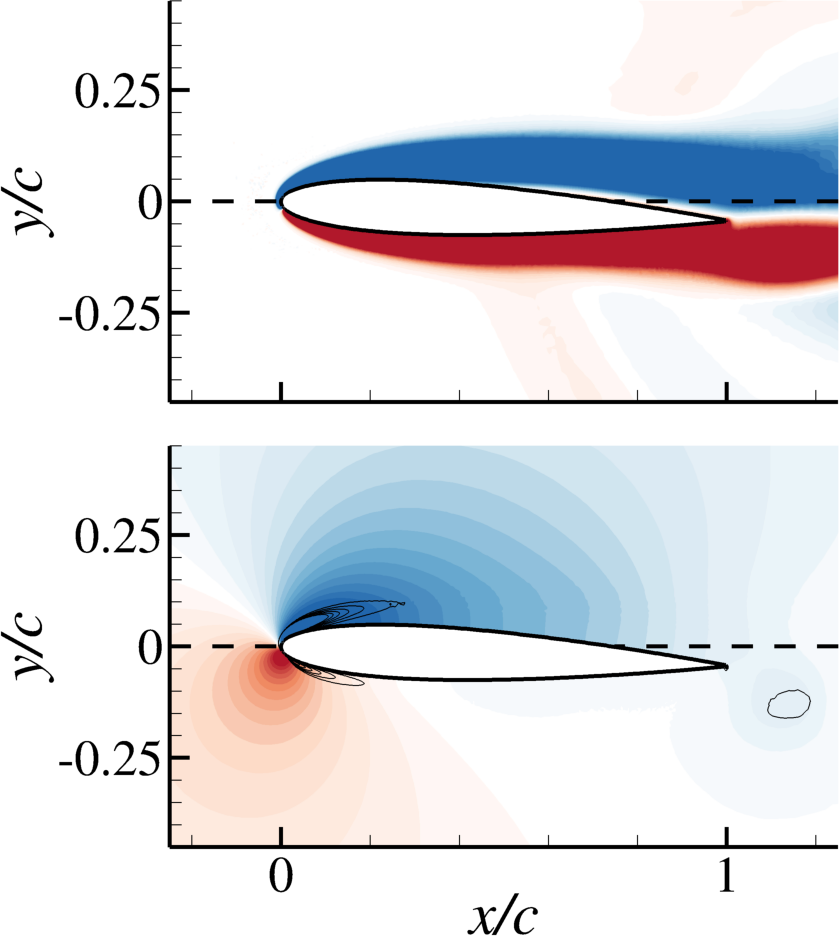}
        \caption{CW2.5-S,~$t=5.54$}
	\end{subfigure}
    \begin{subfigure}[b]{0.24\textwidth}
	\centering
        \includegraphics[width=0.95\textwidth]{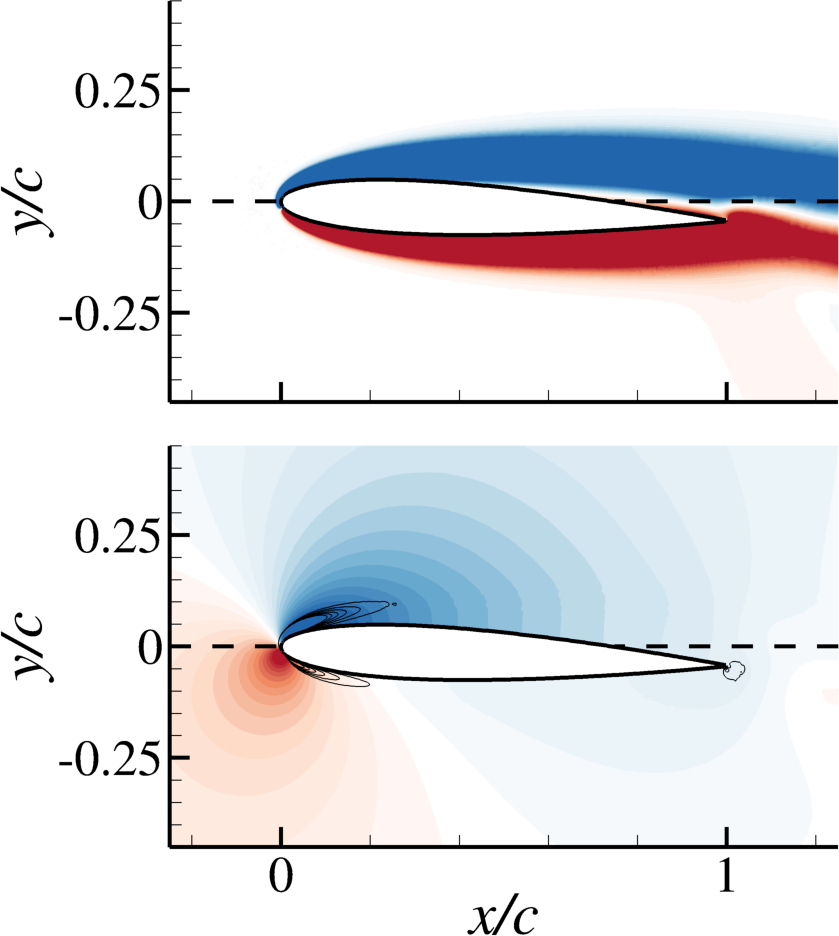}
        \caption{CW2.5-S,~$t=6.04$}
	\end{subfigure}
    \\[0.2cm]   % space between rows
        \begin{subfigure}[b]{0.75\textwidth}
	\centering
        \includegraphics[width=1\textwidth]{figures_new/legend1_cropped.png}
	\end{subfigure}
\\[0.2cm]   % space between rows
        \begin{subfigure}[b]{0.75\textwidth}
	\centering
        \includegraphics[width=1\textwidth]{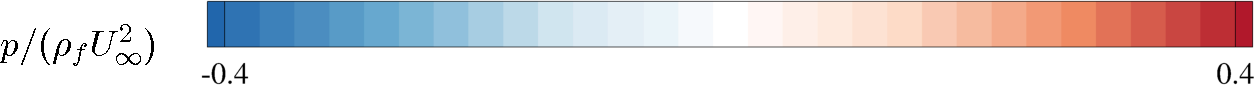}
	\end{subfigure}
    
\caption{Spanwise-vorticity contours (top) and pressure contours (bottom) in the region around the airfoil after vortex impingement for CW2.5 and CW2.5-S. In the pressure panels, contour lines of the \(Q\)-criterion are overlaid to identify vortical structures.}
\label{f:cCW2.5}	
\end{figure}

Fig.~\ref{f:cCW2.5} shows the vorticity and pressure contours after vortex impingement for CW2.5 and CW2.5-S, with prominent vortex structures identified using Q-criterion contour lines. %The first row corresponds to CW2.5 and the second to its stationary counterpart, CW2.5-S. 
In both CW2.5 and CW2.5-S, the vortex core impinges on the leading edge of the airfoil and subsequently splits into upper and lower structures that convect downstream. On the pressure side, a secondary vortex forms, which is stronger in CW2.5-S, as indicated by the larger absolute magnitude of pressure variation in the corresponding region in Fig.~\ref{f:cCW2.5}(e) compared with Fig.~\ref{f:cCW2.5}(a). On the suction side in CW2.5, vorticity accumulates causing shear layer roll up (Fig.~\ref{f:cCW2.5}(b)), eventually forming a concentrated vortex that is subsequently shed. This newly formed vortex generates a low-pressure region that convects over the suction surface, whereas such behavior is not observed in CW2.5-S. These differences explain why the modeled $\widetilde{C}_L$ exceeds the simulated freely-flying $C_L$ during part of the post-impingement stage, and this interpretation is supported by the timing of the flow features. The lower-surface vortex emerges at $t \approx 4.5$, and the suction side vortex exits at $t \approx 6.0$, corresponding closely to the interval over which $\widetilde{C}_L$ is elevated relative to the simulation data. After these induced vortical structures leave the airfoil surface, the simulated freely-flying $C_L$ is rapidly driven back toward its initial loading state and can even overshoot to the opposite sign, whereas $\widetilde{C}_L$ evolves more gradually. The differences in vortex shedding are likely associated with the downward motion of the airfoil immediately prior to vortex formation.
\par
\begin{figure}
\centering
	\begin{subfigure}[b]{0.24\textwidth}
	\centering
        \includegraphics[width=0.95\textwidth]{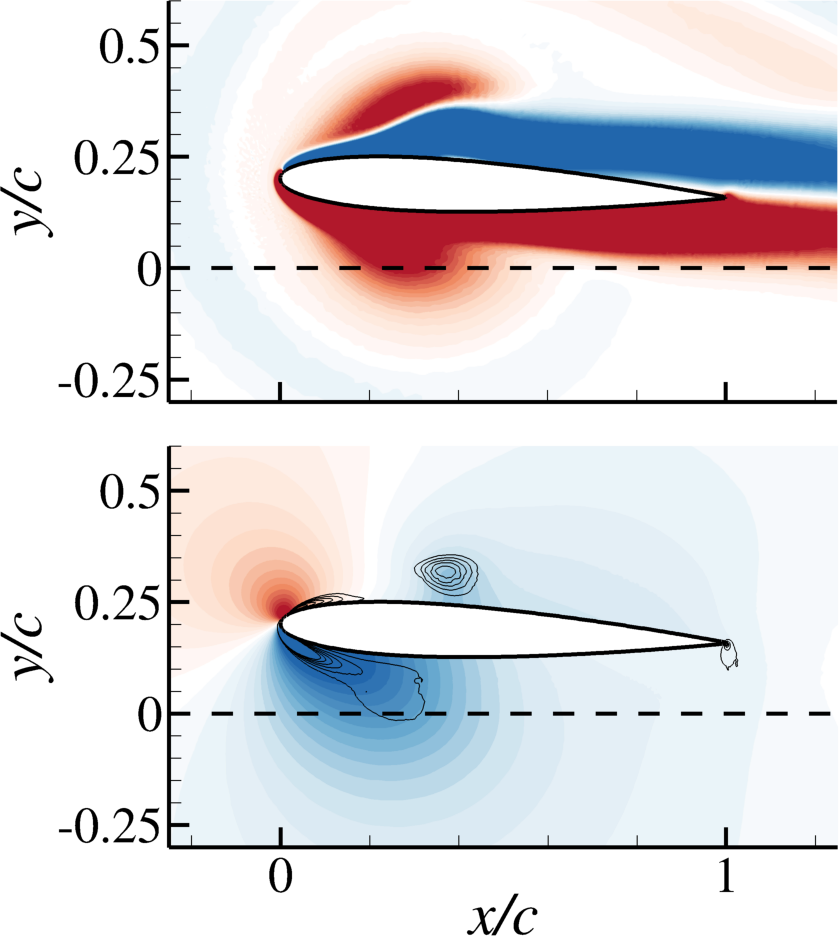}
        \caption{CCW2.5,~$t=5.10$}
	\end{subfigure}
    	\begin{subfigure}[b]{0.24\textwidth}
	\centering
        \includegraphics[width=0.95\textwidth]{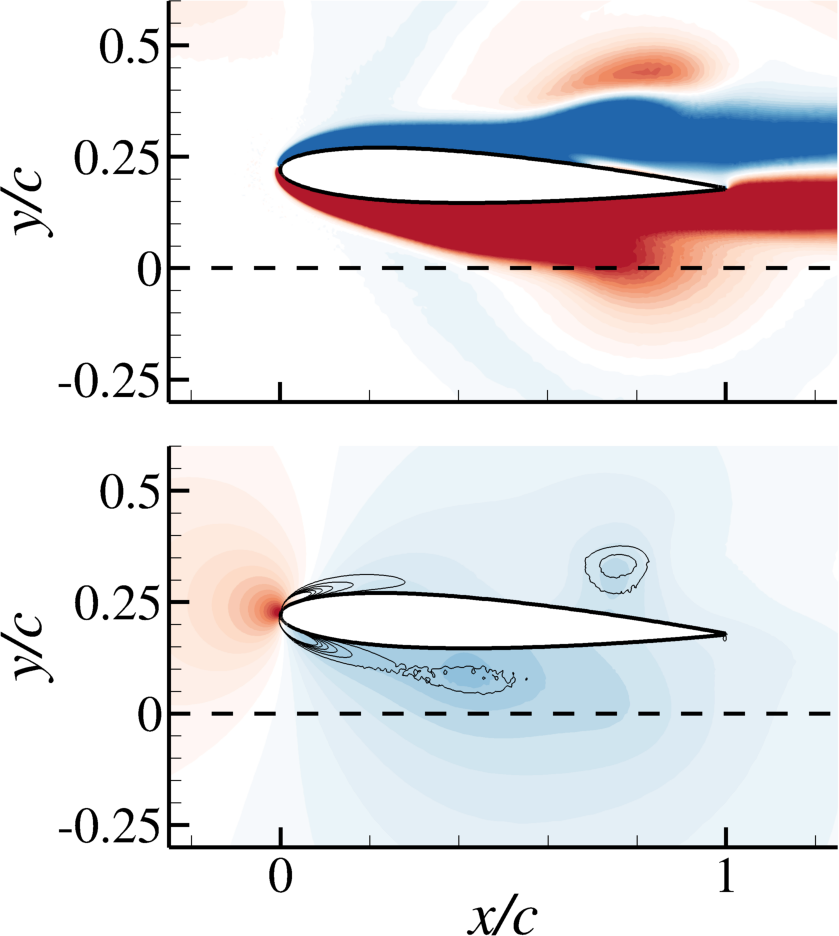}
        \caption{CCW2.5,~$t=5.60$}
	\end{subfigure}
    \begin{subfigure}[b]{0.24\textwidth}
	\centering
        \includegraphics[width=0.95\textwidth]{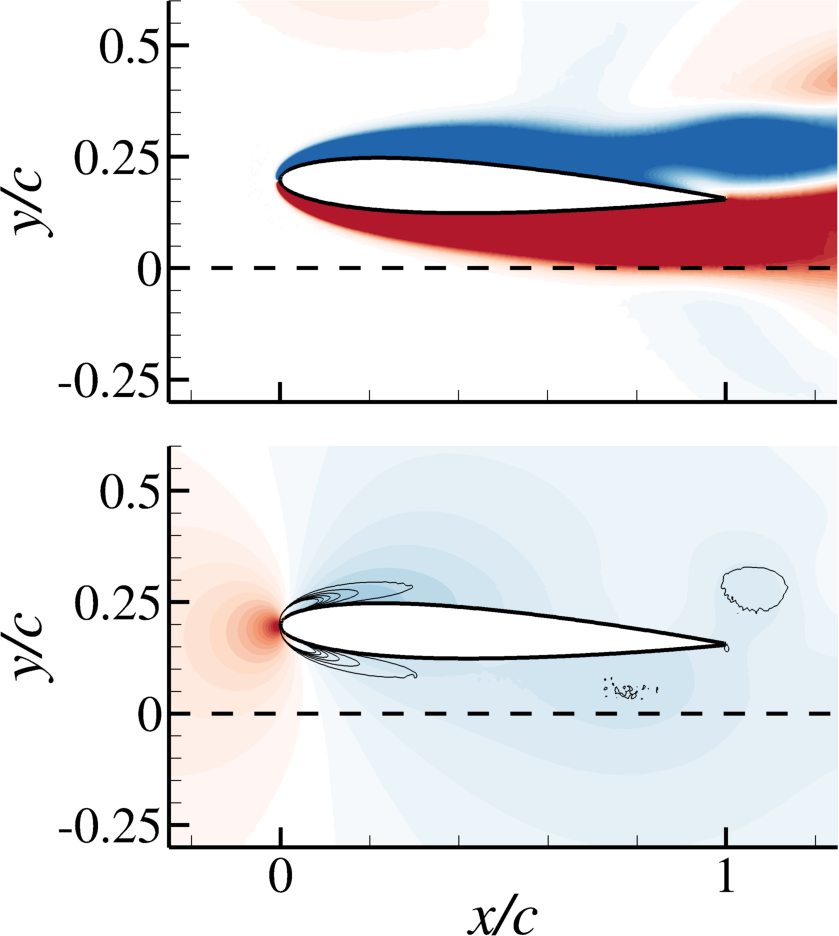}
        \caption{CCW2.5,~$t=6.10$}
	\end{subfigure}
        \begin{subfigure}[b]{0.24\textwidth}
	\centering
        \includegraphics[width=0.95\textwidth]{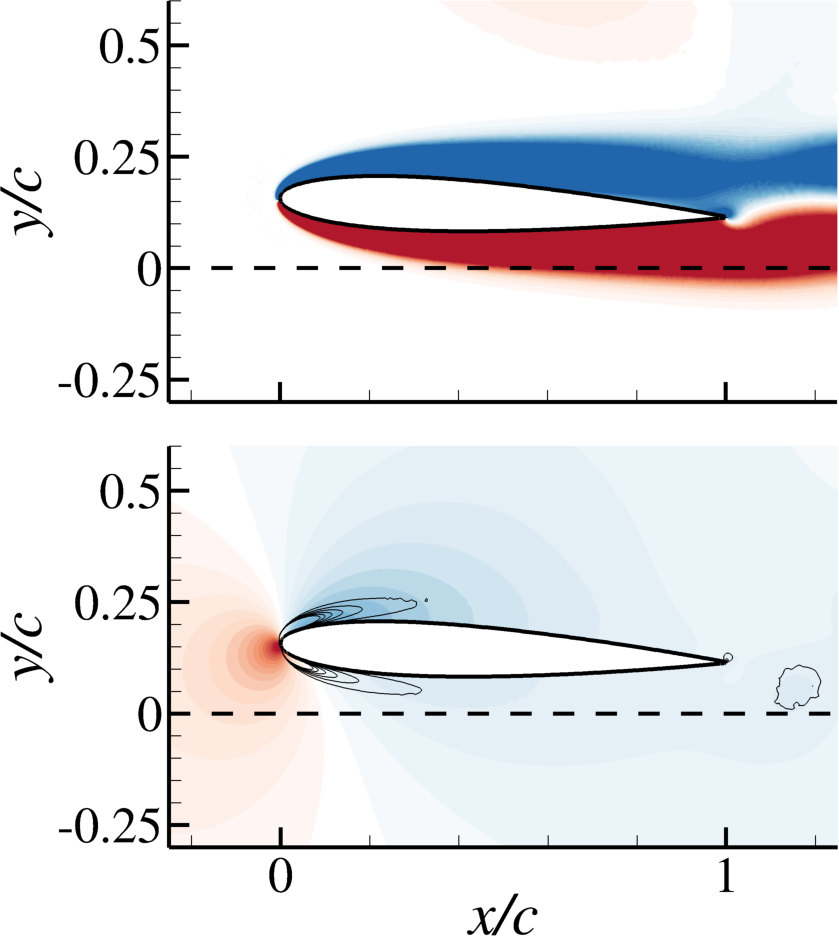}
        \caption{CCW2.5,~$t=6.60$}
	\end{subfigure}
    \begin{subfigure}[b]{0.24\textwidth}
	\centering
        \includegraphics[width=0.95\textwidth]{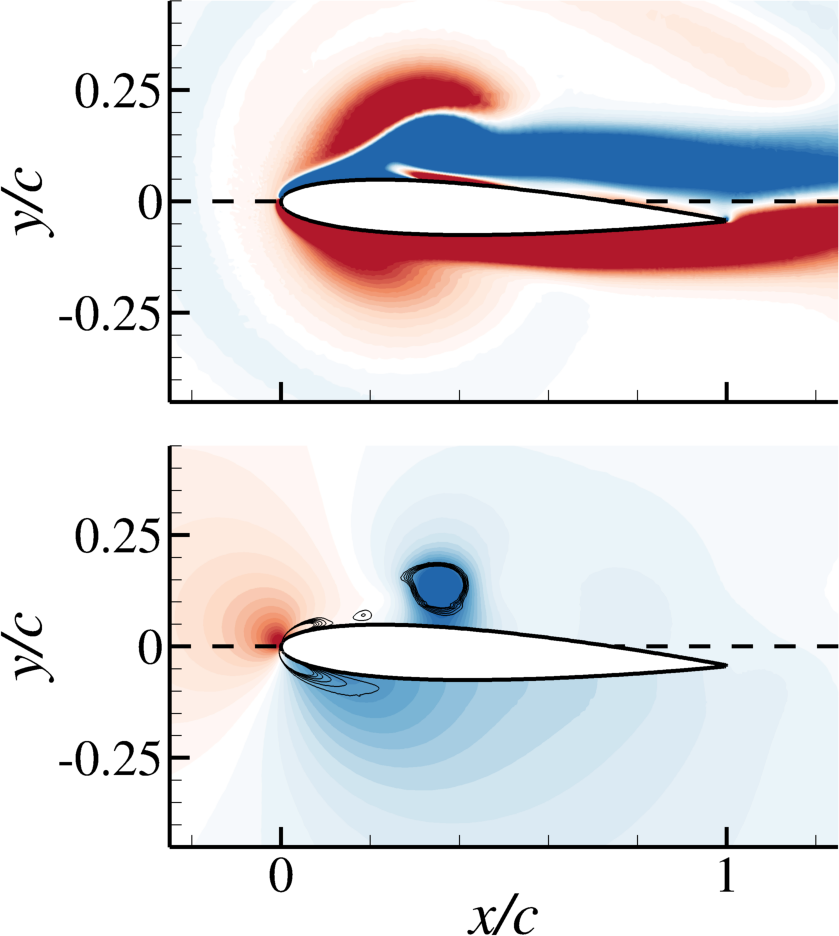}
        \caption{CCW2.5-S,~$t=5.04$}
	\end{subfigure}
    	\begin{subfigure}[b]{0.24\textwidth}
	\centering
        \includegraphics[width=0.95\textwidth]{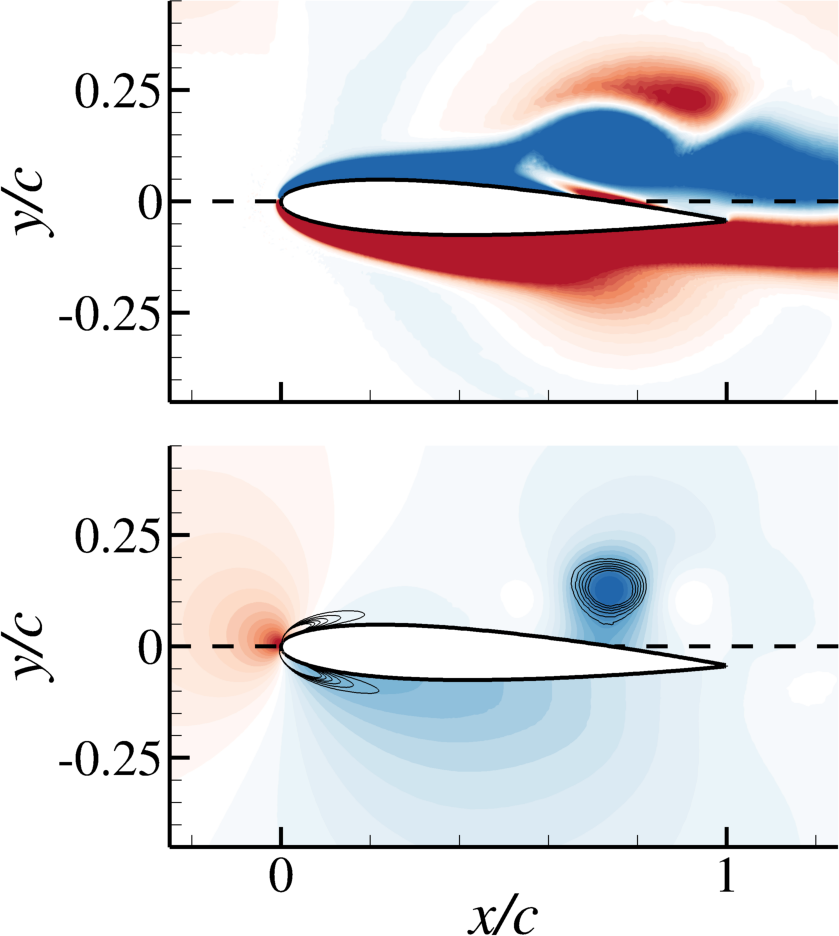}
        \caption{CCW2.5-S,~$t=5.54$}
	\end{subfigure}
    	\begin{subfigure}[b]{0.24\textwidth}
	\centering
        \includegraphics[width=0.95\textwidth]{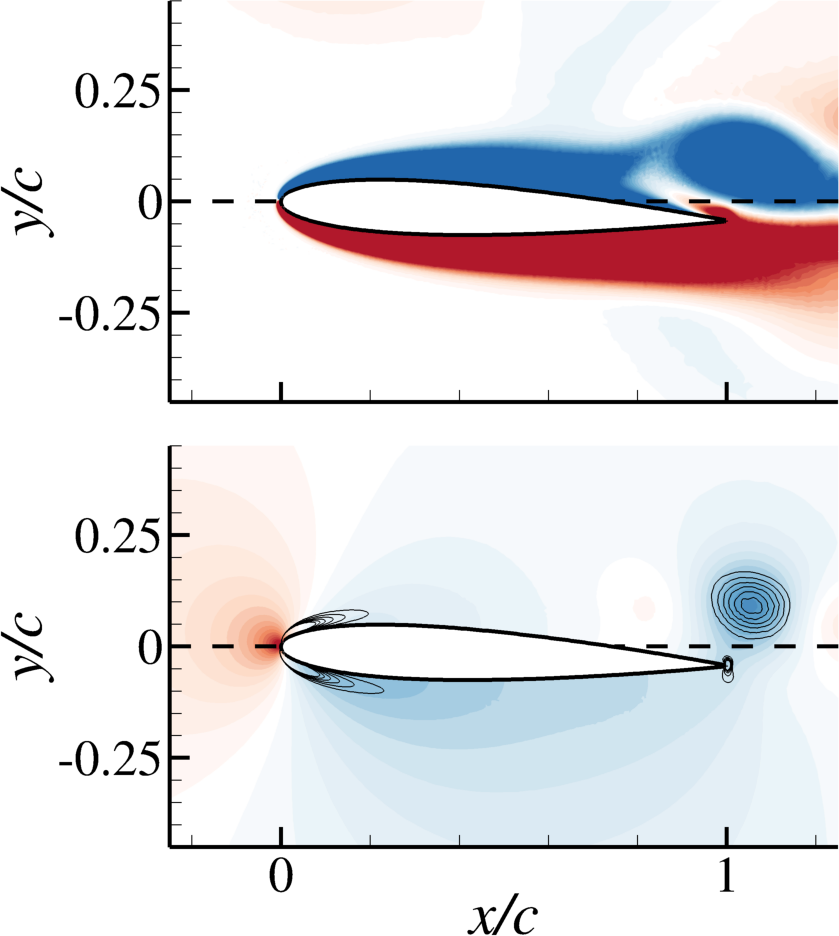}
        \caption{CCW2.5-S,~$t=6.04$}
	\end{subfigure}
    \begin{subfigure}[b]{0.24\textwidth}
	\centering
        \includegraphics[width=0.95\textwidth]{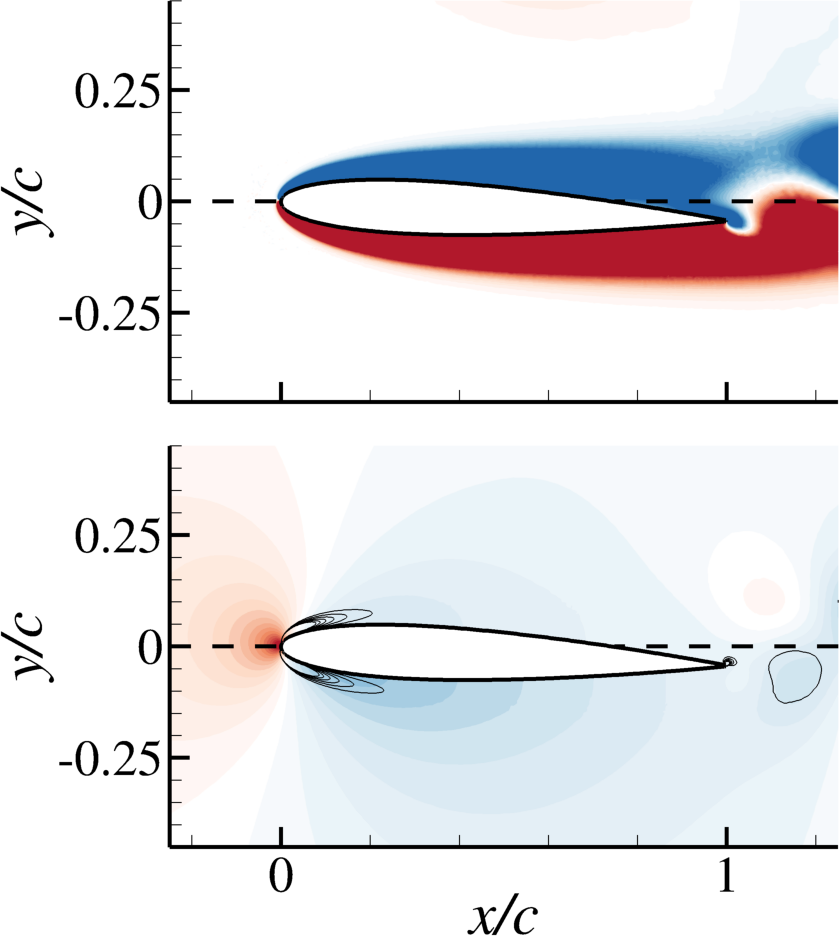}
        \caption{CCW2.5-S,~$t=6.54$}
	\end{subfigure}
     \\[0.2cm]   % space between rows
        \begin{subfigure}[b]{0.75\textwidth}
	\centering
        \includegraphics[width=1\textwidth]{figures_new/legend1_cropped.png}
	\end{subfigure}
\\[0.2cm]   % space between rows
        \begin{subfigure}[b]{0.75\textwidth}
	\centering
        \includegraphics[width=1\textwidth]{figures_new/legend2_cropped.png}
	\end{subfigure}
\caption{Spanwise-vorticity contours (top) and pressure contours (bottom) in the region around the airfoil after vortex impingement for CCW2.5 and CCW2.5-S. In the pressure panels, contour lines of the \(Q\)-criterion are overlaid to identify vortical structures.}
\label{f:cCCW2.5}	
\end{figure}

The CCW interaction also shows a discrepancy between the modeled and simulated freely-flying lift after impingement, but with different timing and magnitude from the CW interaction. As shown in Fig.~\ref{f:reconCCW2.5}(b), $\widetilde{C}_L$ exceeds $C_L$ over a portion of the post-impingement stage, although this interval is shorter than in the CW interaction. Examining the vorticity and pressure contours in Fig.~\ref{f:cCCW2.5} reveals a markedly different vortex-shedding behavior. In particular, vortex shedding along the lower surface is suppressed, in contrast to the enhanced vortex shedding along the upper surface observed for the CW vortex interaction. Furthermore, for the stationary airfoil, a pronounced vortex roll-up occurs within the shear layer along the upper surface following the passage of the primary vortex, leading to a strong low-pressure region and boundary-layer separation. For the freely-flying airfoil, this behavior is significantly alleviated, likely due to the upward motion of the airfoil. Since the dominant difference between CCW2.5 and CCW2.5-S is associated with the vortex dynamics along the upper surface, once this vortex convects downstream past the airfoil, $C_L$ exhibits a rebound and becomes larger in absolute magnitude (i.e., more negative) than $\widetilde{C}_L$. As a result, the duration over which the modeled $\widetilde{C}_L$ exceeds the freely-flying value is reduced to $\Delta t=1$.
In summary, the dominant contribution arises from the upper-surface vortex, which is stronger than in the CW vortex example, but persists over a shorter time interval.

\par
\begin{figure}
 \centering
     \begin{subfigure}[b]{0.49\textwidth}
	\centering
        \includegraphics[width=0.99\textwidth]{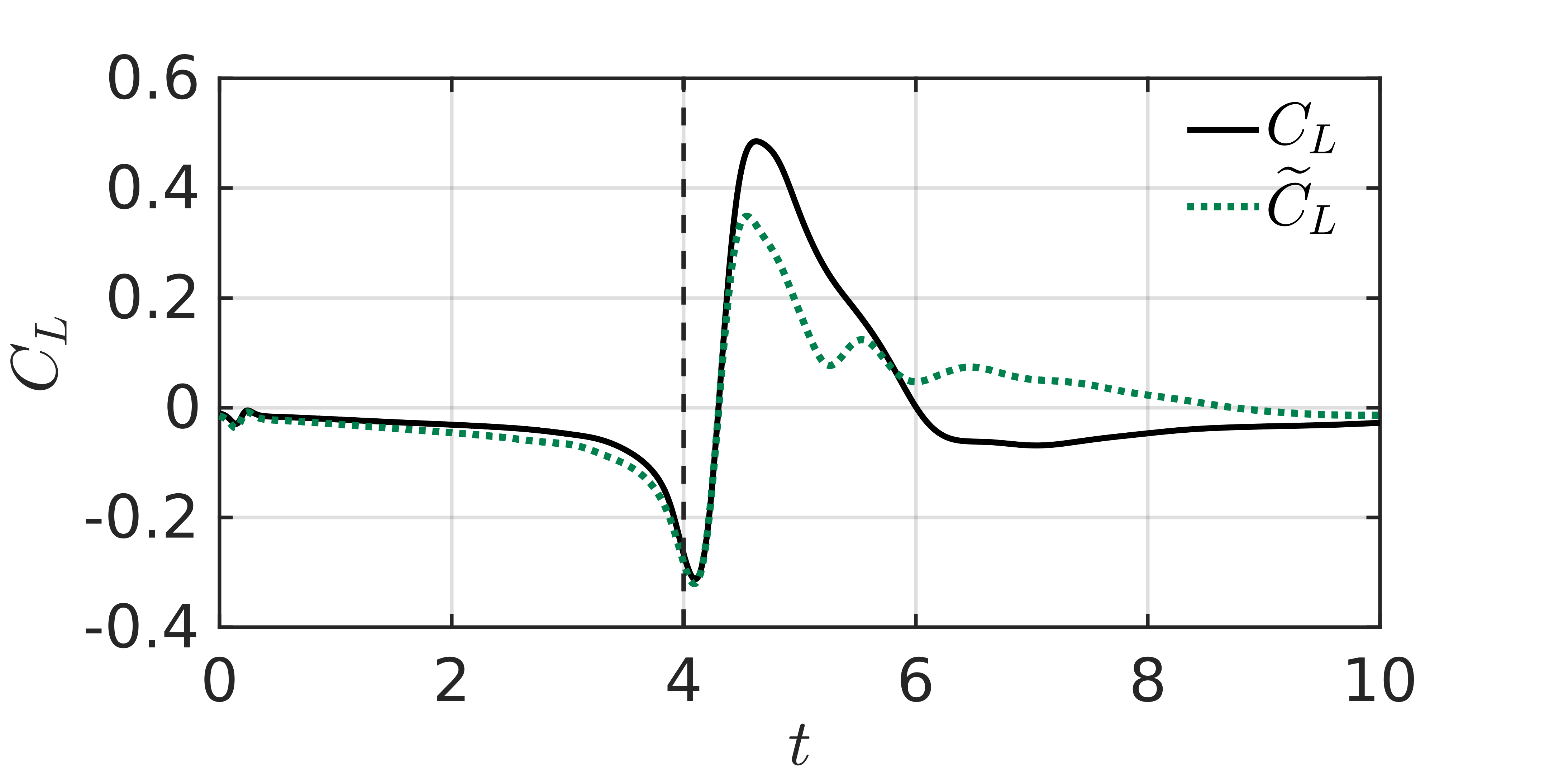}
        \caption{CW0}
	\end{subfigure}
    \begin{subfigure}[b]{0.49\textwidth}
	\centering
        \includegraphics[width=0.99\textwidth]{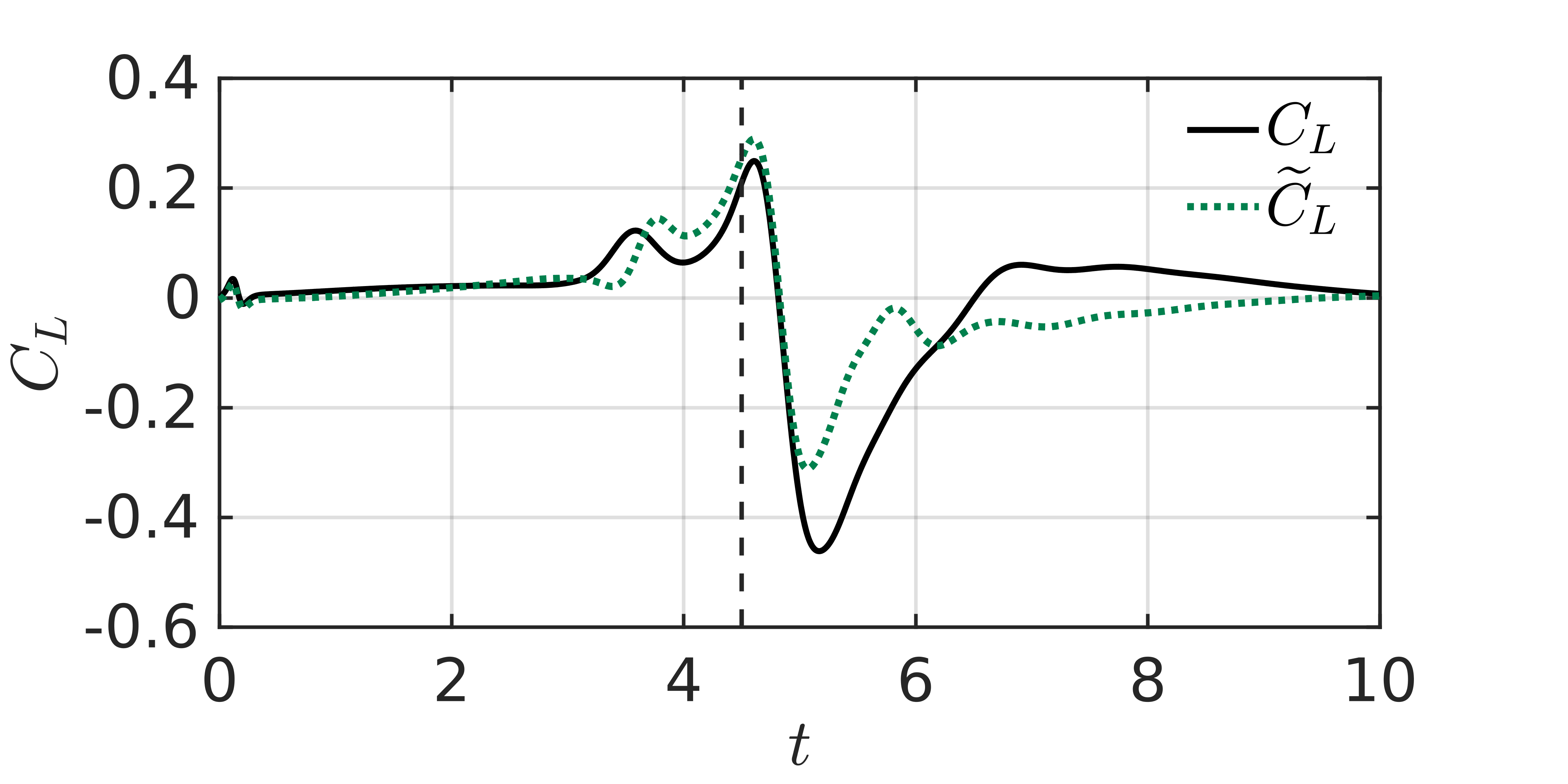}
        \caption{CCW0}
	\end{subfigure}
         \begin{subfigure}[b]{0.49\textwidth}
	\centering
        \includegraphics[width=0.99\textwidth]{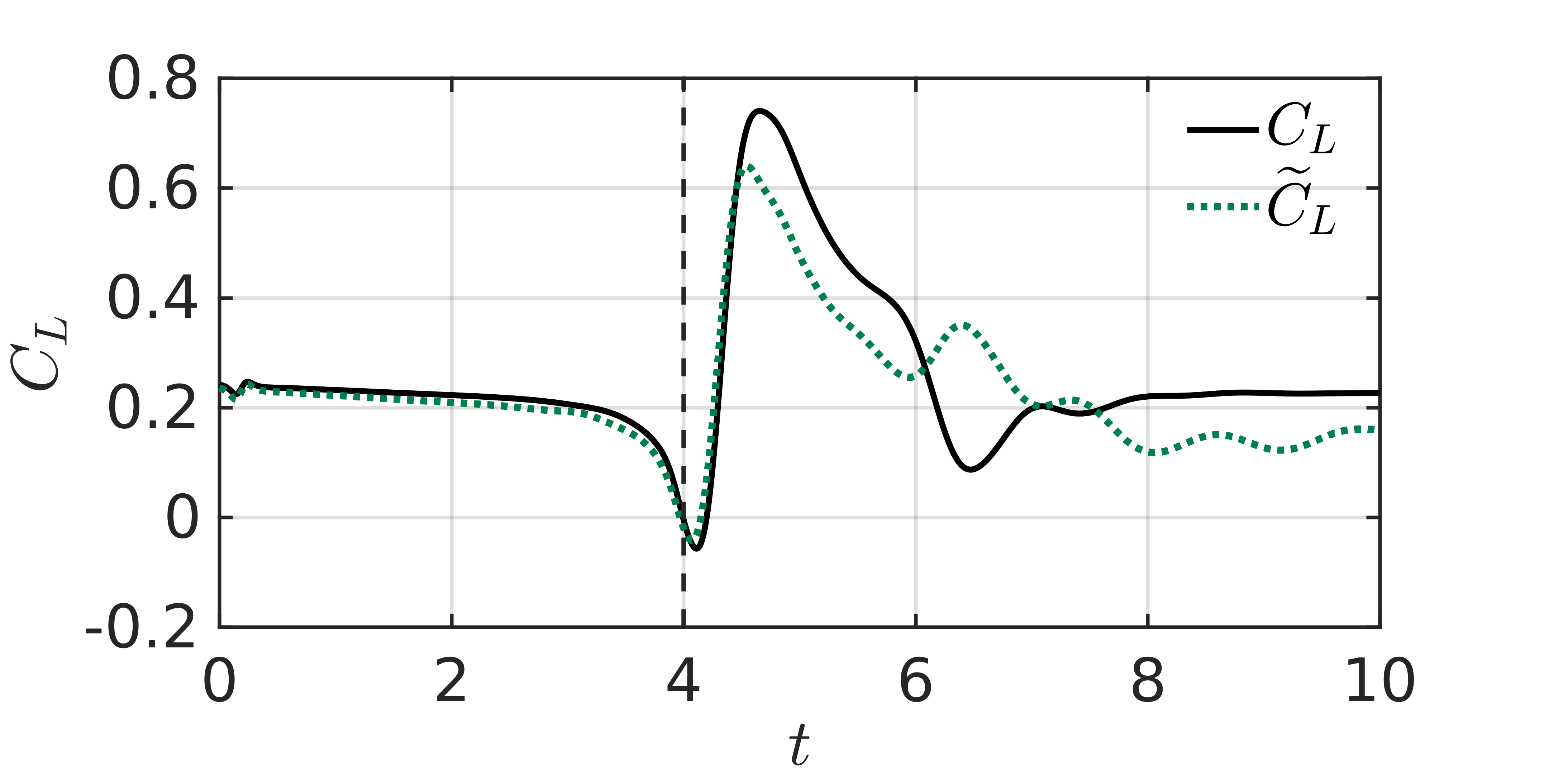}
        \caption{CW5}
	\end{subfigure}
    \begin{subfigure}[b]{0.49\textwidth}
	\centering
        \includegraphics[width=0.99\textwidth]{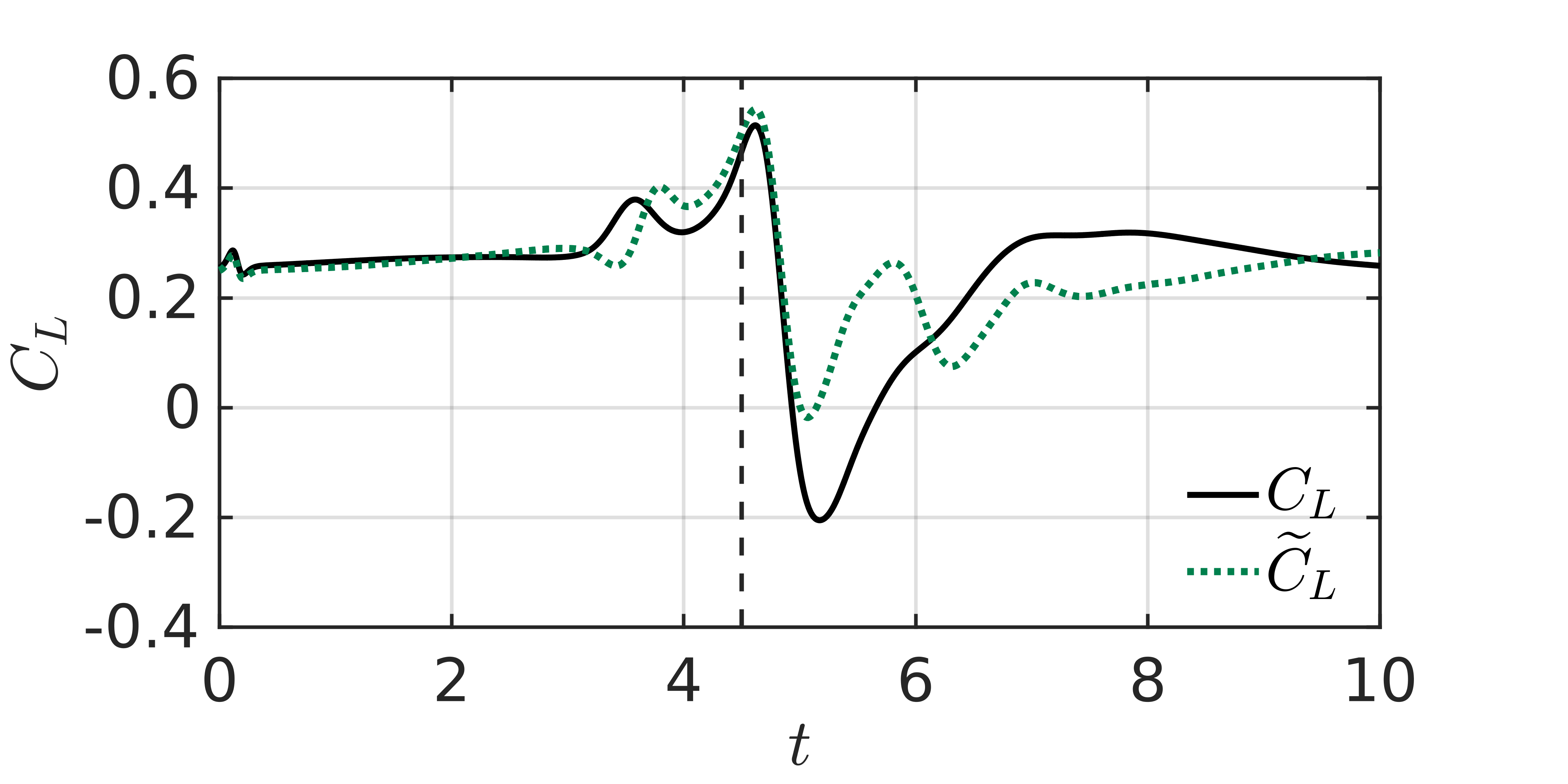}
        \caption{CCW5}
	\end{subfigure}
\caption{Comparison between $\widetilde{C}_L$ and $C_L$ for various direct impingement cases. Dashed lines correspond to approximate impingement time.}
 \label{f:cor4}	
 \end{figure}

Comparisons between $\widetilde{C}_L$ and $C_L$ for airfoils at different angles of attack are shown in Fig.~\ref{f:cor4}, and exhibit similar qualitative trends as previously discussed. In particular, the modeled lift remains in reasonable agreement with the computed freely-flying lift until $0.5$ time units after impingement, and subsequently departs from it as vortex shedding behavior influences the lift. The comparison between CW0 and CW0-S is nearly symmetric to that between CCW0 and CCW0-S, as expected at zero angle of attack, despite the slight asymmetry in $C_L$. In contrast, the comparison between the modeled and actual lift for CW5 and CW5-S exhibits a more prolonged deviation, similar to that observed for CW2.5 and CW2.5-S. These additional cases are not analyzed in detail here, but they support the interpretation developed from CW2.5 and CCW2.5.

\subsection{Effects of airfoil angle of attack and vortex position} 

\subsubsection{Influence of airfoil angle of attack}
\begin{figure}
\centering
	\begin{subfigure}[b]{0.24\textwidth}
	\centering
        \includegraphics[width=0.95\textwidth]{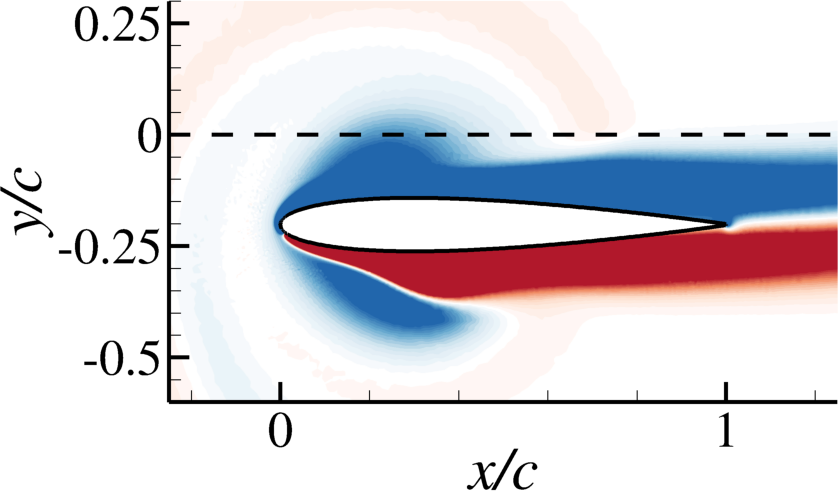}
        \caption{CW0,~$t=4.51$}
	\end{subfigure}
    	\begin{subfigure}[b]{0.24\textwidth}
	\centering
        \includegraphics[width=0.95\textwidth]{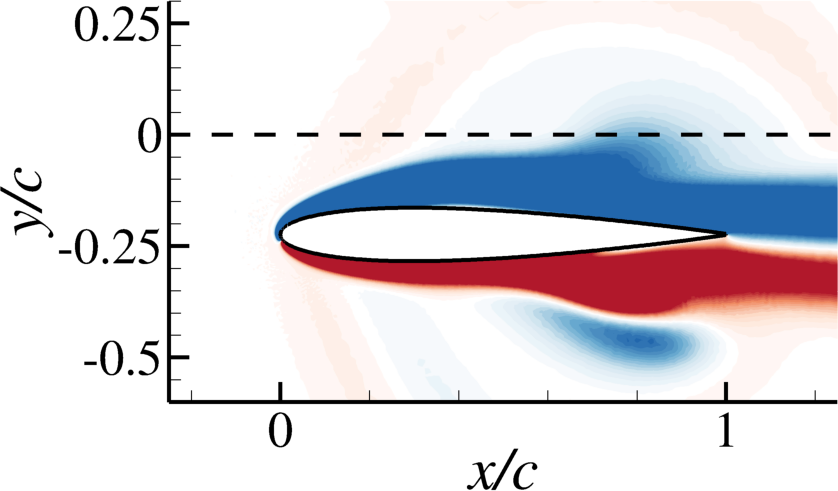}
        \caption{CW0,~$t=5.01$}
	\end{subfigure}
    \begin{subfigure}[b]{0.24\textwidth}
	\centering
        \includegraphics[width=0.95\textwidth]{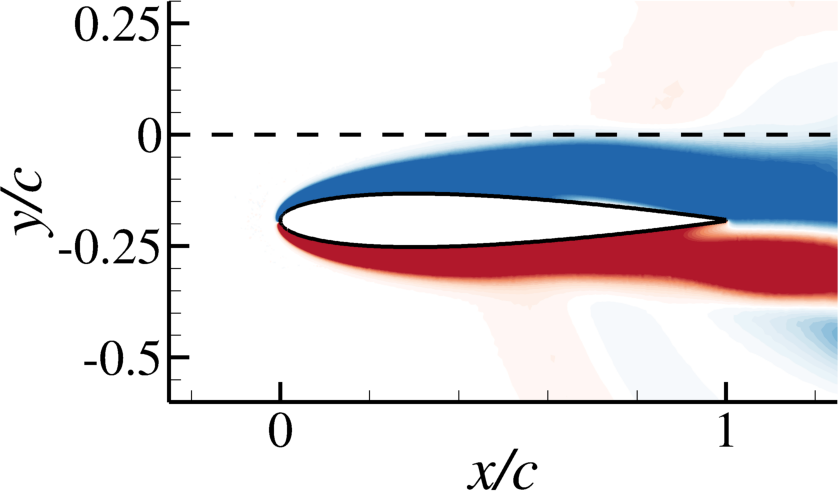}
        \caption{CW0,~$t=5.51$}
	\end{subfigure}
        \begin{subfigure}[b]{0.24\textwidth}
	\centering
        \includegraphics[width=0.95\textwidth]{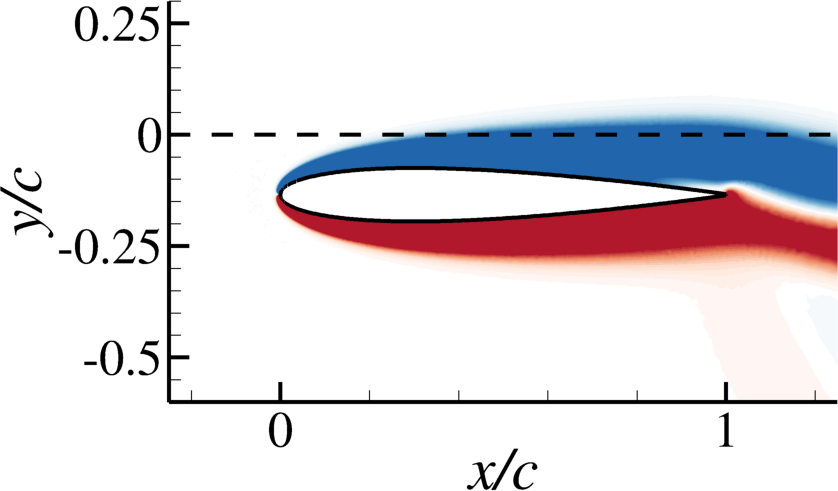}
        \caption{CW0,~$t=6.01$}
	\end{subfigure}
    	\begin{subfigure}[b]{0.24\textwidth}
	\centering
        \includegraphics[width=0.95\textwidth]{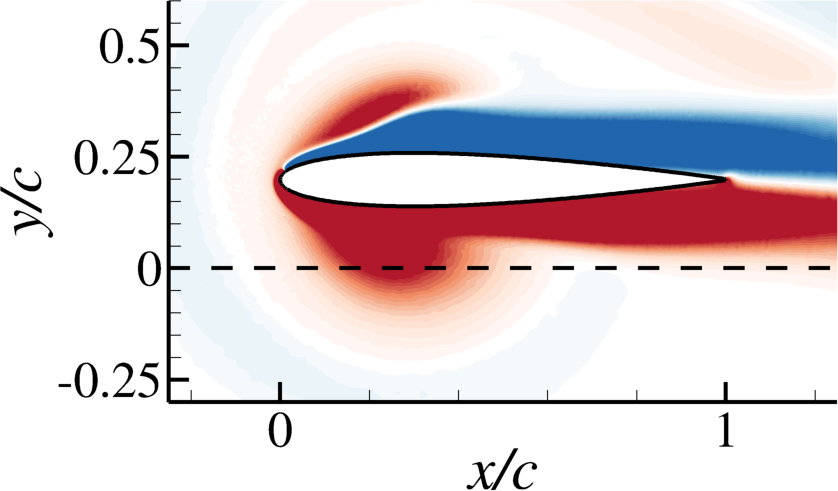}
        \caption{CCW0,~$t=5.07$}
	\end{subfigure}
    	\begin{subfigure}[b]{0.24\textwidth}
	\centering
        \includegraphics[width=0.95\textwidth]{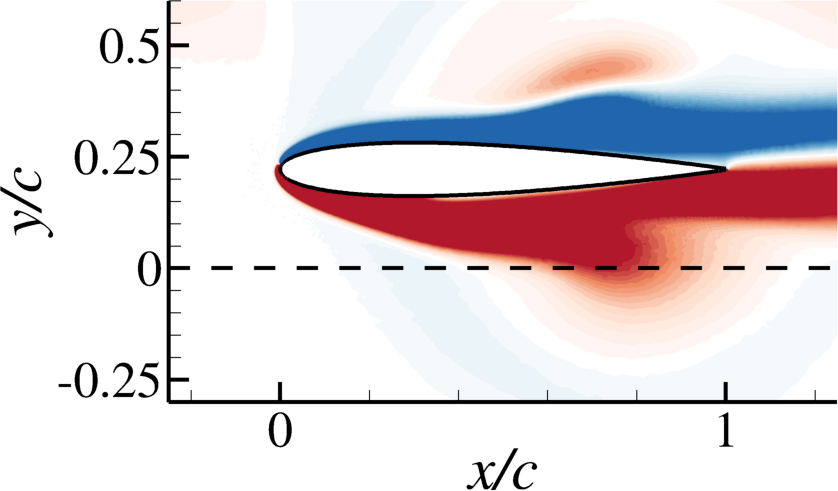}
        \caption{CCW0,~$t=5.57$}
	\end{subfigure}
    \begin{subfigure}[b]{0.24\textwidth}
	\centering
        \includegraphics[width=0.95\textwidth]{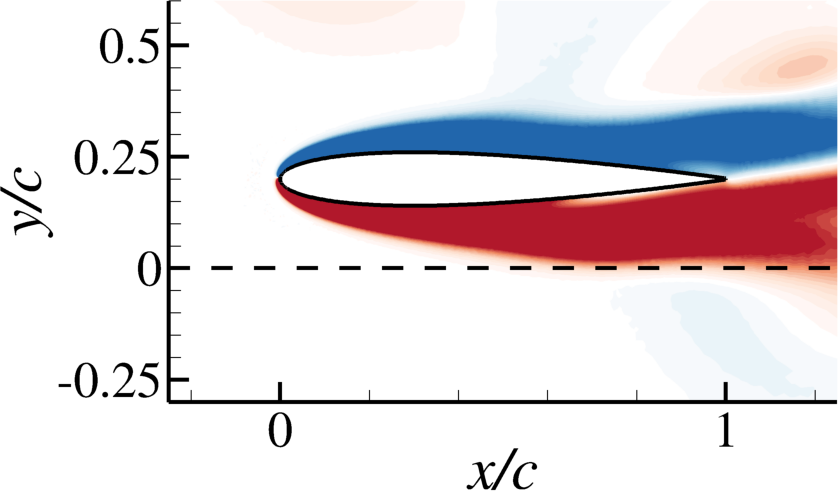}
        \caption{CCW0,~$t=6.07$}
	\end{subfigure}
        \begin{subfigure}[b]{0.24\textwidth}
	\centering
        \includegraphics[width=0.95\textwidth]{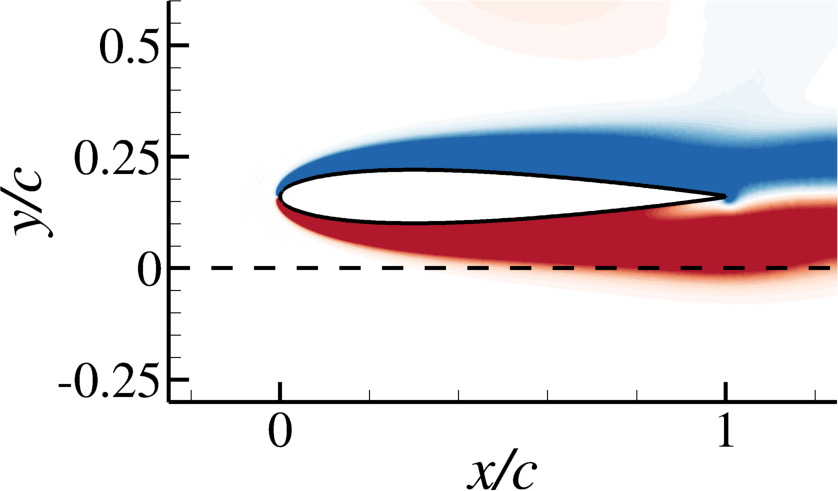}
        \caption{CCW0,~$t=6.57$}
	\end{subfigure}
	\begin{subfigure}[b]{0.24\textwidth}
	\centering
        \includegraphics[width=0.95\textwidth]{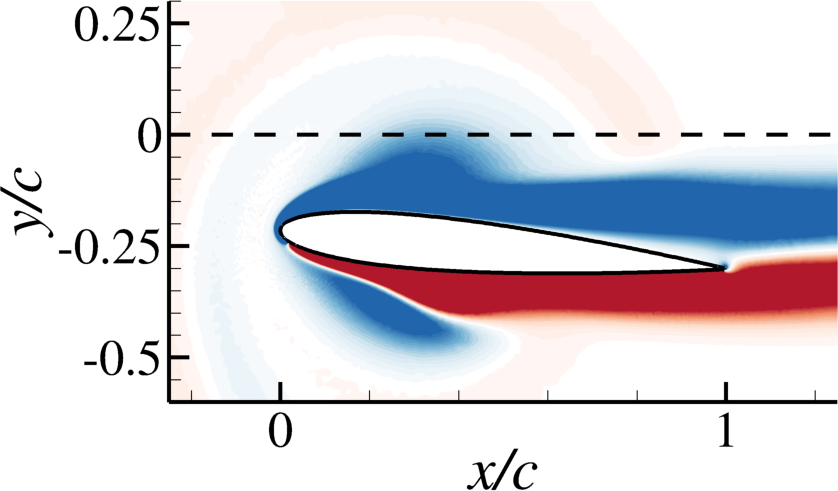}
        \caption{CW5,~$t=4.58$}
	\end{subfigure}
    	\begin{subfigure}[b]{0.24\textwidth}
	\centering
        \includegraphics[width=0.95\textwidth]{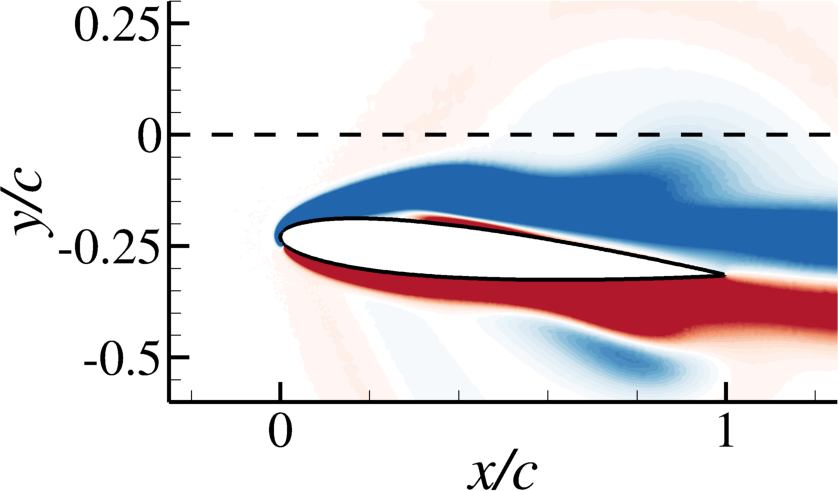}
        \caption{CW5,~$t=5.08$}
	\end{subfigure}
    \begin{subfigure}[b]{0.24\textwidth}
	\centering
        \includegraphics[width=0.95\textwidth]{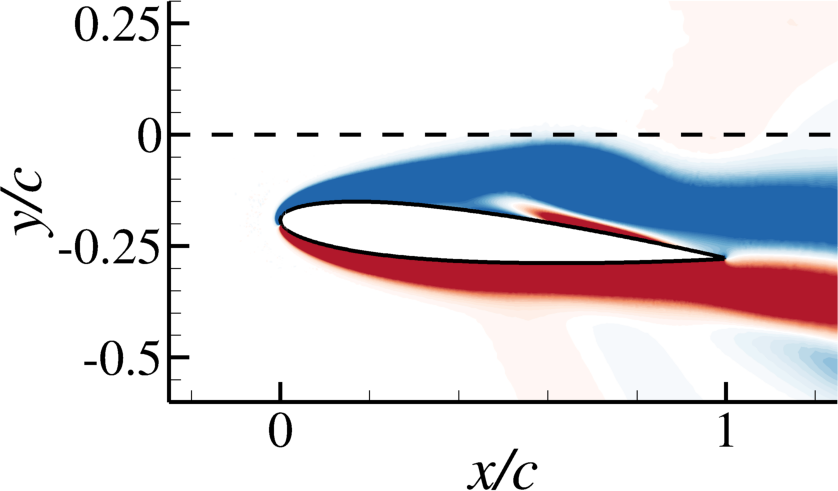}
        \caption{CW5,~$t=5.58$}
	\end{subfigure}
        \begin{subfigure}[b]{0.24\textwidth}
	\centering
        \includegraphics[width=0.95\textwidth]{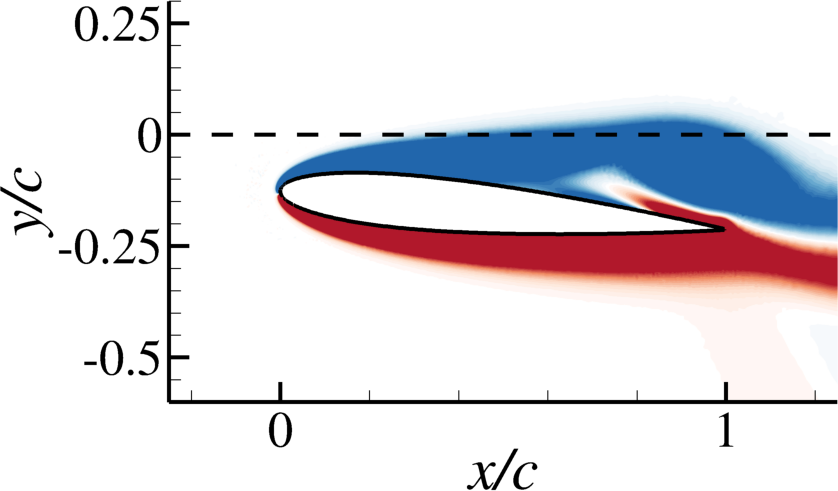}
        \caption{CW5,~$t=6.08$}
	\end{subfigure}
	\begin{subfigure}[b]{0.24\textwidth}
	\centering
        \includegraphics[width=0.95\textwidth]{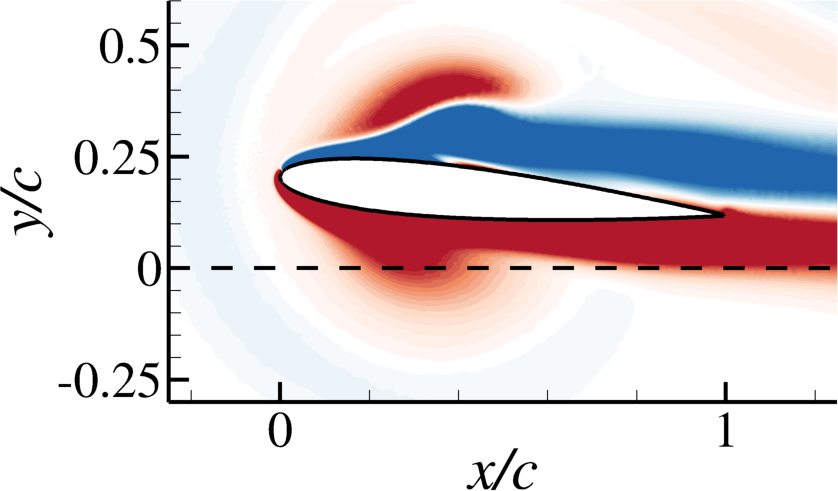}
        \caption{CCW5,~$t=5.13$}
	\end{subfigure}
    	\begin{subfigure}[b]{0.24\textwidth}
	\centering
        \includegraphics[width=0.95\textwidth]{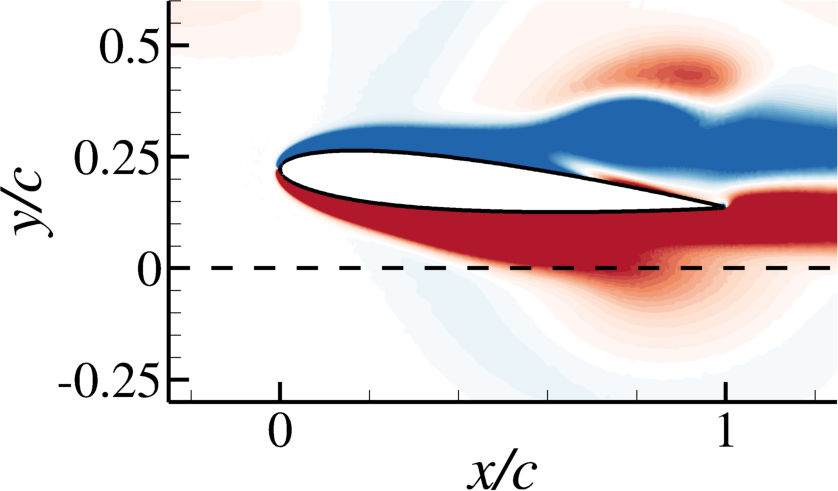}
        \caption{CCW5,~$t=5.63$}
	\end{subfigure}
    \begin{subfigure}[b]{0.24\textwidth}
	\centering
        \includegraphics[width=0.95\textwidth]{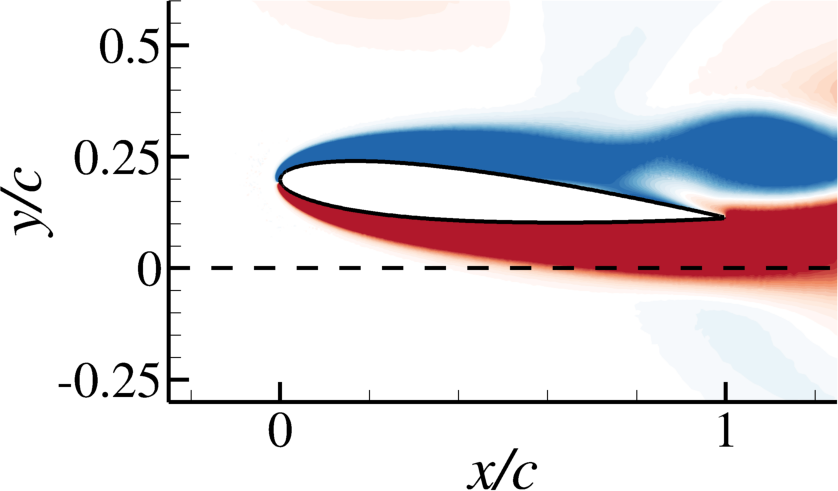}
        \caption{CCW5,~$t=6.13$}
	\end{subfigure}
        \begin{subfigure}[b]{0.24\textwidth}
	\centering
        \includegraphics[width=0.95\textwidth]{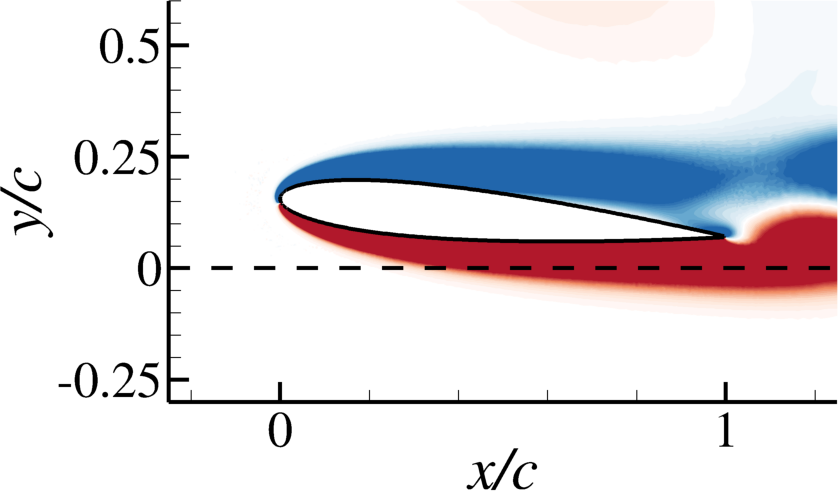}
        \caption{CCW5,~$t=6.63$}
	\end{subfigure}
         \\[0.2cm]   % space between rows
       \begin{subfigure}[b]{0.75\textwidth}
	\centering
        \includegraphics[width=1\textwidth]{figures_new/legend1_cropped.png}
	\end{subfigure}
\caption{Spanwise vorticity contour of the region around the airfoil in the post-impingement stage for CW0, CCW0, CW5, and CCW5.}
\label{f:vcdirect}	
\end{figure}

The angle of attack of the freely-flying airfoil has a strong influence on the vortex-airfoil interaction. 
Thus, Fig.~\ref{f:vcdirect} shows the vorticity contours for CW0, CCW0, CW5, and CCW5 which can be compared to the previous results at angle of attack 2.5 degrees (CW2.5 and CCW2.5). Consistent with the preceding discussion, the overall vortex shedding behavior is nearly symmetric for the $0^\circ$ cases, whereas a much stronger asymmetry is observed for the $5^\circ$ cases. 
\par
From Fig.~\ref{f:vcdirect} and the earlier figures, Fig.~\ref{f:cCW2.5} and Fig.~\ref{f:cCCW2.5}, two main types of vortex shedding behavior can be identified. The first occurs immediately after impingement as a roll-up within the shear layer accompanying the passage of the impinging vortex. This type of shedding arises when the split vortex structure passes through a shear layer whose vorticity has the opposite sign, such as along the upper surface for CCW interactions and along the lower surface for CW interactions. Representative examples can be observed along the upper surface in Figs.~\ref{f:cCCW2.5}(b) and \ref{f:vcdirect}(n), and along the lower surface in Fig.~\ref{f:vcdirect}(b). For the CCW interactions, the upper-surface shedding becomes stronger as the angle of attack increases, whereas for the CW interactions the lower-surface shedding becomes weaker. This trend for CCW interactions is consistent with previous observations for stationary-airfoil vortex interactions at transitional Reynolds number~\cite{barnes2020angle}, although the present study include the additional influence of freely-flying motion. In addition, the heaving motion generally weakens this type of shedding relative to the corresponding stationary airfoil.
\par
The second type of vortex shedding does not occur immediately after impingement, but instead results from the roll-up and subsequent shedding of vorticity accumulated near the leading edge after the split vortex structure has passed. This type of shedding arises when the split vortex passes through a shear layer whose vorticity has the same sign, such as in Figs.~\ref{f:cCCW2.5}(b,c) and \ref{f:vcdirect}(j,k). A similar dependence on angle of attack is observed for this mode, with upper-surface shedding becoming stronger as the angle of attack increases, whereas lower-surface shedding becomes weaker. However, at lower angles of attack this shedding is often so weak that the shed structure is barely identifiable as a vortex. It becomes most evident along the upper surface at larger angles of attack, possibly because of the pre-existing accumulation of vorticity there. Unlike the first shedding mode, the heaving motion tends to promote this delayed shedding when it occurs. Since the shed vortex leaves the airfoil after the primary impinging vortex, it contributes to the extended duration over which the $\widetilde{C}_L$ differs from the heaving result in the comparisons of CW2.5 with CW2.5-S and CW5 with CW5-S.
\par
It is also observed that, at larger angles of attack, the boundary layer along the upper surface is more prone to separate from the airfoil. This is expected, since the baseline flow is already closer to stall and is therefore more susceptible to separation when perturbed by the incoming vortex. In case CW5, this leads to a more oscillatory lift response after the vortex has passed, as shown in Fig.~\ref{f:hclt}.
\subsubsection{Influence of vortex position}
\begin{figure}
\centering
	\begin{subfigure}[b]{0.24\textwidth}
	\centering
        \includegraphics[width=0.95\textwidth]{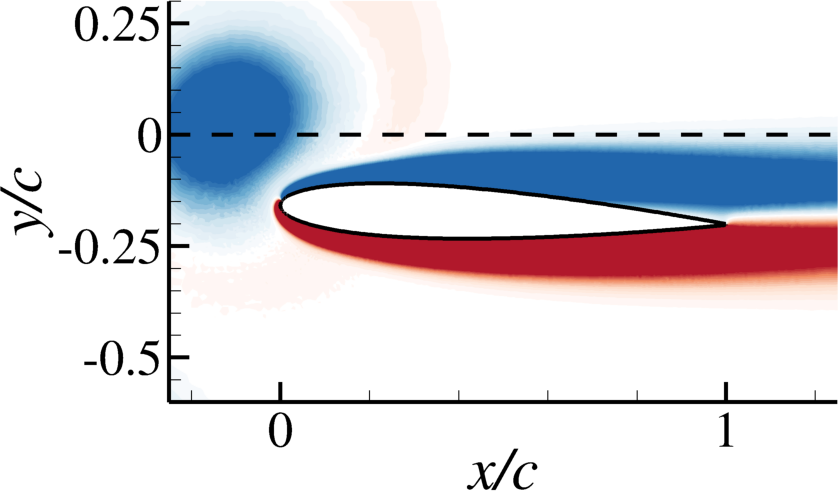}
        \caption{CW2.5a,~$t=4.5$}
	\end{subfigure}
    	\begin{subfigure}[b]{0.24\textwidth}
	\centering
        \includegraphics[width=0.95\textwidth]{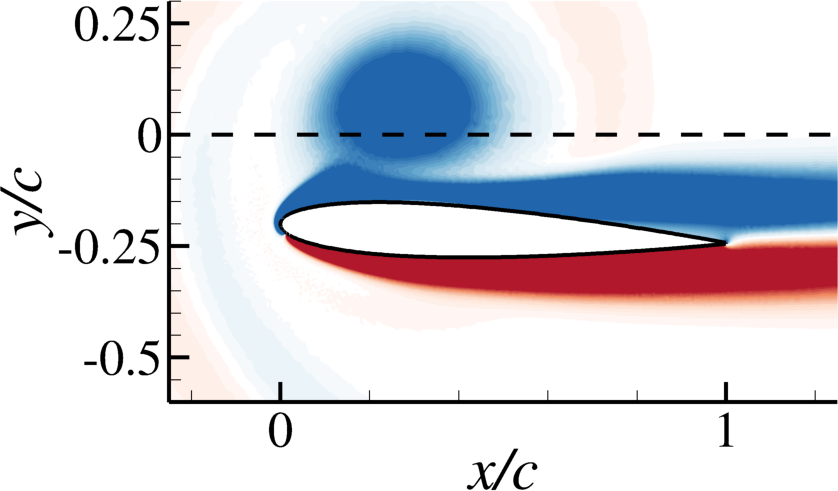}
        \caption{CW2.5a,~$t=5.0$}
	\end{subfigure}
    \begin{subfigure}[b]{0.24\textwidth}
	\centering
        \includegraphics[width=0.95\textwidth]{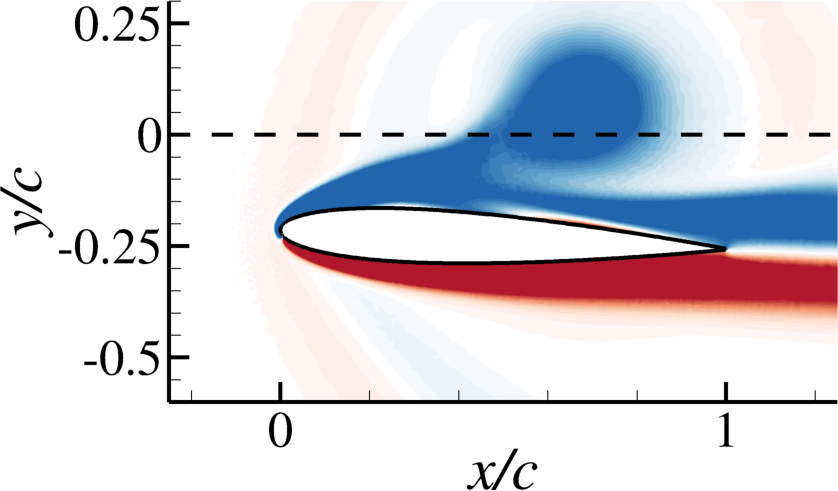}
        \caption{CW2.5a,~$t=5.5$}
	\end{subfigure}
        \begin{subfigure}[b]{0.24\textwidth}
	\centering
        \includegraphics[width=0.95\textwidth]{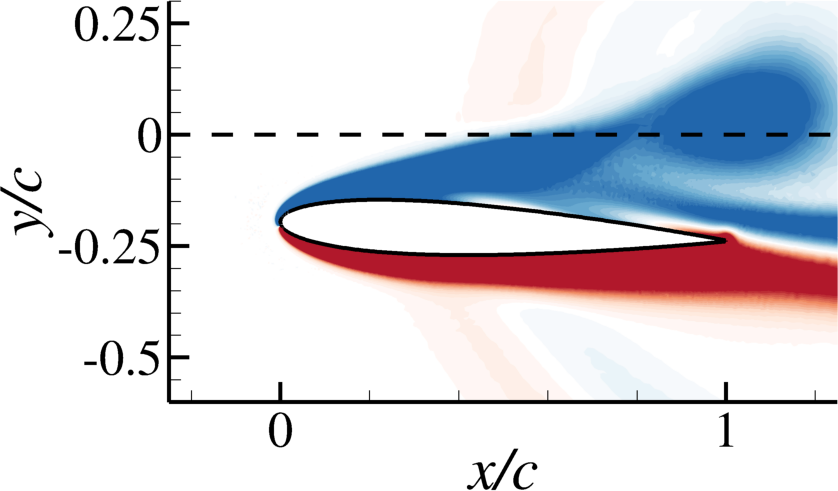}
        \caption{CW2.5a,~$t=6.0$}
	\end{subfigure}
    	\begin{subfigure}[b]{0.24\textwidth}
	\centering
        \includegraphics[width=0.95\textwidth]{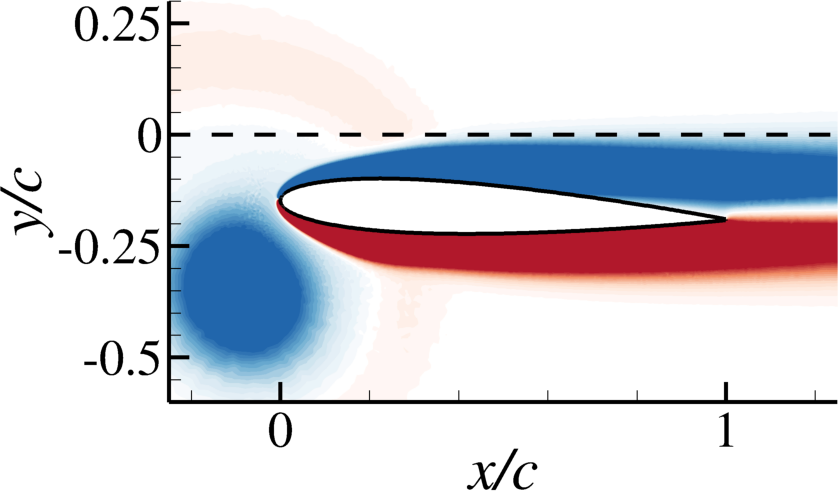}
        \caption{CW2.5b,~$t=4.5$}
	\end{subfigure}
    	\begin{subfigure}[b]{0.24\textwidth}
	\centering
        \includegraphics[width=0.95\textwidth]{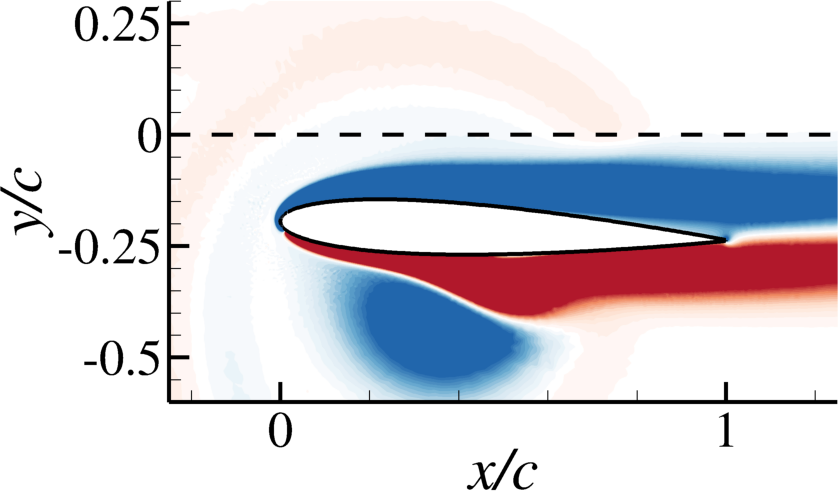}
        \caption{CW2.5b,~$t=5.0$}
	\end{subfigure}
    \begin{subfigure}[b]{0.24\textwidth}
	\centering
        \includegraphics[width=0.95\textwidth]{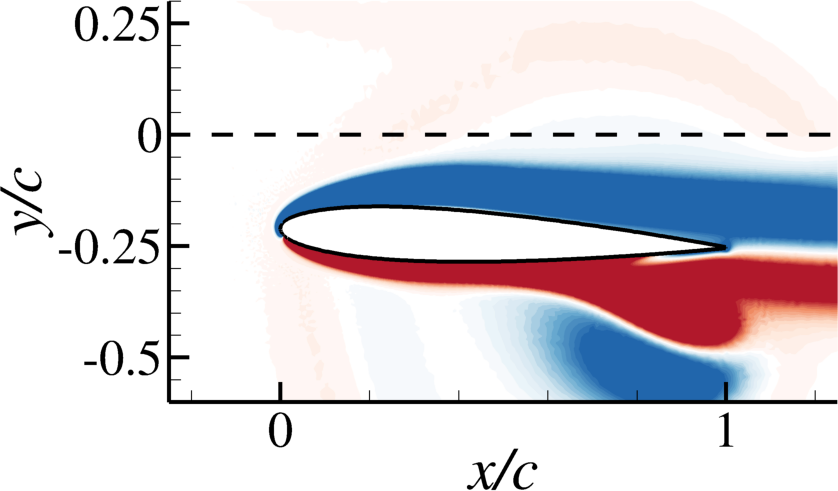}
        \caption{CW2.5b,~$t=5.5$}
	\end{subfigure}
        \begin{subfigure}[b]{0.24\textwidth}
	\centering
        \includegraphics[width=0.95\textwidth]{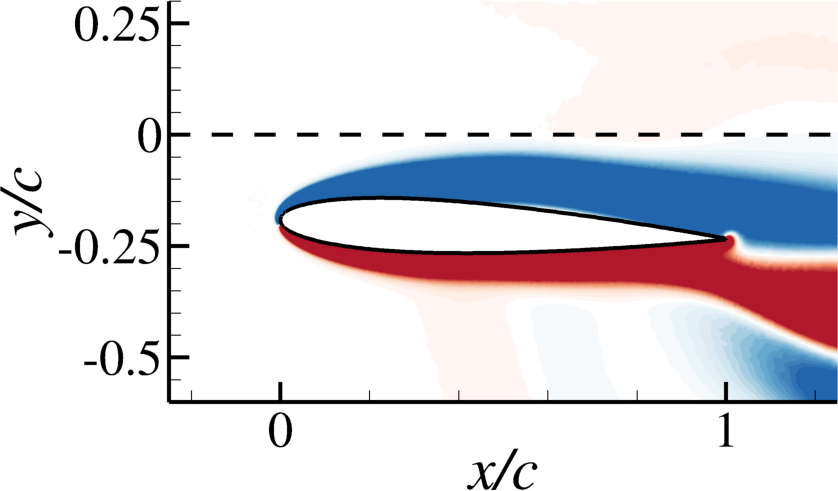}
        \caption{CW2.5b,~$t=6.0$}
	\end{subfigure}
	\begin{subfigure}[b]{0.24\textwidth}
	\centering
        \includegraphics[width=0.95\textwidth]{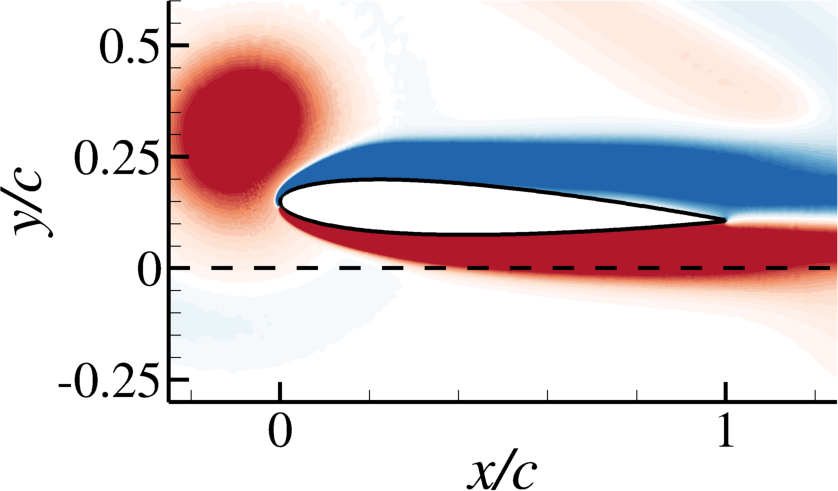}
        \caption{CCW2.5a,~$t=4.5$}
	\end{subfigure}
    	\begin{subfigure}[b]{0.24\textwidth}
	\centering
        \includegraphics[width=0.95\textwidth]{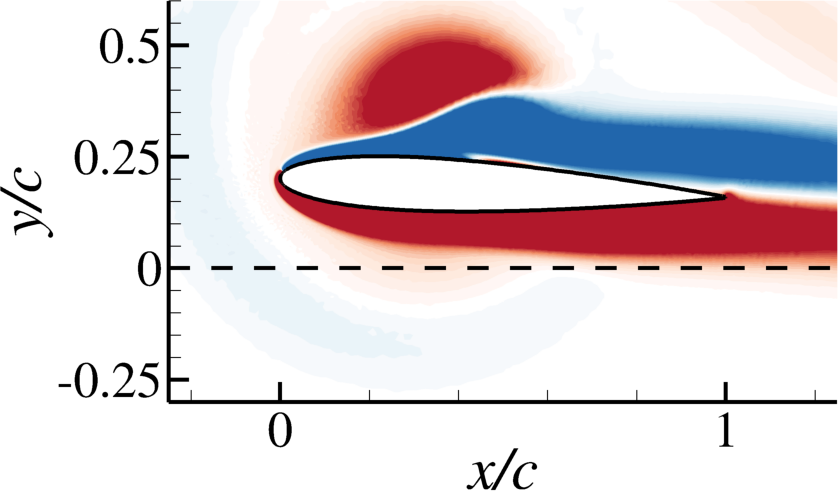}
        \caption{CCW2.5a,~$t=5.0$}
	\end{subfigure}
    \begin{subfigure}[b]{0.24\textwidth}
	\centering
        \includegraphics[width=0.95\textwidth]{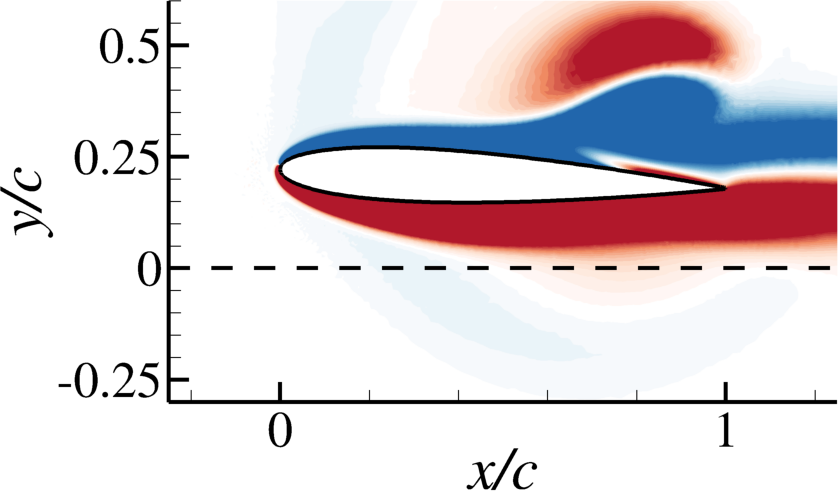}
        \caption{CCW2.5a,~$t=5.5$}
	\end{subfigure}
        \begin{subfigure}[b]{0.24\textwidth}
	\centering
        \includegraphics[width=0.95\textwidth]{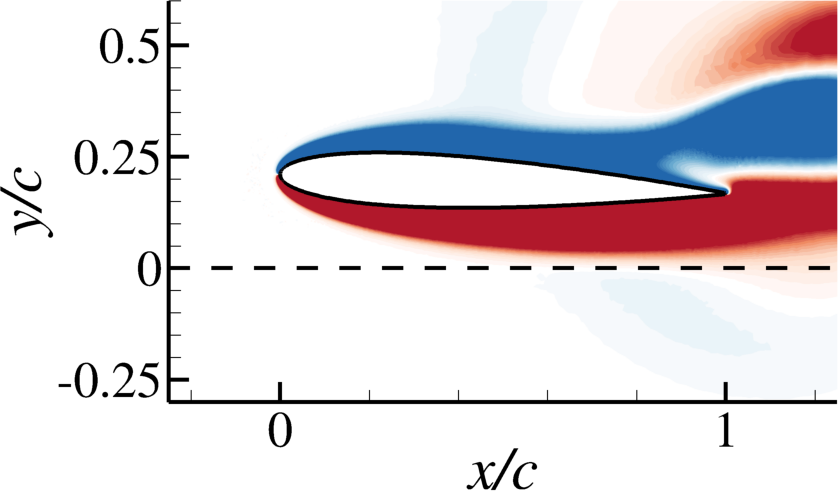}
        \caption{CCW2.5a,~$t=6.0$}
	\end{subfigure}
	\begin{subfigure}[b]{0.24\textwidth}
	\centering
        \includegraphics[width=0.95\textwidth]{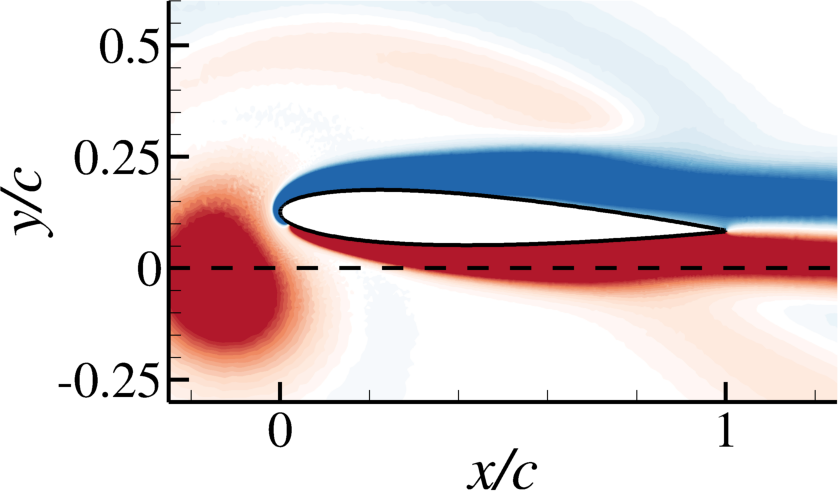}
        \caption{CCW2.5b,~$t=4.5$}
	\end{subfigure}
    	\begin{subfigure}[b]{0.24\textwidth}
	\centering
        \includegraphics[width=0.95\textwidth]{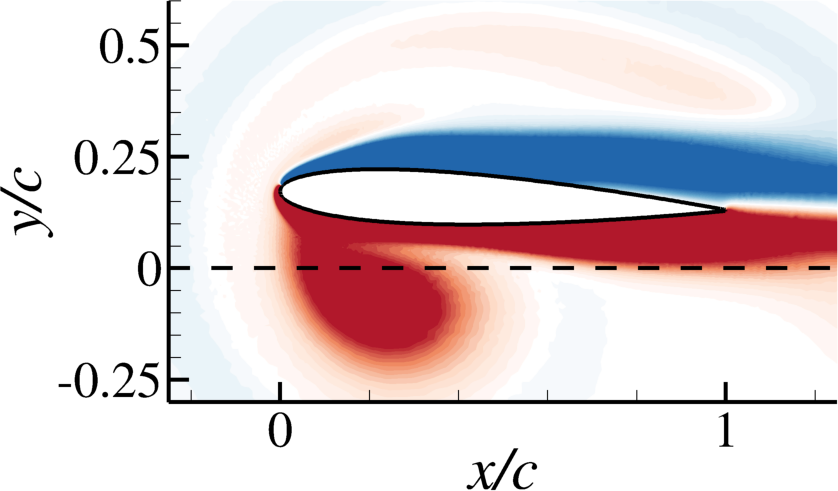}
        \caption{CCW2.5b,~$t=5.0$}
	\end{subfigure}
    \begin{subfigure}[b]{0.24\textwidth}
	\centering
        \includegraphics[width=0.95\textwidth]{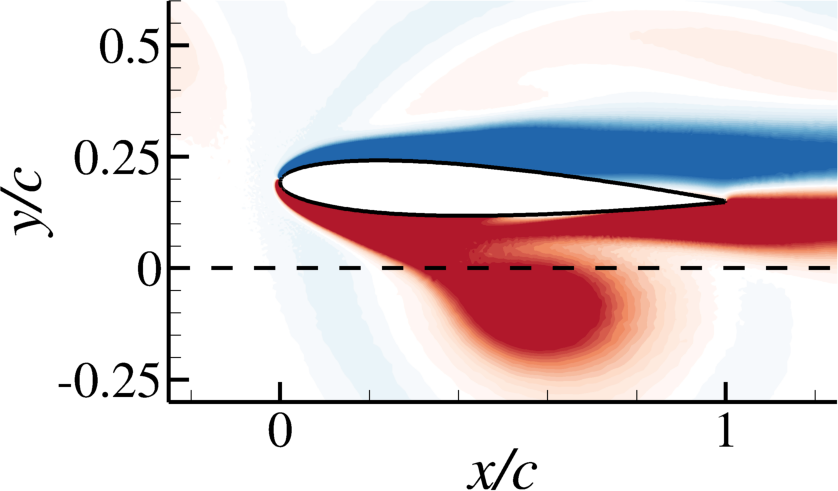}
        \caption{CCW2.5b,~$t=5.5$}
	\end{subfigure}
        \begin{subfigure}[b]{0.24\textwidth}
	\centering
        \includegraphics[width=0.95\textwidth]{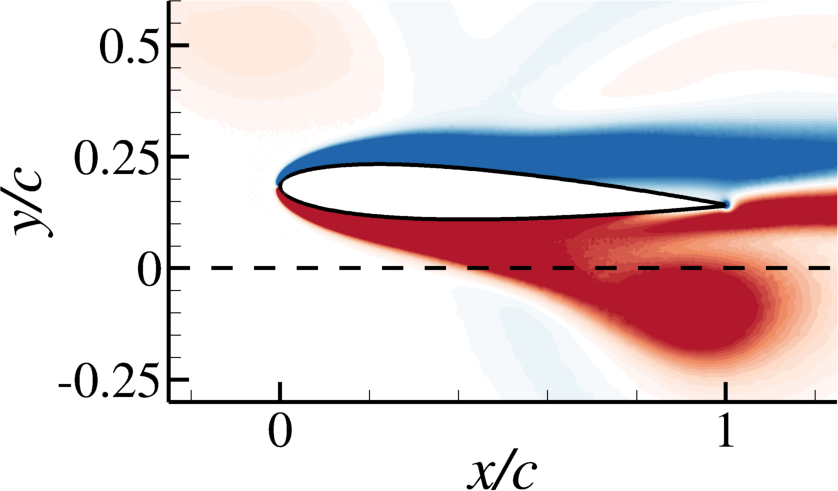}
        \caption{CCW2.5b,~$t=6.0$}
	\end{subfigure}
     \\[0.2cm]   % space between rows
       \begin{subfigure}[b]{0.75\textwidth}
	\centering
        \includegraphics[width=1\textwidth]{figures_new/legend1_cropped.png}
	\end{subfigure}
\caption{Spanwise vorticity contour of the region around the airfoil in the post-impingement stage for CW2.5a, CW2.5b, CCW2.5a, and CCW2.5b.}
\label{f:vcindirect}	
\end{figure}

Fig.~\ref{f:vcindirect} shows the vorticity contours for CW2.5a, CW2.5b, CCW2.5a, and CCW2.5b. Unlike for direct-impingement, when the vortex passes along only one side of the airfoil, its influence is primarily confined to that side.
\par
In CW2.5b and CCW2.5a, the vortex passes along the side where the boundary-layer vorticity has the opposite sign. As a result, a vortex shedding behavior similar to the first shedding mode identified in the previous subsection is observed. Although the airfoil heaves toward the vortex as it convects over the surface, the vortex is nevertheless displaced away from the airfoil, likely because the newly generated shed vortex grows between the primary vortex and the airfoil surface. In CW2.5a and CCW2.5b, by contrast, the vortex passes along a shear layer of the same sign. In these cases, the airfoil also moves farther away from the vortex during the convection process because it heaves away from the vortex. This is likely the reason why case CW2.5a has the least rebound among the CW cases as can be seen in Fig.~\ref{f:hclt}.
\par
Among these cases, stronger boundary-layer separation is observed in CW2.5a and CCW2.5a, where the vortex passes over the upper surface of the airfoil. This is consistent with the trend identified in the previous subsection, namely that the upper-surface boundary layer is more prone to separation because of the positive angle of attack of the airfoil. However, the coupled effects of angle of attack and off-center vortex passage are not examined in detail here, and the effect of varying the separation distance between the vortex and the airfoil is also beyond the scope of the present study.

\subsection{Predicting the heave trajectory from the stationary airfoil lift data}

The heave trajectory prediction from the stationary lift coefficient, developed in Sec.~\ref{sec:theory}, is applied to direct-impingement interactions at $\alpha=2.5^\circ$ for density ratios, $\rho_s/\rho_f=50$, $20$, $10$, and $5$. The density ratio is varied because it enters the heave equation directly through the airfoil mass and provides a simple way to modify the amplitude of the heave response. This allows the prediction accuracy to be assessed over different levels of motion induced coupling between the airfoil and the surrounding flow.

\begin{figure}
\centering
	\begin{subfigure}[b]{0.48\textwidth}
	\centering
        \includegraphics[width=0.95\textwidth]{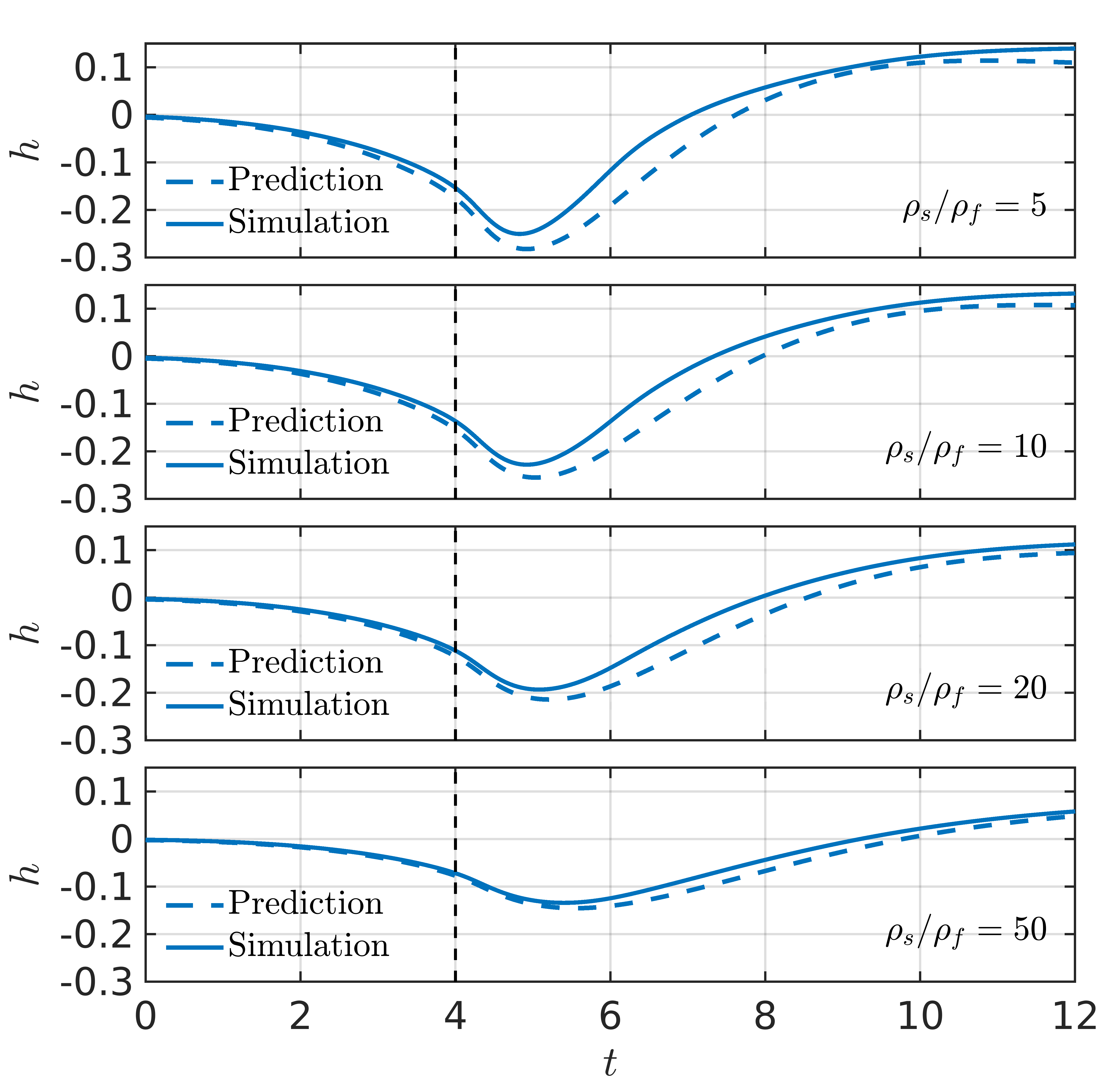}
        \caption{CW predictions}
	\end{subfigure}
    	\begin{subfigure}[b]{0.48\textwidth}
	\centering
        \includegraphics[width=0.95\textwidth]{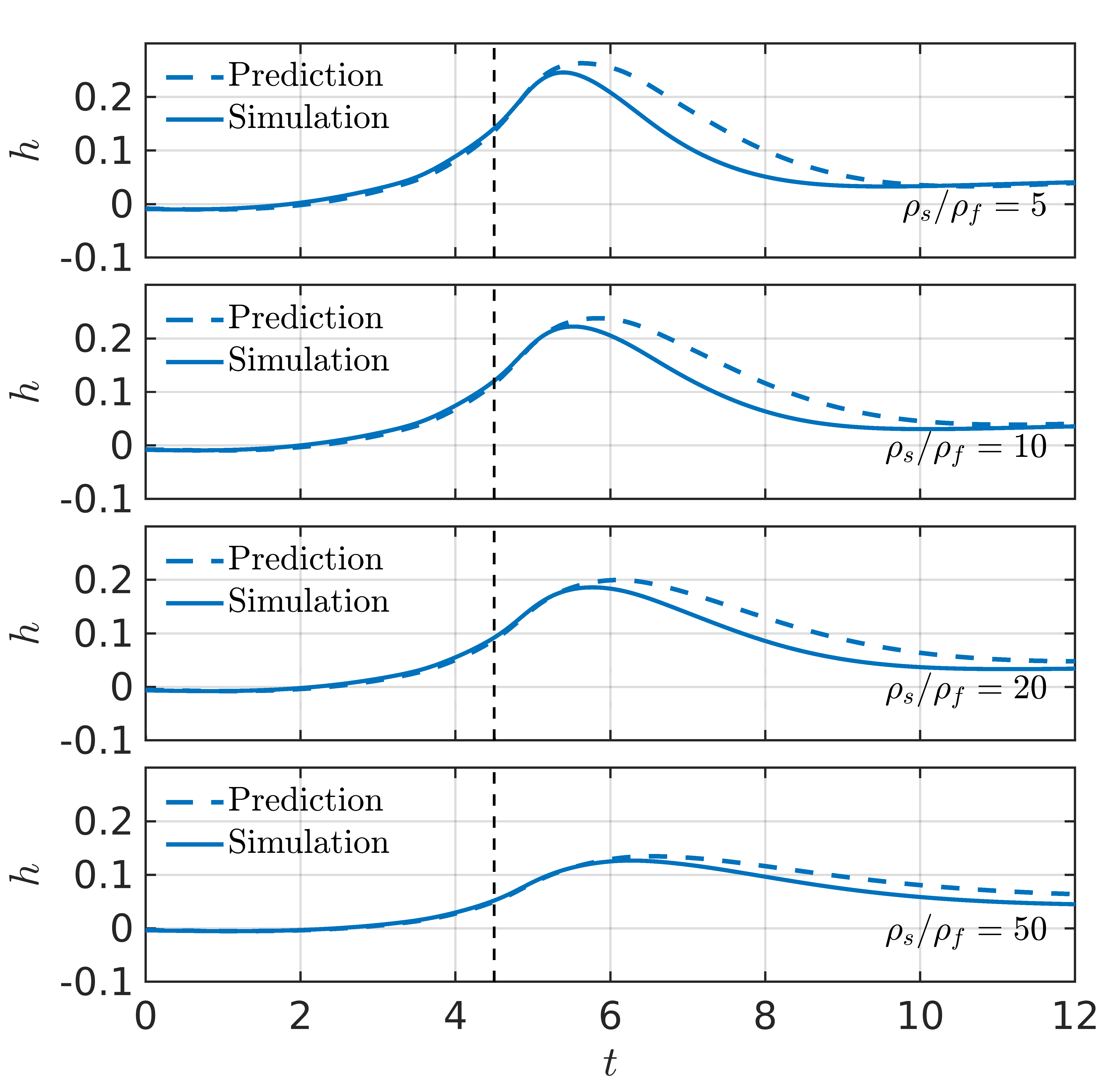}
        \caption{CCW predictions}
	\end{subfigure}

\caption{Predicted vs. actual heave trajectory for cases at different density ratio. Dashed lines correspond to instance of vortex impingement.}
\label{f:pred_DR}	
\end{figure}

Fig.~\ref{f:pred_DR} compares the predicted and simulated heave trajectories for the CW and CCW direct-impingement interactions. The maximum heave amplitude decreases with increasing density ratio for both vortex orientations; in the simulations, it decreases from approximately \(0.25\) at \(\rho_s/\rho_f=5\) to \(0.13\) at \(\rho_s/\rho_f=50\). The prediction captures this decreasing trend, but slightly overpredicts the amplitude by \(8\%\)--\(13\%\) for CW interactions and \(6\%\)--\(7\%\) for the CCW interactions. 
Although heave amplitude is expected to decrease with increasing density ratio, the reduction is far smaller than would be expected from a simple inverse scaling with density ratio or effective mass. This is because the force driving the heave motion is itself influenced by the airfoil motion. In particular, even before impingement, the induced-angle-of-attack term contributes substantially to the lift and, as shown in the previous subsection, and dominates over the added-mass term during this stage. The heave dynamics therefore cannot be interpreted simply as the response of different masses to the same prescribed forcing.

After the first heave extremum, both the predicted and simulated trajectories exhibit rebound behavior. However, the rebound is generally weaker in the prediction, leading to an increase in the discrepancy between the predicted and simulated heave. This behavior is consistent with the post-impingement lift comparison discussed above, where the modeled lift captures the basic rebound tendency but does not fully reproduce the additional lift contribution associated with induced vortex shedding. At later times, the discrepancy between the predicted and simulated heave trajectories tends to decrease. This later convergence is also consistent with the lift comparison. After the main vortical structures have left the airfoil surface, the simulated lift curve changes more rapidly, whereas the modeled lift remains offset for a longer time. The resulting change in the sign of the force discrepancy partially compensates for the earlier error in the predicted heave response.
\par
The prediction error at impingement decreases as the density ratio increases with the largest error occurring for the lowest density ratio CW interaction, with \(|h_{model}-h_{sim}| = 0.021\). The error decreases to about \(0.005\) for the highest density ratio CW interaction, and for the CCW interactions, all remain below \(0.005\) across all density ratios tested. Because the relative error also decreases with density ratio, this trend is not only a consequence of the weaker heave amplitude, but also suggests that higher density ratio leads to weaker coupling between the incoming vortex and the airfoil motion before impingement. 

For all density ratios considered, the prediction remains very close to the actual heave trajectory until approximately the stage at which the vortex reaches the leading edge, demonstrating that the early-time response is captured robustly by the present quasi-steady framework. Recall that the $C_{L}$ of the non direct-impingement CW2.5a and CW2.5b, in which the vortex passes above and below the airfoil, remain nearly coincident with the corresponding direct-impingement case before the vortex reaches the leading-edge region (as shown in Fig.~\ref{f:hclt}). Similar behavior is also observed for CCW2.5a and CCW2.5b. This suggests that the present framework provides a useful approximation for close-interaction cases even when direct impingement does not ultimately occur. Provided that the vortex trajectory is already known, the framework can therefore indicate whether impingement is likely to occur as the airfoil repositions itself due to the induced velocity.

\section{Conclusion}

In this study, we numerically investigated the interaction between an isolated vortex gust and a freely-flying airfoil at \(Re=1000\). Comparisons with corresponding stationary airfoil interactions were utilized to isolate the effects of airfoil motion and to develop a framework for predicting the heave trajectory.
\par
Before impingement, the freely-flying airfoil moves in the direction induced by the incoming vortex, upward for a CCW vortex and downward for a CW vortex. This pre-impingement response is governed primarily by vortex rotation direction and is only weakly affected by angle of attack or vortex transverse position. After impingement, vortex position has a stronger influence, while the angle-of-attack is secondary. The heave motion generally reaches an extremum shortly after impingement and then rebounds, which is associated with a rapid change in lift, a post-impingement vortex--shear layer interaction, and vortex shedding.
\par
For the direct-impingement freely-flying airfoil, the motion induced terms play a dominant role in the transient lift response. In particular, they strongly attenuate the first lift peak, while only weakly attenuating the second peak, thereby contributing to a strong rebound in heave position. Flow-field analysis demonstrated that vortex shedding also makes an important contribution to the rebound, strengthening it over a time interval of approximately \(1\) to \(1.5\) convective time units. Two shedding mechanisms were identified. The first is associated with the counterpart of the impinging vortex, whereas the second occurs later and is linked to the accumulation of vorticity. Comparison shows that, relative to the stationary airfoil, the freely-flying airfoil tends to suppress the first mechanism and promote the second. A larger angle of attack was also found to promote vortex formation on the suction side and thereby intensify the associated boundary-layer separation. For cases without direct impingement, the vortex exerts only limited influence on the opposite side of the airfoil and affects the suction side more strongly.
\par
Finally, a model was introduced that predicts the heave trajectory of a freely-flying airfoil in response to a vortex gust, assuming that one knows the equivalent stationary airfoil-gust interaction lift coefficient. This is performed by superimposing the motion induced lift terms, and solving for the new equation of motion. The prediction slightly overestimates the maximum heave, but comparisons across different density ratios show excellent agreement prior to impingement. After impingement, the model captures the rebound with a slight offset.  The full rebound is underestimated likely due to the the motion induced changes in vortex shedding that are not captured by the model. 

More broadly, once the stationary lift of a gust interaction is available, the proposed model can predict the heave motion efficiently across different density ratios, without the need for coupled FSI simulations. The approach may also be extended to other vortex-gust conditions, Reynolds numbers, or body geometries by updating the relevant motion induced parameters, such as the effective lift slope and added-mass coefficient. Improved agreement may be achieved by incorporating transient aerodynamic effects or by modeling the modified vortex shedding that develops after impingement, although such extensions would greatly increase the complexity of the model.

\appendix
\section{Details on the kinematic profile of the vortex-generating airfoil}
\label{app:vortex_generation}
The motion of the vortex-generating airfoil is prescribed in terms of the pitch angle \(\theta(t)\), with the airfoil pitching about the quarter-chord point, and the effective angle of attack \(\alpha_{\mathrm{eff}}(t)\),
\begin{equation}
\alpha_{\mathrm{eff}}(t) = \theta(t) - \arctan\left(\frac{\dot{h}_{\mathrm{v}}(t)}{U_\infty}\right),
\label{eqn:alphaeff}
\end{equation}
where \(\dot{h}_{\mathrm{v}}(t)\) is the heave velocity and \(U_\infty\) is the freestream velocity. This relation may be inverted to determine the heave velocity \(\dot{h}_{\mathrm{v}}(t)\), from which the heave displacement \(h_{\mathrm{v}}(t)\) is obtained by time integration.
\par
The temporal evolution of \(\theta(t)\) and \(\alpha_{\mathrm{eff}}(t)\) is prescribed through a sequence of smooth variations,
\begin{equation}
\label{eqn:vortex_profile_theta}
\theta(t) = \theta_0 \, r(t; t_{s1},t_{d1}) + \Delta \theta \, r(t; t_{s2},t_{d2}) - (\theta_0 + \Delta \theta) \, r(t; t_{s3},t_{d3}),
\end{equation}
\begin{equation}
\label{eqn:vortex_profile_alpha}
\alpha_{\mathrm{eff}}(t) = \alpha_{\mathrm{eff},0} \, r(t; t_{s1},t_{d1}) + \Delta \alpha_{\mathrm{eff}} \, r(t; t_{s2},t_{d2}) - (\alpha_{\mathrm{eff},0} + \Delta \alpha_{\mathrm{eff}}) \, r(t; t_{s3},t_{d3}),
\end{equation}
where \(r(t; t_s, t_d)\) is a smooth ramp function defined as
\begin{equation}
r(t; t_s,t_d) = \frac{1}{2} \left[ 1 + \tanh\left(K\left(\frac{t - t_s}{t_d} - \frac{1}{2}\right)\right) \right].
\end{equation}
Here, \(t_s\) and \(t_d\) denote the start time and duration of each variation, respectively, and \(K = 2\,\mathrm{arctanh}(0.975)\) ensures that 97.5\% of each variation is completed within \(t_d\).
\par
The rapid variation over the interval \((t_{s2}, t_{d2})\) corresponds to the vortex-generating maneuver, during which the effective angle of attack changes most abruptly and determines the strength of the resulting vortex. As shown in \citet{yan2026generation}, the vortex strength is primarily governed by \(\Delta \theta\) and \(\Delta \alpha_{\mathrm{eff}}\), while the timing parameter \(t_{s2}\) shifts the transverse position of the vortex with minimal influence on its structure. In the present study, \(\Delta \theta\) and \(\Delta \alpha_{\mathrm{eff}}\) are held fixed within each group of cases corresponding to clockwise and counterclockwise vortices, and only the timing parameters are varied to control the vortex location. The full set of motion parameters used to generate the vortex in each case is provided in Tab.~\ref{tab:detailed_param}.
\begin{table}
\centering
\begin{tabular}{lcccccccccc}
\toprule
Label
& $t_{\mathrm{s1}}$
& $t_{\mathrm{d1}}$
& $t_{\mathrm{s2}}$
& $t_{\mathrm{d2}}$
& $t_{\mathrm{s3}}$
& $t_{\mathrm{d3}}$
& $\theta_0$
& $\alpha_{\mathrm{eff},0}$
& $\Delta\theta$
& $\Delta\alpha_{\mathrm{eff}}$ \\
\hline
CW0       & 3.00 & 1.50 & 6.49 & 0.35 & 8.90  & 1.50 & 45 & 5    & $-15$ & $-15$ \\
CW2.5     & 3.00 & 1.50 & 6.46 & 0.35 & 8.90  & 1.50 & 45 & 5    & $-15$ & $-15$ \\
CW2.5a    & 3.00 & 1.50 & 6.66 & 0.35 & 8.90  & 1.50 & 45 & 5    & $-15$ & $-15$ \\
CW2.5b    & 3.00 & 1.50 & 6.32 & 0.35 & 8.90  & 1.50 & 45 & 5    & $-15$ & $-15$ \\
CW5       & 3.00 & 1.50 & 6.42 & 0.35 & 8.90  & 1.50 & 45 & 5    & $-15$ & $-15$ \\
\hline
CCW0      & 3.00 & 1.50 & 8.33 & 0.30 & 10.60 & 1.50 & 30 & $-5$ & 15 & 15 \\
CCW2.5    & 3.00 & 1.50 & 8.30 & 0.30 & 10.60 & 1.50 & 30 & $-5$ & 15 & 15 \\
CCW2.5a   & 3.00 & 1.50 & 8.50 & 0.30 & 10.60 & 1.50 & 30 & $-5$ & 15 & 15 \\
CCW2.5b   & 3.00 & 1.50 & 8.05 & 0.30 & 10.60 & 1.50 & 30 & $-5$ & 15 & 15 \\
CCW5      & 3.00 & 1.50 & 8.27 & 0.30 & 10.60 & 1.50 & 30 & $-5$ & 15 & 15 \\
\hline
CW0-S     & 3.00 & 1.50 & 6.69 & 0.35 & 8.90  & 1.50 & 45 & 5    & $-15$ & $-15$ \\
CW2.5-S   & 3.00 & 1.50 & 6.66 & 0.35 & 8.90  & 1.50 & 45 & 5    & $-15$ & $-15$ \\
CW5-S     & 3.00 & 1.50 & 6.65 & 0.35 & 8.90  & 1.50 & 45 & 5    & $-15$ & $-15$ \\
\hline
CCW0-S    & 3.00 & 1.50 & 8.09 & 0.30 & 10.60 & 1.50 & 30 & $-5$ & 15 & 15 \\
CCW2.5-S  & 3.00 & 1.50 & 8.06 & 0.30 & 10.60 & 1.50 & 30 & $-5$ & 15 & 15 \\
CCW5-S    & 3.00 & 1.50 & 8.00 & 0.30 & 10.60 & 1.50 & 30 & $-5$ & 15 & 15 \\
\bottomrule
\end{tabular}
\caption{Kinematic parameters in Eqns.~\ref{eqn:vortex_profile_theta}
and~\ref{eqn:vortex_profile_alpha} for all vortex--airfoil configurations.
Angles are given in degrees.}
\label{tab:detailed_param}
\end{table}

\section{Code Validation}
\label{app:validation}
\begin{figure}
\centering
\includegraphics[width=0.35\textwidth]{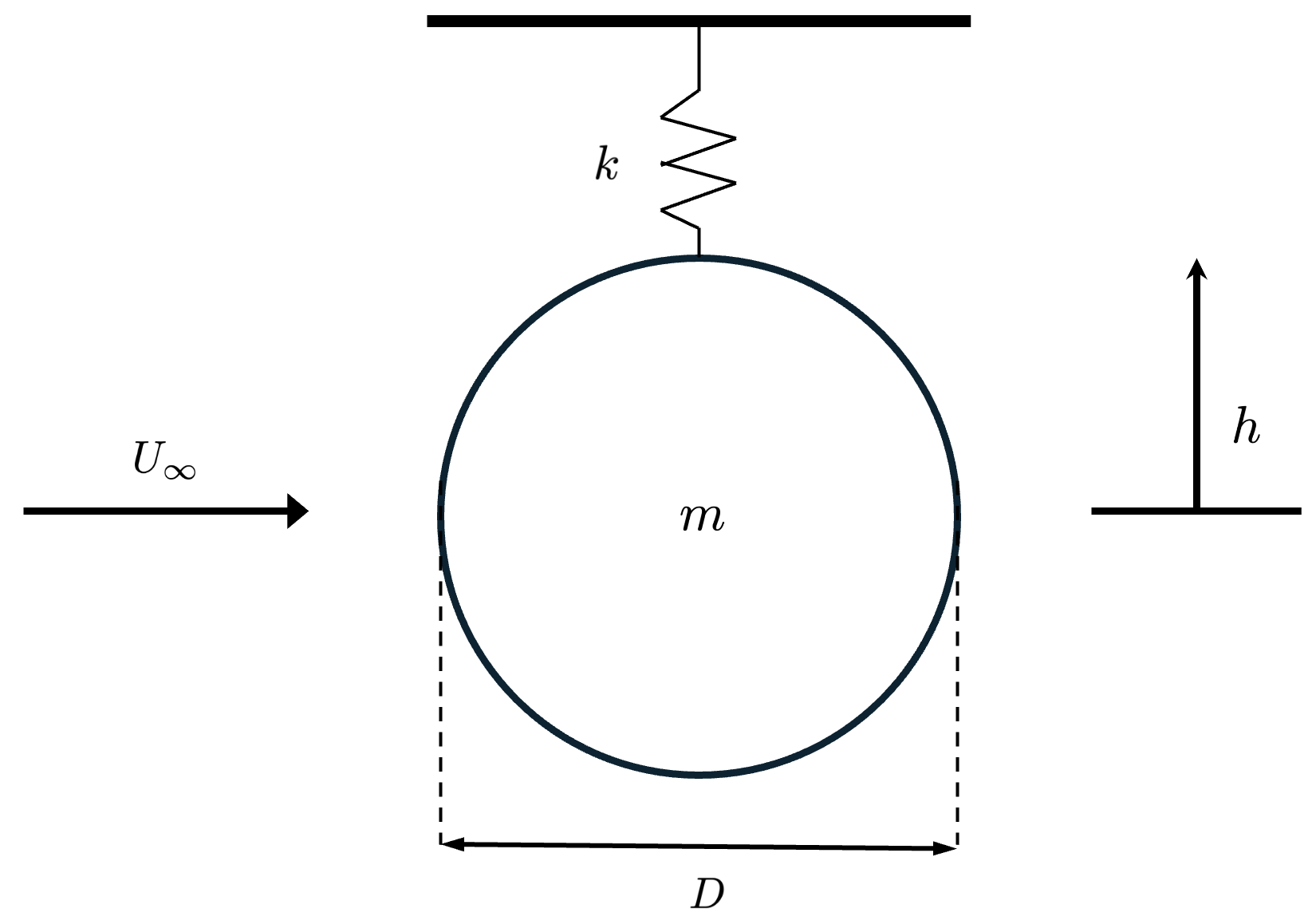}
\caption{Schematic of the FSI validation demonstrating an elastically mounted heaving circular cylinder in incoming flow.}
\label{fig:sch_cy}
\end{figure}

To verify the FSI implementation, the flow-induced vibration of an elastically mounted heaving circular cylinder at a diameter-based Reynolds number of 100 is simulated, as illustrated in Fig.~\ref{fig:sch_cy}. Tab.~\ref{tab:veri} compares the computed nondimensional heaving amplitudes and frequencies against reference results from~\citet{shiels2001flow}. For all cases, the current amplitudes are within $4\%$ of the established results, and the current frequencies are within $3\%$ of the established results, confirming the reliability of the coupled solver.
\begin{table}
\centering
\begin{tabular}{ccccccc}
\toprule
Case & $m^*$ & $k^*$ & $A^*_{\text{present}}$ & $A^*_{\text{ref}}$ & $f^*_{\text{present}}$ & $f^*_{\text{ref}}$ \\
\hline
1    & 4.0   & 0.00  & 0.05                 & 0.05              & 0.157                 & 0.158 \\
2    & 5.0   & 4.74  & 0.47                 & 0.46              & 0.153                 & 0.156 \\
3    & 2.5   & 4.96  & 0.56                 & 0.58              & 0.191                 & 0.196 \\
4    & 5.0   & 8.74  & 0.57                 & 0.57              & 0.192                 & 0.194 \\
\bottomrule
\end{tabular}
\caption{
Comparison of nondimensional heaving amplitude \(A^*\) and frequency \(f^*\) between the present results and those reported by Shiels et al.~\cite{shiels2001flow}. The mass ratio is defined as \(m^* = m/(\rho D^2/2)\), the stiffness as \(k^* = k/(\rho U_{\infty}^2/2)\), the amplitude as \(A^* = A/D\), and the frequency as \(f^* = fD/U_{\infty}\), where \(m\) is the mass of cylinder per unit depth, \(k\) is the linear stiffness per unit depth, \(\rho\) is the fluid density, \(D\) is the cylinder diameter, and \(U_{\infty}\) is the freestream velocity. There is no structural damping for all cases.
}
\label{tab:veri}
\end{table} 
\section*{Acknowledgments}
This work is funded by AFOSR Award Number: FA9550-23-1-0478. Computational time is provided by the Center for High-Throughput Computing (CHTC) at UW-Madison. The authors thank Kenny Breuer and Eric Handy-Cardenas for their insights and complementary experimental work on vortex-gust interactions that inspired much of this work.

\bibliographystyle{jfm}
\bibliography{vorticalgust_freelyflying}

@article{jones2020gust,
  title={Gust encounters of rigid wings: Taming the parameter space},
  author={Jones, Anya R},
  journal={Physical Review Fluids},
  volume={5},
  number={11},
  pages={110513},
  year={2020},
  publisher={APS}
}

@article{barnes2018counterclockwise,
  title={Counterclockwise vortical-gust/airfoil interactions at a transitional Reynolds number},
  author={Barnes, Caleb J and Visbal, Miguel R},
  journal={AIAA Journal},
  volume={56},
  number={7},
  pages={2540--2552},
  year={2018},
  publisher={American Institute of Aeronautics and Astronautics}
}

@article{barnes2018clockwise,
  title={Clockwise vortical-gust/airfoil interactions at a transitional Reynolds number},
  author={Barnes, Caleb J and Visbal, Miguel R},
  journal={AIAA Journal},
  volume={56},
  number={10},
  pages={3863--3874},
  year={2018},
  publisher={American Institute of Aeronautics and Astronautics}
}

@article{martinez2020analysis,
  title={Analysis of vortical gust impact on airfoils at low Reynolds number},
  author={Mart{\'\i}nez-Muriel, Cayetano and Flores, O},
  journal={Journal of Fluids and Structures},
  volume={99},
  pages={103138},
  year={2020},
  publisher={Elsevier}
}

@article{peng2015vortex,
  title={Vortex dynamics during blade-vortex interactions},
  author={Peng, Di and Gregory, James W},
  journal={Physics of Fluids},
  volume={27},
  number={5},
  year={2015},
  publisher={AIP Publishing}
}

@article{hufstedler2019vortical,
  title={Vortical gusts: experimental generation and interaction with wing},
  author={Hufstedler, Esteban AL and McKeon, Beverley J},
  journal={AIAA Journal},
  volume={57},
  number={3},
  pages={921--931},
  year={2019},
  publisher={American Institute of Aeronautics and Astronautics}
}

@article{chen2020aeroelastic,
  title={Aeroelastic interactions and trajectory selection of vortex gusts impinging upon Joukowski airfoils},
  author={Chen, Huansheng and Jaworski, Justin W},
  journal={Journal of Fluids and Structures},
  volume={96},
  pages={103026},
  year={2020},
  publisher={Elsevier}
}

@article{barnes2018gust,
  title={Gust response of rigid and elastically mounted airfoils at a transitional Reynolds number},
  author={Barnes, Caleb J and Visbal, Miguel R},
  journal={Aerospace Science and Technology},
  volume={74},
  pages={112--119},
  year={2018},
  publisher={Elsevier}
}

@article{shiels2001flow,
  title={Flow-induced vibration of a circular cylinder at limiting structural parameters},
  author={Shiels, D and Leonard, A and Roshko, A},
  journal={Journal of Fluids and Structures},
  volume={15},
  number={1},
  pages={3--21},
  year={2001},
  publisher={Elsevier}
}

@techreport{theodorsen1949general,
  title={General theory of aerodynamic instability and the mechanism of flutter},
  author={Theodorsen, Theodore},
  year={1949}
}

@article{yan2026generation,
  title={Generation of an isolated vortex gust through a heaving and pitching foil},
  author={Yan, Bingfei and Handy-Cardenas, Eric and Breuer, Kenny and Franck, Jennifer A},
  journal={arXiv preprint arXiv:2603.21336},
  year={2026}
}

@article{barnes2020angle,
  title={Angle of attack and core size effects on transitional vortical-gust--airfoil interactions},
  author={Barnes, Caleb J and Visbal, Miguel R},
  journal={AIAA Journal},
  volume={58},
  number={7},
  pages={2881--2898},
  year={2020},
  publisher={American Institute of Aeronautics and Astronautics}
}

@article{fukami2025extreme,
  title={Extreme vortex-gust airfoil interactions at Reynolds number 5000},
  author={Fukami, Kai and Smith, Luke and Taira, Kunihiko},
  journal={Physical Review Fluids},
  volume={10},
  number={8},
  pages={084703},
  year={2025},
  publisher={APS}
}

@article{peng2017asymmetric,
  title={Asymmetric distributions in pressure/load fluctuation levels during blade-vortex interactions},
  author={Peng, Di and Gregory, James W},
  journal={Journal of Fluids and Structures},
  volume={68},
  pages={58--71},
  year={2017},
  publisher={Elsevier}
}

@article{qian2022interaction,
  title={Interaction of quasi-two-dimensional vortical gusts with airfoils, unswept and swept wings},
  author={Qian, Yuanzhi and Wang, Zhijin and Gursul, Ismet},
  journal={Experiments in Fluids},
  doi={10.1007/s00348-022-03477-8},
  volume={63},
  number={8},
  pages={124},
  year={2022},
  publisher={Springer}
}

@article{biler2021experimental,
  title={Experimental investigation of transverse and vortex gust encounters at low Reynolds numbers},
  author={Biler, H{\"u}lya and Sedky, Girguis and Jones, Anya R and Saritas, Murat and Cetiner, Oksan},
  journal={AIAA Journal},
  doi={10.2514/1.J059658},
  volume={59},
  number={3},
  pages={786--799},
  year={2021},
  publisher={American Institute of Aeronautics and Astronautics}
}

@article{wang2024airfoil,
  title={Airfoil response to periodic vertical and longitudinal gusts},
  author={Wang, Tong and Feng, Li-Hao and Cao, Yu-Tian and Wang, Jin-Jun},
  journal={Journal of Fluid Mechanics},
  volume={979},
  pages={A35},
  year={2024},
  publisher={Cambridge University Press}
}

@article{ma2021unsteady,
  title={The unsteady lift of an oscillating airfoil encountering a sinusoidal streamwise gust},
  author={Ma, Ruwei and Yang, Yang and Li, Mingshui and Li, Qiusheng},
  journal={Journal of Fluid Mechanics},
  volume={908},
  pages={A22},
  year={2021},
  publisher={Cambridge University Press}
}

@article{goldstein1976complete,
  title={A complete second-order theory for the unsteady flow about an airfoil due to a periodic gust},
  author={Goldstein, Marvin E and Atassi, H},
  journal={Journal of Fluid Mechanics},
  volume={74},
  number={4},
  pages={741--765},
  year={1976},
  publisher={Cambridge University Press}
}

@article{andreu2023controlling,
  title={Controlling upwards and downwards gust loads on aerofoils by pitching},
  author={Andreu-Angulo, Ignacio and Babinsky, Holger},
  journal={Experiments in Fluids},
  volume={64},
  number={7},
  pages={129},
  year={2023},
  publisher={Springer}
}

@techreport{greenberg1947airfoil,
  title={Airfoil in sinusoidal motion in a pulsating stream},
  author={Greenberg, J Mayo},
  year={1947}
}

@article{ribeiro2021wake,
  title={Wake-foil interactions and energy harvesting efficiency in tandem oscillating foils},
  author={Ribeiro, Bernardo Luiz R and Su, Yunxing and Guillaumin, Quentin and Breuer, Kenneth S and Franck, Jennifer A},
  journal={Physical Review Fluids},
  doi={10.1103/PhysRevFluids.6.074703},
  volume={6},
  number={7},
  pages={074703},
  year={2021},
  publisher={APS}
}

\end{document}